\pdfoutput=1

\documentclass[a4paper,11pt]{article}
\usepackage[affil-sl,auth-sc]{authblk}
\usepackage{amssymb,amsmath,amsbsy}
\usepackage{mathrsfs}
\usepackage{mathbbol}
\usepackage{dsfont}
\usepackage{appendix}
\usepackage{hyperref}
\usepackage{pinlabel}
\usepackage[top=1.2in, bottom=1.2in,left=0.9in,includefoot]{geometry}               % See geometry.pdf to learn the layout options. There are lots.
\usepackage{graphicx}
\usepackage{cite}
\usepackage{float}
\usepackage{caption}
\usepackage{subfig}
\usepackage{color}
\usepackage{axodraw}

\def\theequation{\arabic{section}.\arabic{equation}}
\def\eq#1{eq.~(\ref{#1})}
\def\Eq#1{Eq.~(\ref{#1})}

\def\eqs#1#2{eqs.~(\ref{#1}) and (\ref{#2})}
\def\eqss#1#2#3{eqs.~(\ref{#1}), (\ref{#2}) and (\ref{#3})}

\def\Ref#1{ref.~\cite{#1}}
\def\Rref#1{Ref.~\cite{#1}}
\def\Refs#1{refs.~\cite{#1}}
\newcommand{\bra}[1]{\langle #1|}
\newcommand{\ket}[1]{|#1\rangle}

\newcommand{\ie}{{\it i.e., }}

\title{Doublet-Triplet Fermionic Dark Matter}
\author{Athanasios Dedes\footnote{email: {\tt adedes@cc.uoi.gr}}~  and 
Dimitrios Karamitros\footnote{email: {\tt dkaramit@cc.uoi.gr}}}
\affil{\small Department of Physics, Division of Theoretical Physics, \\
 University of Ioannina,  GR 45110, Greece}
\date{\today}                                           % Activate to display a given date or no date
\begin{document}

\maketitle

\begin{abstract}
%Inspired from the Dark Matter (DM) sector structure of the Minimal Supersymmetric
%Standard Model (MSSM) 
We extend the Standard Model (SM) by adding a pair of fermionic $SU(2)$-doublets
with opposite hypercharge and a fermionic $SU(2)$-triplet with zero hypercharge. We impose a
discrete ${Z}_{2}$-symmetry that distinguishes the SM fermions from the new ones. 
Then, gauge invariance allows for two renormalizable Yukawa couplings
between the new fermions and the SM Higgs field, as well as for
direct masses for the doublet ($M_{D}$) and the triplet ($M_{T}$). 
After electroweak symmetry breaking, this model contains, in addition to  SM particles,
two  charged Dirac fermions and 
a set of three neutral Majorana fermions, the lightest of which 
 contributes to Dark Matter (DM).   
We consider a case where the lightest neutral fermion is  an equal admixture 
of the two doublets with mass $M_{D}$ close to the $Z$-boson mass. 
This state remains stable under radiative corrections
thanks to a custodial $SU(2)$-symmetry and 
is consistent with the experimental data from oblique electroweak corrections.
Moreover, the  amplitudes relevant to spin-dependent or independent 
nucleus-DM particle scattering cross section {\em both} vanish at tree level. They 
arise at one loop at a level that may be observed in near future DM 
direct detection experiments.
For Yukawa couplings comparable to the top-quark,
the DM particle relic abundance is consistent with observation, not relying on co-annihilation or resonant effects and has a  mass at the electroweak scale.
Furthermore, the heavier fermions decay to the DM particle and to electroweak gauge
bosons making this model easily testable at the LHC. 
In the regime of interest, 
the charged fermions suppress the Higgs decays to diphoton by 45-75\%
relative to SM prediction.

\end{abstract}

\newpage
%%%%%%%%%%%%%%%%%%%%%%%%%%%%%%%%%
\section{Introduction}
\label{intro}

Motivated by  astrophysical observations that suggest the existence of 
Dark Matter~\cite{Bertone:2004pz}, we would like 
to propose a  model with 
a fermionic Weakly Interacting Massive Particle (WIMP)
 ($\chi_{1}^{0}$)  whose mass and couplings
are directly associated to electroweak scale providing the universe with 
the right thermal relic density abundance,  not ``tuned''
by co-annihilation or  resonance effects. 
Today, as opposed to five years ago, attempts of this
sort immediately face difficulties due to strong experimental 
bounds~\cite{XENON100,LUX}\footnote{There are of course tantalising hints 
from DAMA, CoGeNT, CRESST-II and CDMS-Si experiments but these face stringent
constraints from recent null result experiments like XENON100 and LUX making puzzling 
any theoretical interpretation of them all. For a recent review, see \Ref{Panci:2014gga}.}  
from direct 
searches on nucleus recoiling energy in  WIMP-nucleus scattering 
processes~\cite{Goodman:1984dc}.
As a result,    $Z$- and Higgs- boson couplings to  $\chi_{1}^{0}$-pairs are strongly
constrained  and usually come into conflict with values of couplings required from
the observed~\cite{Ade:2013zuv} DM relic abundance.
We therefore seek for a model at which, at least at tree level, 
these couplings vanish by a symmetry and at the same time 
the  observed relic density is reproduced.  
We then discuss further consequences 
of this idea at Large Hadron Collider (LHC).

We consider a minimal model which realises this situation, hence,
in addition to Standard Model (SM) particles, we add  a pair of Weyl-fermion doublets 
$\mathbf{\bar{D}_{1} \sim (1^{c}, 2)_{-1}}$ and 
$\mathbf{\bar{D}_{2} \sim (1^{c}, 2)_{+1}}$ with opposite hypercharges,  
and a Weyl-fermion triplet $\mathbf{T\sim (1^{c}, 3)_{0}}$ with zero hypercharge.
The new Yukawa interactions allowed by gauge invariance and renormalizability 
are given by\footnote{All gauge group indices are suppressed in this equation. 
Its detailed form 
is given below in \eq{LDM}.} 
%%%%%%%%%%%%%%%%%
\begin{eqnarray}
\mathcal{L}_{\mathrm{Yuk}} \ \supset \ 
Y_{1}\: \mathbf{T\, H\, \tau \,  \bar{D}_{1}} \ + \  Y_{2}\:  \mathbf{T\, H^{\dagger}\ \tau \, \bar{D}_{2}}
 \ - \  M_{D}\:  \mathbf{\bar{D}_{1}\, \bar{D}_{2}} \  -  \ \frac{1}{2}\, M_{T} \: \mathbf{T \,T} \;,
 \label{lag1}
\end{eqnarray}
%%%%%%%%%%%%%%%%%%%
with $\mathbf{\tau}$ being the Pauli matrices. A $Z_{2}$-discrete parity symmetry 
has been employed
to guarantee that the new fermions interact always in pairs. 
Clearly, $\mathcal{L}_{\mathrm{Yuk}}$ is invariant 
under the interchange symmetry  $\mathbf{H\leftrightarrow H^{\dagger}}$ and 
$\mathbf{\bar{D}_{1} \leftrightarrow \bar{D}_{2}}$ 
when $Y_{1} = Y_{2}\equiv Y$.  Then, it is very easy to see that in this limit, one eigenvalue 
with mass $M_{D}$,
of the neutral ($3\times 3$) mixing mass 
matrix,  decouples from the two heavier ones and the latter
is degenerate with the two eigenvalues of the ($2\times 2$) charged fermion
mass matrix.
At tree level approximation, except for the lightest neutral fermion ($\chi^{0}_{1}$), all other 
masses are controlled by the Yukawa coupling $Y$.  
The state with $m_{\chi_{1}^{0}} = M_{D}$ is our DM candidate particle. This 
particle state contains an equal admixture of the two doublets 
 but has \emph{no} triplet component, 
 %(in supersymmetry (SUSY) jargon it is a pure higgsino state),
%%%%%%%%%%%%%%%%
\begin{eqnarray} 
\ket{\chi_{1}^{0}} = 0 \cdot \ket{\mathbf{T}} \ + \ \frac{1}{\sqrt{2}}\: \ket{\mathbf{\bar{D}_{1}}} \ + \  
\frac{1}{\sqrt{2}}\: \ket{\mathbf{\bar{D}_{2}}} \;.
\end{eqnarray}
%%%%%%%%%%%%%%%%%%
Because the neutral component of the triplet does not participate in $\ket{\chi_{1}^{0}}$,
the latter does not couple to the Higgs boson at tree level. It does not couple to the
$Z$-gauge boson neither because of its equal admixture of neutral particles with opposite weak isospin.
The situation here is analogous to the custodial symmetry~\cite{Sikivie:1980hm} 
imposed in strongly coupled EW scenarios, where 
the ``custodian'' new particles are inserted in a similar way to protect certain quark-gauge 
boson couplings to obtain large radiative corrections~\cite{delAguila:2010es,Agashe:2006at,SekharChivukula:2009if,Carmona:2013cq}. 

\begin{figure}[t]
   \centering
   \includegraphics[height=3in]{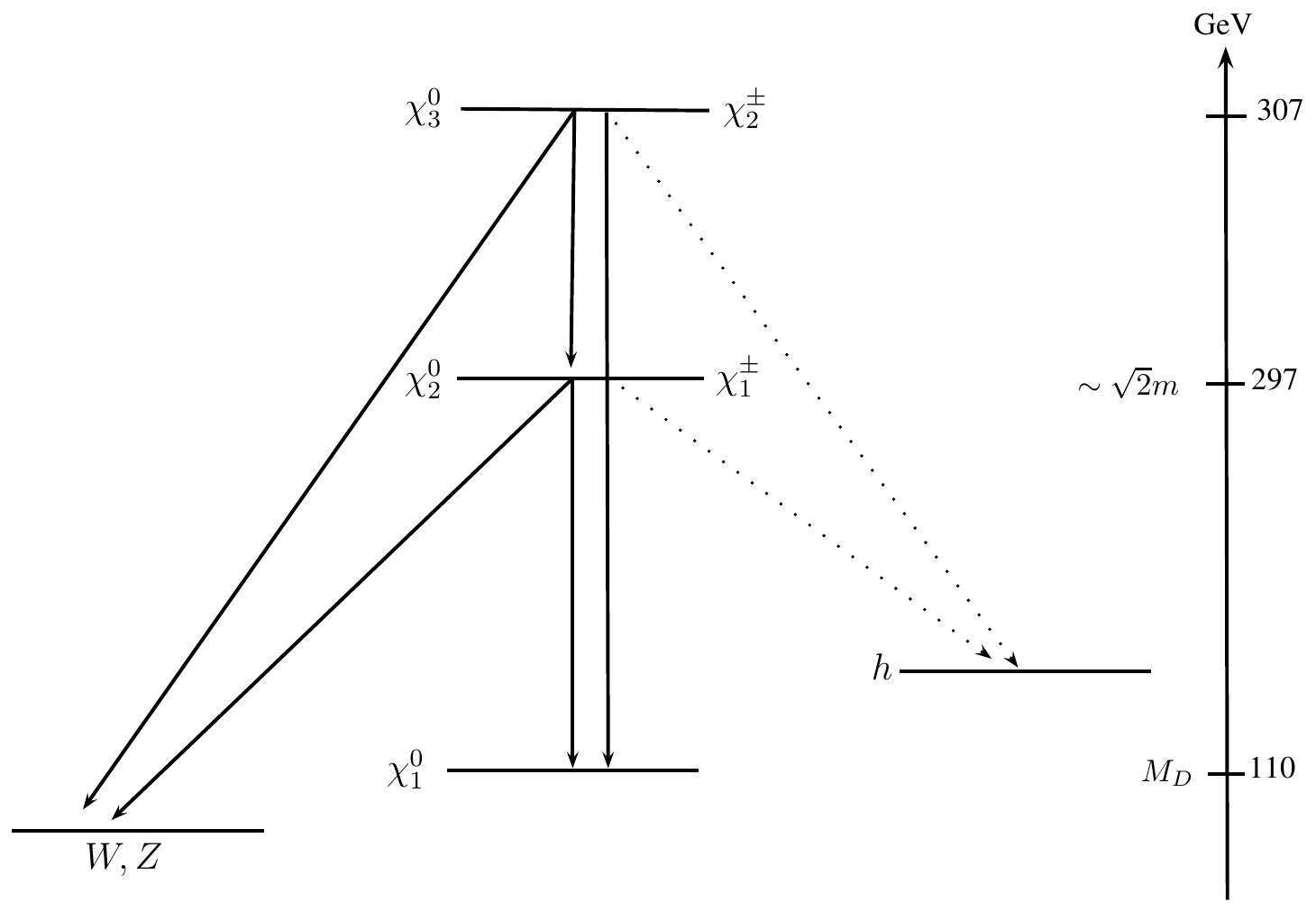} % requires the graphicx package
   \caption{\sl A sketch for the mass spectrum and decays of the new physical 
    doublet and triplet fermions. The lightest neutral particle, $\chi_{1}^{0}$ is an equal admixture
    of the two doublets and has mass $M_{D}$. Particles $\chi_{2}^{0}$ ($\chi_{3}^{0}$) 
    and $\chi_{1}^{\pm}$ ($\chi_{2}^{\pm}$) are
    mass degenerate. For the spectrum masses written to the right 
    we have chosen $M_{D}=110$ GeV, $M_{T}=100$ GeV and $m=Y v=200$ GeV. 
    It provides the
     correct relic density abundance for dark matter [see Section~\ref{sec:relic}] and is currently 
     about $\sim 10$ times less  sensitive to  current direct detection searches
      [see Section~\ref{sec:direct}].}
   \label{fig:spec}
\end{figure}

The couplings $h \chi_{1}^{0} \chi_{1}^{0}$ and $Z \chi_{1}^{0} \chi_{1}^{0}$ 
vanish at tree level, and as a result there are no $s$-channel amplitudes contributing 
to the annihilation cross section.
However, there  are off-diagonal interactions such as e.g., $Z \chi_{1}^{0} \chi_{2}^{0}$
that render the $t,u$-channel amplitudes non-zero but yet suppressed enough
to obtain the right relic density $\Omega_{\chi}$ 
for $M_{D}  \approx 100$ GeV and $Y \approx 1$.  
Roughly speaking, the  spectrum of the model  where this happens 
is shown schematically in Fig.~\ref{fig:spec}.
%%%%%%%%%%%%%%%%%%%%%%%%%%%%%%%%%%%%%%
%%%%%%%%%%%%%%%%%%%%%%%%%%%%%%%%%%%%%%%
Typically, the lightest stable new particle ($m_{\chi_{1}^{0}} \approx 110$ GeV)
is in the vicinity of the EW scale
while all other neutral and charged fermions are above  $m \equiv Y v$ 
which is  taken around the top quark mass. The splitting of the charged
fermions is also controlled by the triplet mass ($M_{T}$). Therefore, the
 parameters of the model are just three: $M_{D}, M_{T}$ and $m$.

  Naively, one may think that this model is similar to the ``wino-higgsino''
  sector of the MSSM~\cite{HK} 
 or it is  an extended variant of  the singlet-doublet DM
model of~\Refs{Carena:2004ha,Mahbubani:2005pt,D'Eramo:2007ga,Cohen:2011ec}.
Another obvious question is, why does one want to 
introduce several new fermions, since a single one (for example the triplet, as in 
minimal DM~\cite{Cirelli:2005uq} models) suffices?
The answer to these questions arise from our wish to construct a model
with WIMP mass \emph{at the EW scale}, and hides inside the  model building details, 
 namely: 
%
%
%%%%%%%%%%%%%%%%%%%%%
\begin{enumerate}
\item The off-diagonal entries of the ``chargino'' or ``neutralino'' mass
matrix contain general Yukawa couplings ($Y_{1}$ and $Y_{2}$) that can be enhanced 
as opposed to the fixed-value gauge couplings of the MSSM.
Evenmore, they can be equal here {\it i.e., }$Y_{1}=Y_{2}\equiv Y\sim g$, satisfying a custodial symmetry, 
a realisation which is only phenomenologically allowed in the 
so called Split-SUSY scenarios~\cite{ArkaniHamed:2004fb,Giudice:2004tc}. Therefore,
this fermionic doublet-triplet DM sector generalises the corresponding DM sector of the 
Minimal Supersymmetric Standard Model (MSSM).
\item In the region where the common Yukawa coupling is comparable, say,
to the top Yukawa coupling  there are heavy charged leptons decaying to
the lightest new fermion $\chi_{1}^{0}$. This mass pattern, shown in
Fig.~\ref{fig:spec}, is different from the singlet-doublet DM model (at least
from the minimal version) where the  lightest neutral particle 
is, up to radiative corrections, 
degenerate with the charged particle a situation which is highly constrained from
long lived charged particle searches at LHC~\cite{Chatrchyan:2013oca}.
\item In the limit of equal Yukawa couplings ($Y$) in \eq{lag1},
there is a custodial $SU(2)$-symmetry that guaranties vanishing couplings at tree level
between the lightest neutral particle and the $Z$-boson ($Z\chi_{1}^{0} \chi_{1}^{0}$) 
and also to the Higgs-boson  ($h\chi_{1}^{0} \chi_{1}^{0}$).
This is a certain ``pass''  for this model, at least to leading order, from the current strong direct
detection experimental contraints~\cite{XENON,XENON100,LUX}.  Moreover, 
as we shall see below, 
$h\chi_{1}^{0} \chi_{1}^{0}$-coupling arises radiatively at one-loop order providing us 
with certain model predictions. Note that 
``blind spots'' of this kind have been studied in \Ref{Cheung:2012qy} for Split-SUSY and in 
\Ref{Cheung:2013dua} for the
singlet-doublet and singlet-triplet fermionic DM models.
\item Similar to the case here, 
the dominant annihilation channel in the higgsino DM-phase of  MSSM~\cite{Drees:1996pk}, 
is into gauge bosons. 
But in the higgsino case and due to smallness of the gauge 
coupling, the lightest charged and neutral fermion states are degenerate so
co-annihilation effects~\cite{Griest:1990kh} are very important. It turns out that, that for 
 higgsino mass $\mu\sim 100$ GeV 
the cross section 
$<\sigma v> \approx \frac{g^{4}}{16\pi \mu^{2}}$ is large which results in  $\Omega_{DM}$ that is 
too low unless  $\mu$ is in the TeV range. In the doublet-triplet fermonic DM 
model we consider here,  the lightest neutral
state decouples from the heavy ones, and in the limit of large $m= Y v$  the difference
in mass between the lightest neutral fermion and the lightest charged or the second lightest
neutral one is normally of the order of 100 GeV (see Fig.~\ref{fig:spec} for an example).
The annihilation cross section  now goes through the $t,u$-channels and, relative to higgsino case,
 is suppressed
by a factor $(m_{\chi}/m_{\chi_{j}})^{4}\sim 10-100$ 
where $m_{\chi_{j}}$ are the heavy fermion masses ($\chi_{2,3}^{0}, \chi_{1,2}^{\pm}$),
allowing a WIMP mass, $m_{\chi}$, naturally of the order
of 10-100 GeV.\footnote{In this article, 
we are only interested in DM mass of the order of the electroweak scale.}
\item Our attempt here is to find a DM candidate particle consistent with the
astrophysical and collider data but with  
mass around the electroweak scale.
Vector-like gauge multiplets that are engaged  here  have also been used to construct 
minimal DM Models (MDM) in \Rref{Cirelli:2005uq}. It has been found that the masses 
$M_{D}$ or $M_{T}$
should lie in the few-TeV region. In our scenario, it is the chiral (Dirac) mass terms  in \eq{lag1}
that play the most important role. The latter are constrained from perturbativity to 
be  several hundreds of GeV while the lower vector-like masses, $M_{D}$ and $M_{T}$, 
are protected by an accidental symmetry. Finally, the production and decay 
 phenomenology of the new fermions is very 
distinct  from the ones in MDM models and it is relatively easy to be tested 
with current and near future LHC data.   
\end{enumerate}
%%%%%%%%%%%%%%%%%%%%%%%%%%%%%%%%%%%%%%%%%%%%%
Within this framework of doublet-triplet fermionic DM model that we describe in
Section~\ref{model},
and in particular in the region where the  custodial symmetry is applied,
we discuss and check constraints that include:  
\begin{itemize}
\item An 
estimate of oblique corrections to electroweak observables  ($S,T,U$  parameters) 
[Section~\ref{sec:STU}].
\item DM thermal relic density calculation at tree level [Section~\ref{sec:relic}].
\item Direct DM detection prospects through nucleus-DM particle scattering at 1-loop
[Section~\ref{sec:direct}] 
\item Decay rate of the Higgs boson to two photons ($h\to \gamma\gamma$) 
[Section~\ref{sec:h2gg}]
%and to $Z$-boson and a photon ($h\to Z\gamma$) ({\bf to be checked})
%
\item Vacuum stability and perturbativity  [Section~\ref{sec:Landau}]
\item LHC signatures, production and decays of the new fermions.

%check decays and cross section in the process 
%$p +  p \rightarrow W^{+}  \rightarrow \chi_{1}^{+} + \chi_{2}^{0}$ [Section~\ref{sec:LHC}]
%
\end{itemize}
Our conclusions and various  ways to extend this work
are discussed in  Section~\ref{sec:conclusions}. An appendix 
with the explicit one-loop corrections to the $h\chi_{1}^{0}\chi_{1}^{0}$-vertex
is given.
Beyond the articles we have already mentioned, 
there is a reach literature regarding minimal DM extensions of the SM.
%beyond the one we have mentioned above. 
A partial list is given in 
\Refs{Pospelov:2011yp,Pal:1987mr,Araki:2011hm,Law:2012mj,Aoki:2011he,Kim:2006af,Kim:2008pp,Ma:2008cu,Hamaguchi:2008ta,Kim:2009ke,deS.Pires:2010fu,Zheng:2010js,Fukushima:2011df,Bellazzini:2011et,Chen:2011bc,Ma:2012gb,Chao:2012sz,LopezHonorez:2012kv,Yang:2012jd,Okada:2012np,Baek:2012uj}.

%%%%%%%%%%%%%%%%%%%%%%%%
\section{Model Details}
\label{model}
%%%%%%%%%%%%%%%%%%%%%%%%%%%

As a result of  what we have already mentioned in the introduction, 
we scan chiral fermion matter extensions of the SM gauge group according to
the following, rather obvious, assumptions for the new set of fermions:
\begin{enumerate}
\item they must have vectorial electromagnetic interactions,
\item they must be colour singlets with integer charges,
\item their interactions must be  gauge (and gravitational) anomaly free,
\item their masses are obtained after $SU(2)_{W}\times U(1)_{Y}$ gauge symmetry breaking,
with only the SM Higgs doublet, and
if gauge symmetry allows, directly, and
\item there is a parity symmetry, ${\mathbf Z}_{2}$, under which the SM
fermions transform as $+1$ while the new fermions
as $-1$. 
\end{enumerate}
The most minimal model, not containing pure 
singlet fields,\footnote{However, see comments below.} consists of three fields 
arranged in colour singlets and representations of $SU(2)_{W}$, 
with quantum numbers  denoted as
$\mathbf{(1^{c}, 2I_{W}+1)^{Y}_{L,R}}$, where $\mathbf{I_{W}}$ is the weak
$SU(2)_{W}$ isospin and $Y$ is the hypercharge related to the electric charge by 
$Q=I_{3W} +\frac{Y}{2}$. These new fields are:
\begin{eqnarray}
\mathbf{T \sim (1^{c}, 3)^{0}_{L}\;, \qquad D_{1}\sim (1^{c} , 2)_{R}^{+1}\;, 
\qquad D_{2}\sim (1^{c},2)^{-1}_{R} \;.}
\end{eqnarray}
One can easily check that this is a gauge and gravitational anomaly free 
set of chiral fermions. They sit in adjacent representations of $SU(2)_{W}$
with weak isospin difference $\Delta I_{W} = \frac{1}{2}$. This matches
with the only spinless field of the SM, the 
Higgs field,  with gauge labels $\mathbf{H \sim (1^{c}, 2)_{+1}}$. 
%As we shall see below, all new fermions receive masses 
%when the neutral component of the
%Higgs field acquires a vacuum expectation value ($v\approx 174$ GeV) 
%which breaks the SM gauge group down to $U(1)_{\mathrm em}$.

It is convenient to represent 
all fermions, \ie SM quarks and leptons plus new fermions that belong to the DM sector,
 with two component,  left handed, Weyl
fields~\cite{Dreiner:2008tw}, namely\footnote{The bar symbol over the Weyl fields 
 is part of their names.} 
%%%%%%%%%%%%%%%%%%%%%%%%%%%%%%
\begin{align}
\mathrm{SM~quarks} :\qquad & 
\mathbf{Q = \left ( \begin{array}{c} u \\ d \end{array} \right )\sim (3^{c}, 2)_{+1/3}\;, \quad 
\bar{u} \sim (3^{c}, 1)_{-4/3}\;, \quad \bar{d} \sim (3^{c},1)_{+2/3} } \;, \label{SMq}\\[3mm]
\mathrm{SM~leptons} :\qquad &
\mathbf{L = \left ( \begin{array}{c} \nu \\ e \end{array} \right )\sim (1^{c}, 2)_{-1}\;, \quad 
\bar{\nu} \sim (1^{c}, 1)_{0}\;, \quad \bar{e} \sim (1^{c},1)_{+2} } \;, \label{SMl} \\[3mm]
\mathrm{DM~fermions} :\qquad &
\mathbf{T = \left ( \begin{array}{c} T_{1} \\ T_{2} \\ T_{3} \end{array} \right )\sim (1^{c}, 3)_{0}\;, }
\nonumber \\[3mm]
&\hspace*{-0.15in} \mathbf{\quad  \bar{D}_{1} = 
\left( \begin{array}{c} \bar{D}_{1}^{1} \\ \bar{D}_{1}^{2} \end{array} \right )\sim (1^{c}, 2)_{-1}\;, 
\quad  \bar{D}_{2} = 
\left( \begin{array}{c} \bar{D}_{2}^{1} \\ \bar{D}_{2}^{2} \end{array} \right ) \sim (1^{c}, 2)_{+1} }\;\;\;.
\label{DMf} 
\end{align}
%%%%%%%%%%%%%%%%%%%%%%%%%%%%%%%%%%%%%%
SM fermions come in three copies of (\ref{SMq}) and (\ref{SMl}) sets of fields.
 We have added a left-handed antineutrino Weyl field in the SM field content 
 in order to account for light neutrino masses via the seesaw mechanism. 
 Although there may be interesting links between the neutrino and DM sector fields 
 we shall scarcely  refer to neutrinos in this article. 
 We assume only one copy of the DM-sector fields in (\ref{DMf}).
 Of course, we could also add more singlet fermions either in
 the SM or in the DM-sector but our intention is to keep the model
 as minimal as possible.
 
Physical masses are obtained from the gauge invariant form 
of Yukawa interactions.
% and possible extra, \emph{ad-hoc} symmetries . 
Under the assumption-5 above, the whole Yukawa Lagrangian of the model is
%%%%%%%%%%%%%%%%%%%
\begin{equation}
\mathscr{L}_{\rm Yuk} = \mathscr{L}_{\rm Yuk}^{\rm SM} \ + \ \mathscr{L}_{\rm Yuk}^{\rm DM}  \;,
\end{equation}
%%%%%%%%%%%%%%%%%%%%%
where the SM part reads (flavour indices are suppressed): 
\begin{align}
\mathscr{L}_{\rm Yuk}^{\rm SM} &= Y_{u} \epsilon^{ab} H_{a} Q_{b} \bar{u} -
Y_{d} H^{\dagger\, a} Q_{a} \bar{d} - Y_{e} H^{\dagger \, a} L_{a} \bar{e} 
\nonumber \\[3mm]
& +Y_{\nu} \epsilon^{ab} H_{a} L_{b} \bar{\nu} - \frac{1}{2} M_{N} \bar{\nu} \bar{\nu} + {\rm H.c.}\;,
 \label{LSM}
 \end{align}
 %%%%%%%%%%%%%%%%%%%%%
 and the available DM-sector interactions are
\begin{align}
\mathscr{L}_{\rm Yuk}^{\rm DM} &= 
Y_{1}\: \epsilon^{ab}\, T^{A}\, H_{a}\,  (\tau^{A})_{b}^{c}\, \bar{D}_{1\, c}
\ - \ Y_{2}\:  T^{A}\, H^{\dagger \, a}\,  (\tau^{A})_{a}^{c} \, \bar{D}_{2\, c} \nonumber \\[3mm]
&- M_{D}\: \epsilon^{ab}  \bar{D}_{1\,a} \bar{D}_{2\, b} \ - \ \frac{1}{2} M_{T}\: T^{A} T^{A} 
\ + \ {\rm H.c.} \;.
\label{LDM}
\end{align}   
%%%%%%%%%%%%%%%%%%%%%%%%%%
By choosing appropriate field redefinitions and without loss of generality  
we can make the parameters
$Y_{1}, Y_{2}$, and  $M_{T}$ real and positive, while
 leaving $M_{D}$ to be a general complex parameter. 
This is the only source of $CP$-violation\footnote{Electron and Neutron EDMs will arise 
first at two-loop level. Similarly for the anomalous magnetic moments of SM leptons.
See relevant discussion in \Ref{Fan:2013qn}.} arising from the DM-sector in this model.
If not stated otherwise, we  consider real $M_{D}$ values in our numerical results.
The parity symmetry 
assumption-5 removes the following renormalizable  operators:
%%%%%%%%%%%%%%%%%%%%
\begin{equation}
H^{\dagger}\: \bar{D}_{2}\: \bar{\nu}\;, \quad H\: \bar{D}_{1}\: \bar{\nu}\;, \quad L \: \bar{D}_{2} \;, \quad H\: T\: L \quad  {\rm and} \quad H^{\dagger} \: \bar{D}_{1}\:  \bar{e} \;.
 \end{equation}
 %%%%%%%%%%%%%%%%%
 Note that apart from the first two, the rest will not be allowed under the custodial symmetry.
 Finally, 
 we assume that possible non-renormalizable operators that are allowed by the discrete symmetry 
 are Planck scale suppressed and do not play any particular role in what follows.
 
 %%%%%%%%%%%%%%%%%%%%%%%%%%
 \subsection{The spectrum}
 \label{spec}
 %%%%%%%%%%%%%%%%%%%%%%
 
 Since there is no-mixing between the mass terms of the SM fermions and the
 DM sector ones, we solely concentrate on the non-SM Yukawa interactions of \eq{LDM}.
 After electroweak symmetry breaking and the shift of the neutral component of the
 only Higgs field, $H^{0} = v + h/\sqrt{2}$, we obtain the following  mass terms
%%%%%%%%%%%%%%%%%%%%%%% 
\begin{align}
\mathscr{L}^{\rm DM}_{\rm Y \, (mass)} 
&= - \left ( \mathcal{\tau}_{1}\quad  \bar{D}_{2}^{1} \right )^{T}
\: \mathcal{M}_{C} \: \left ( \begin{array}{c} \mathcal{\tau}_{3} \\[2mm] \bar{D}_{1}^{2} \end{array} \right ) - \frac{1}{2} \left ( \mathcal{\tau}_{2}\quad  \bar{D}_{1}^{1} \quad \bar{D}_{2}^{2} \right )^{T}
\mathcal{M}_{N} \left ( \begin{array}{c}
\mathcal{\tau}_{2} \\[2mm] \bar{D}_{1}^{1} \\[2mm] \bar{D}_{2}^{2} \end{array}\right )
\ + \ {\rm H.c.}
\nonumber \\
& = - \sum_{i=1}^{2} m_{\chi_{i}^{\pm}} \chi_{i}^{-} \: \chi_{i}^{+} \ - \ 
\frac{1}{2} \sum_{i=1}^{3} m_{\chi^{0}_{i}} 
\chi_{i}^{0} \chi_{i}^{0} \ + \ {\rm H.c.} \;,
\label{Lmass}
\end{align}    
%%%%%%%%%%%%%%%%%%%
where $\mathcal{\tau}_{1} \equiv (T_{1} - i T_{2})/\sqrt{2}$, 
$\mathcal{\tau}_{3} \equiv (T_{1} + i T_{2})/\sqrt{2}$ 
and $\mathcal{\tau}_{2} \equiv T_{3}$.  The charged ($\mathcal{M}_{C}$) and the 
neutral ($\mathcal{M}_{N}$) fermion mass matrices in \eq{Lmass} are given by
%%%%%%%%%%%%%%%%%
\begin{equation}
\mathcal{M}_{C} = \left(\begin{array}{cc}M_{T} & \sqrt{2}\, m_{1} \\[2mm] \sqrt{2}\, m_{2} & -M_{D}
\end{array}\right) \;, \qquad \mathcal{M}_{N} = 
\left(\begin{array}{ccc}M_{T} & m_{1} & -m_{2} \\ m_{1} & 0 & M_{D} 
\\ -m_{2} & M_{D} & 0\end{array}\right) \;, \label{mcmn}
\end{equation}
%%%%%%%%%%%%%%
where $m_{1,2} \equiv Y_{1,2} \: v$. Matrices $\mathcal{M}_{C}$ and $\mathcal{M}_{N}$
are diagonalized following the singular value decomposition  and the Takagi factorisation 
theorems~\cite{Horn}
into $m_{\chi^{\pm}} =(2\times 2)$  and $m_{\chi^{0}}= (3\times 3)$ diagonal matrices,
%with positive definite eigenvalues arranged in ascending order,
%%%%%%%%%%%%%%%
\begin{equation}
U_{L}^{T}\:  \mathcal{M}_{C} \: U_{R} = m_{\chi^{\pm}}\;, 
\quad O^{T}\: \mathcal{M}_{N} \: O = m_{\chi^{0}}\;, \label{omat}
\end{equation} 
respectively, after rotating the current eigenstate fields into their mass eigenstates  
$\chi^{\pm}_{i} , \chi^{0}_{i}$ with  unitary matrices, $U_{L}, U_{R}$ and $O$, as 
%%%%%%%%%%%%%%%%%%%%%%%
\begin{equation}
\left(\begin{array}{c}\mathcal{\tau}_{3} \\[2mm] \bar{D}_{1}^{2} \end{array}\right) = U_{R} \:
\left(\begin{array}{c}\chi_{1}^{-} \\[2mm] \chi_{2}^{-} \end{array}\right)\;, \quad
\left(\begin{array}{c}\mathcal{\tau}_{1} \\[2mm] \bar{D}_{2}^{1} \end{array}\right) = U_{L} \:
\left(\begin{array}{c}\chi_{1}^{+} \\[2mm] \chi_{2}^{+} \end{array}\right)\;, \qquad
\left(\begin{array}{c}\mathcal{\tau}_{2} \\[2mm] \bar{D}_{1}^{1} \\[2mm] 
\bar{D}_{2}^{2} \end{array}\right) = 
O \:
\left(\begin{array}{c}\chi_{1}^{0} \\[2mm] \chi_{2}^{0} \\[2mm] \chi_{3}^{0} \end{array}\right)\;.
\label{rots}
\end{equation}
%%%%%%%%%%%%%%%%%%%%
Therefore the spectrum of this model contains, apart from the SM masses for quarks
and leptons, two additional charged Dirac fermions and three neutral Majorana particles.
It is the lightest Majorana  particle $\chi_{1}^{0}$ with  mass  $m_{\chi_{1}^{0}}$,
that, perhaps, supplies the universe  with cold Dark Matter. 

It is crucial for what follows  and also  enlightening,
 to discuss the decoupling of the $M_{D}$-eigenvalue from
the particle spectrum.
First, $\mathcal{M}_{N}$, is a real symmetric matrix, under the assumption of real $M_{D}$.
Then, consider the following unitary matrix 
$\Sigma$, having as columns orthonormal vectors,
%%%%%%%%%%%%%%%%%
\begin{eqnarray}
\Sigma = \frac{1}{\sqrt{2}}\:
\left(\begin{array}{ccc}\sqrt{2} & 0 & 0 \\0 & 1 & 1 \\0 & -1 & 1\end{array}\right)\;,
\end{eqnarray}
%%%%%%%%%%%%%%%%
which by a similarity transformation,
 brings the lower right $2\times 2 $ sub-block of $\mathcal{M}_{N}$ 
into a diagonal form, 
%%%%%%%%%%%%%%%%
\begin{eqnarray}
\mathcal{M}_{N}^{\prime}  \ = \Sigma^{\dagger}\: \mathcal{M}_{N}\:  \Sigma = 
\left(\begin{array}{ccc}M_T & (m_1 + m_2)/\sqrt{2} &  (m_1 - m_2)/\sqrt{2} \\ 
(m_1 + m_2)/\sqrt{2} & -M_D & 0 \\ (m_1 - m_2)/\sqrt{2} & 0 & M_D\end{array}\right) \;.
\label{mnp}
\end{eqnarray}
%%%%%%%%%%%%%%%
Note that since $\Sigma$ is unitary matrix, the eigenvalues of $\mathcal{M}_{N}$
and $\mathcal{M}_{N}^{\prime}$ are equal. 
We therefore obtain, 
 that for $m_{1} = m_{2}$ the charged fermion mass matrix 
$\mathcal{M}_{C}$ becomes the upper-left sub-block of the $\mathcal{M}_{N}^{\prime}$
in \eq{mnp}. Therefore
the eigenvalue, $M_{D}$, decouples from the neutral fermion mass matrix \ie it is independent
of any mixing and therefore any v.e.v, while the
rest of eigenvalues of both matrices, $\mathcal{M}_{C}$ and $\mathcal{M}_{N}$,
 are one to one degenerate. 

%%%%%%%%%%%%%%%%%%%%%%%%%%%%%
\subsection{The interactions}
\label{sec:interactions}
%%%%%%%%%%%%%%%%%%%%%%%%%%

%
We now turn to the interactions between  the new fermions and 
the SM gauge-bosons or the  SM Higgs-boson. 
The latter  can be read from \eq{LDM} after
rotating fields  by exploiting  the relations in (\ref{rots}). After a little bit of algebra we 
obtain\footnote{We use  Weyl notation for fermions~\cite{Dreiner:2008tw} throughout.}
%%%%%%%%%%%%%%%%%%%%
\begin{equation}
\mathscr{L}_{\rm Y (int)}^{\rm DM} = - Y^{h\chi_{i}^{-} \chi_{j}^{+}} \: h \: \chi_{i}^{-} \: \chi_{j}^{+}
\  - \ \frac{1}{2}\: Y^{h\chi_{i}^{0} \chi_{j}^{0}} \: h \: \chi_{i}^{0} \: \chi_{j}^{0} \ + \  {\rm H.c.} \;,
\label{lyint}
 \end{equation}
 %%%%%%%%%%%%%%%%%%
 where
%%%%%%%%%%%%%%%%%%%%
\begin{align}
 Y^{h\chi_{i}^{-} \chi_{j}^{+}} & \equiv \frac{1}{v} \left ( m_{1}\: U_{R\, 2i} \:U_{L\, 1j} + 
 m_{2}\: U_{R\, 1i} \: U_{L\, 2j} \right ) \;, \label{yhpp}\\[2mm]
 Y^{h\chi_{i}^{0} \chi_{j}^{0}} & \equiv \frac{O_{1i}}{\sqrt{2}\:  v}\: \left (
 m_{1} \: O_{2j} - m_{2}\: O_{3j} \right ) \ + \ (i \leftrightarrow j)\;.\label{yhxx}
 \end{align}
 %%%%%%%%%%%%%%% 
 For completeness and especially for loop calculations,
  we append here the interactions between 
 Goldstone bosons and the new fermions:
%%%%%%%%%%%%%%%%%%%
 \begin{align}
\mathscr{L}_{\rm G\chi\chi } &= -\frac{i\: O_{1i}}{\sqrt{2} v} \, 
(m_{1}\:  O_{2j}  +  m_{2} \:  O_{3j})\, G^{0} \chi_{i}^{0} \chi_{j}^{0} 
% \nonumber \\
 -\frac{i}{v}\, (m_{1}\: U_{R\,2i} \:U_{L\, 1j}  -  m_{2} \: U_{R\, 1i} \: U_{L\, 2j})\, 
 G^{0} \chi_{i}^{-} \chi_{j}^{+}
 \nonumber \\[2mm]
 &+ \frac{m_{1}}{v} (\sqrt{2} \: U_{R\,1i}\: O_{2j} - U_{R\,2i} \: O_{1j})\, G^{+} \chi^{-}_{i} \chi_{j}^{0}
 -\frac{m_{2}}{v} \: (\sqrt{2} \: U_{L\, 1i}\: O_{3j} + U_{L\, 2i}\: O_{1j})\, G^{-} \chi_{i}^{+} \chi_{j}^{0}
 \nonumber \\[2mm] &+ \mathrm{H.c} \;.\label{LGB}
 \end{align}
 %%%%%%%%%%%%%%%%%%%
 %
 Interactions among the new fermions and gauge bosons arise from the respective fermion
 kinetic terms. Interactions between $\chi^{\pm}$ and the photon are purely vectorial,
 %%%%%%%%%%%%%%%
 \begin{equation}
 \mathscr{L}_{\rm KIN (int)}^{\gamma - \chi^{\pm}} = - 
 (+e) \:(\chi_{i}^{+})^{\dagger} \bar{\sigma}^{\mu}
 \chi_{i}^{+} \: A_{\mu} - (-e)\: (\chi_{i}^{-})^{\dagger} \bar{\sigma}^{\mu}
 \chi_{i}^{-} \: A_{\mu} \;,
 \end{equation} 
 %%%%%%%%%%%%%%%%%%%%%
 where $A_{\mu}$ is the photon field and $(-e)$ 
 the electron electric charge. The $Z$-gauge boson couplings to both
 charged and neutral fermions 
 can be read from\footnote{Our  notation resembles closely 
 the one in Appendix E of \Ref{Dreiner:2008tw}
 \ie
 $U\to U_{L}^{\dagger}$, $V\to U_{R}^{\dagger}$ and $N\to O^{\dagger}$.},
 %%%%%%%%%%%%%%%%%%%
 \begin{equation}
 \mathscr{L}_{\rm KIN (int)}^{Z - \chi} = \frac{g}{c_{W}} O_{ij}^{\prime \, L} \: (\chi_{i}^{+})^{\dagger}
\: \bar{\sigma}^{\mu}\: \chi_{j}^{+} \:Z_{\mu} -
\frac{g}{c_{W}} O_{ij}^{\prime \, R} \: (\chi_{j}^{-})^{\dagger}
\: \bar{\sigma}^{\mu}\: \chi_{i}^{-} \:Z_{\mu}  +
\frac{g}{c_{W}} O_{ij}^{\prime\prime \, L} \: (\chi_{i}^{0})^{\dagger}
\: \bar{\sigma}^{\mu}\: \chi_{j}^{0} \:Z_{\mu} \;, 
 \end{equation} 
 %%%%%%%%%%%%%%%%%%%%%%%
 where
 %%%%%%%%%%%%%%%%%%
 \begin{align}
O_{ij}^{\prime\, L} &= -U^{*}_{L1i} \: U_{L1j} - \frac{1}{2}\: U_{L 2i}^{*}\: U_{L2j} + 
s_{W}^{2} \delta_{ij}\;, \label{eq:OLp}\\
O_{ij}^{\prime\, R} &= -U_{R1i} \: U^{*}_{R1j} - \frac{1}{2} \: U_{R 2i}\: U_{R2j}^{*} + 
s_{W}^{2} \delta_{ij}\;, \label{eq:ORp} \\
O_{ij}^{\prime\prime\, L} &= \frac{1}{2}\: \left ( O_{3i}^{*} \: O_{3j} - O_{2i}^{*} \: O_{2j} \right ) \;. 
\label{gZxx}
\end{align}
%%%%%%%%%%%%%%%%%%%%%%%%
with $s_{W}, c_{W}$ the $\sin$ and $\cos$
 of the weak mixing angle and $g$ the $SU(2)_{W}$ gauge
coupling. Finally, interactions between $\chi$'s  and $W$-bosons 
are described by the terms
%%%%%%%%%%%%%%%%
\begin{align}
\mathscr{L}_{\rm KIN (int)}^{W^{\pm}-\chi^{0}-\chi^{\mp}} \ &= \ 
g\: O_{ij}^{L} \: (\chi_{i}^{0})^{\dagger} \: \bar{\sigma}^{\mu} \: \chi^{+}_{j} \: W_{\mu}^{-}
-g \: O_{ij}^{R} \: (\chi_{j}^{-})^{\dagger} \: \bar{\sigma}^{\mu} \: \chi^{0}_{i} \: W_{\mu}^{-} 
\nonumber \\[2mm]
& + g \: O_{ij}^{L*} \: (\chi_{j}^{+})^{\dagger} \: \bar{\sigma}^{\mu}\: \chi^{0}_{i}\: W_{\mu}^{+}
-g\: O_{ij}^{R*} \: (\chi_{i}^{0})^{\dagger} \: \bar{\sigma}^{\mu}\: \chi^{-}_{j} \: W_{\mu}^{+} \;,
\end{align}
%%%%%%%%%%%%%%%%%
where the mixing matrices $O^{L}$ and $O^{R}$ are given by 
%%%%%%%%%%%%%%%
\begin{subequations}
\begin{align}
O_{ij}^{L} &= O_{1i}^{*} \: U_{L1j} \ - \ \frac{1}{\sqrt{2}} \: O_{3i}^{*} \: U_{L2j} \;, 
\\[2mm]
O_{ij}^{R} &= O_{1i} \: U_{R1j}^{*} \ + \ \frac{1}{\sqrt{2}} \: O_{2i} \: U_{R2j}^{*} \;.
\end{align}
\label{eq:OLOR}
\end{subequations}
%%%%%%%%%%%%%%%%

We open a parenthesis here to discuss 
a comparison with MSSM: mass matrices for neutral and charged fermion  in \eq{mcmn}
remind those of neutralinos and charginos in the MSSM. 
It is of course trivially understood why this happens: the doublet
and the triplet fields possess  
the same gauge quantum numbers as the  higgsino and wino  fields,
respectively. However, there are two crucial differences: first there is no restriction
to add a bino singlet and therefore the minimal $\mathcal{M}_{N}$ is a  $3\times 3$, instead of $4\times 4$, simpler matrix 
and second, and more important, the off-diagonal entries in $\mathcal{M}_{N}$ and 
$\mathcal{M}_{C}$, are not
proportional to  gauge couplings but to, 
Yukawa couplings, $Y_{1}$ and $Y_{2}$. The latter entries ($\sim Y v$) can be
substantially bigger than the corresponding ones ($\sim g v$) in the 
neutralino mass matrix of MSSM. 
Furthermore,
since $\tan\beta = 1$ is not, in general, 
a phenomenologically viable case in MSSM, there should always
be a factor of  hierarchy between the off diagonal entries. This is not necessarily the case here.
In fact, the $\tan\beta=1$ ``blind spot''~\cite{Cheung:2012qy}, 
is a point in parameter space protected by a custodial symmetry.
%The MSSM higgsino/wino interactions is only a subset from those in  doublet/triplet 
%fermionic model. 

%%%%%%%%%%%%%%%%%%%
\subsection{A custodial symmetry}
\label{sec:symmetry}
%%%%%%%%%%%%%%%%%%%%
It is well known that the Higgs sector in the SM obeys, in addition to the standard
electroweak gauge symmetry, a custodial $SU(2)_{R}$ global symmetry. This
symmetry is broken explicitly by the hypercharge gauge coupling $g'$, and by the difference
between the top- and bottom-quark Yukawa couplings.  
Similarly,  the fermionic DM sector, described by  \eq{LDM}, 
obeys also such a symmetry if $Y_{1}=Y_{2}\equiv Y$. 
More explicitly, \eq{LDM} can be written in a $SU(2)_{L}\times SU(2)_{R}\times U(1)_{X}$
invariant form as 
%%%%%%%%%%%%%%%%
\begin{equation}
\mathscr{L}_{\rm Yuk}^{\rm DM} = -Y \: T^{A} \: \mathscr{H}^{x,a} \: (\tau^{A})_{a}^{b} \:\bar{\mathscr{D}}_{x,b} - \frac{1}{2}\: M_{D} \:\epsilon^{xy}\: \epsilon^{ab} \:
\bar{\mathscr{D}}_{x,a} \: \bar{\mathscr{D}}_{y,b} - \frac{1}{2}\: M_{T} \: T^{A} \: T^{A}  
\ + \ {\rm H.c.}\;,
\label{LDMsym}
\end{equation}
%%%%%%%%%%%%%%%
where $x,y$ denote $SU(2)_{R}$ group indices and
%%%%%%%%%%%%%%%%%%%
\begin{equation}
\mathscr{H}^{x,a} =  \left(\begin{array}{c}  H^{a} \\ H^{\dagger\, a}\end{array}\right)
%\;, \qquad 
%\bar{H}_{a} = \epsilon_{ab} \: H^{*b}
\;, \qquad
\bar{\mathscr{D}}_{x,a} =  \left(\begin{array}{c}  \bar{D}_{1a} \\  
\bar{D}_{2a} \end{array}\right) \;, \label{eq:sym}
\end{equation}
%%%%%%%%%%%%%%%%%%%%
with $H^{a} = \epsilon^{ab} H_{b}$. This extra global symmetry stands for 
the rotations between $H \leftrightarrow H^{\dagger}$ and $\bar{D}_{1} \leftrightarrow \bar{D}_{2}$.
Although this symmetry is broken by the hypercharge gauge symmetry,
it is  natural to study  interactions among extra fermions $(\bar{\mathscr{D}}, T)$
and SM-bosons under the assumption that $SU(2)_{R}$ is approximately preserved in the DM sector, that is, 
%%%%%%%%%%%%%%
\begin{equation}
Y_{1}\ = \ Y_{2}  \ \Rightarrow \ m_{1}\ = \ m_{2}  \;. \label{ass}
\end{equation}
%%%%%%%%%%%%%%%
In addition, \eq{eq:sym} is invariant under a global $U(1)_{X}$ fermion number symmetry, 
under which 
only $\bar{\mathscr{D}}$ and $T$ fields are charged with 
$[\bar{\mathscr{D}}]= [D_{1}] = [D_{2}]=-[T]=1$.
In that case $M_{D}$ and  $M_{T}$ are not allowed. We therefore conclude that the
limit where $Y\equiv Y_{1}=Y_{2}$ and $M_{D} = M_{T} \to 0$ is radiatively stable
and this fact motivates us to study it in more detail.
%It is exactly this limit we want to investigate in this article and we follow from now on. 
Note again that, both $SU(2)_{R}$ and $U(1)_{X}$ symmetries 
are broken explicitly by  hypercharge symmetry.

%Note also  that certain discrete symmetries can be exploited so that 
 %$M_{T} = M_{D}=0$ (``Dirac Dominance'')
%or $Y_{1}=Y_{2}=0$ (``Majorana dominance'').
 
%%%%%%%%%%%%%%%%%%%%%%%%%%%%%%%%%%%%%%% 
\subsection{Lightest Neutral fermion interactions under the symmetry}
\label{sec:LP}
%%%%%%%%%%%%%%%%%%%%%% %%%%%%%%%%%%%%%%
Let's introduce the mass difference, $\Delta m \equiv m_{1} -m_{2}$, between the chiral
masses (or between Yukawa  couplings, $Y_{1}$ and $Y_{2}$, if you wish). 
If $SU(2)_{R}$ symmetry is approximately preserved, \ie \eq{ass} approximately holds, $\Delta m$ must be treated as perturbation compared to $m_{1}$
or $m_{2}$ masses, which collectively denoted by $m=m_{1}$, \ie $\Delta m\ll m$. 
We can then write the neutral fermion mass matrix in  a suggestive perturbative 
form 
%%%%%%%%%%%%%%%%%%%%%
\begin{equation}
\mathcal{M}_{N} \ = \ \mathcal{M}_{N}^{(0)} + Q \;,
\end{equation}
%%%%%%%%%%%%%%%%%%%%%%%%%
where
%%%%%%%%%%%%%%%%%%%
\begin{equation}
\mathcal{M}_{N}^{(0)}=
\left(\begin{array}{ccc}M_T & m & -m \\m & 0 & M_D \\-m & M_D & 0\end{array}\right)
\;, \qquad Q = 
\left(\begin{array}{ccc}0 & 0 & \Delta m \\0 & 0 & 0 \\\Delta m & 0 & 0\end{array}\right)\;.
\end{equation}
%%%%%%%%%%%%%%%%%%%%%%
The zeroth order eigenvalues of $\mathcal{M}_{N}^{(0)}$ read
%%%%%%%%%%%%%%%%%%%%%
\begin{subequations}
\begin{align}
m_{\chi_{1}^{0}} &= M_{D}\;,  \label{eig1}\\
 m_{\chi_{2}^{0}} &= \frac{1}{2} \left [ M_{T} - M_{D} - 
\sqrt{8 m^{2} + (M_{T} + M_{D})^{2}} \right ]\;,  \label{eig2}\\
m_{\chi_{3}^{0}} &= \frac{1}{2} \left [ M_{T} - M_{D} + 
\sqrt{8 m^{2} + (M_{T} + M_{D})^{2}} \right ]\;,  \label{eig3}
\end{align}
\label{eigen}
\end{subequations}
%%%%%%%%%%%%%%%%%
while the corresponding eigenvectors are 
%%%%%%%%%%%%%%%%%%%%%%%%%
\begin{equation}
\ket{1}^{(0)} = \frac{1}{\sqrt{2}}\: \left(\begin{array}{c}0 \\1 \\ 1 \end{array}\right)\;,
\quad 
\ket{2}^{(0)} = \frac{-1}{\sqrt{2+a^{2}}} \: \left(\begin{array}{c}a \\1 \\ -1\end{array}\right) 
\;, \quad \ket{3}^{(0)} = \frac{1}{\sqrt{2+a^{2}}} \: 
\left(\begin{array}{c}\sqrt{2} \\-\frac{a}{\sqrt{2}} \\\frac{a}{\sqrt{2}}\end{array}\right) \;,
\label{vecs}
\end{equation}
%%%%%%%%%%%%%%%%%%%%%
where the  parameter $a$ is given by
%%%%%%%%%%%%%%%%%%
\begin{equation}
a = \frac{m_{\chi_{1}^{0}} + m_{\chi_{2}^{0}}}{m} \;. \label{eq:a}
\end{equation}
%%%%%%%%%%%%%%%
The parameter $a$, varies in the interval $[-\sqrt{2} , 0]$ for positive $M_{D}$.
A little examination of the eigenvalues show that unless, $M_{D} \gg M_{T}>0$ where the 
Lightest Particle (LP) becomes the triplet, in the rest of the parameter space the LP is
a  ``very well tempered''  mixed doublet fermion, $\ket{\chi_{1}^{0}} = \frac{1}{\sqrt{2}} 
(\ket{\bar{D}_{1}^{1}} + \ket{\bar{D}_{2}^{2}})$, with mass
$m_{\chi_{1}^{0}} = M_{D}$.\footnote{It is easy to 
show that since $^{(0)}\bra{1}Q\ket{1}^{(0)} = 0$, there is no 
correction, up to $(\Delta m)^{2}$, on  $m_{\chi_{1}^{0}} = M_{D}$ LP mass.}
The DM particle $(\chi_{1}^{0})$ has then vanishing
coupling to the Higgs boson because in \eq{yhxx} it is  $O_{11}=0$.
Note that, every neutral fermion has always 
vanishing diagonal couplings to $Z$-gauge boson, $|O_{2i}|=|O_{3i}|$, since the
two doublets, $\bar{D}_{1}$ and $\bar{D}_{2}$ couple to $Z$ with opposite weak isospin.
It is therefore worth examining how eigenvalues
and  eigenvectors are corrected after switching on to $\Delta m \ne 0$.

Obviously, 
in order to find how $\chi^{0}_{1}$ couples to $Z$ or $h$ non-trivially,
\ie to find the couplings $Y^{h\chi_{1}^{0}\chi_{1}^{0}}$
and $g^{Z\chi_{1}^{0}\chi_{1}^{0}} = g O_{11}^{\prime\prime \, L}/c_{W}$ in \eqs{yhxx}{gZxx}, 
respectively, we need to know the $O(\Delta m)$ corrections, 
in  eigenvector the $O_{i1}$. 
%This is because as we observe from $\ket{1}^{(0)}$ in \eq{vecs},
%both couplings exactly vanish  at zeroth order (or in the custodial $SU(2)_{R}$ limit).
%
The corrected eigenvector, $\ket{1} = \ket{1}^{(0)} + \ket{1}^{(1)} + O[(\Delta m)^{2}]$, 
which is nothing else but the first column of the matrix $O$ in \eq{omat} is
 found to be,
%%%%%%%%%%%%%
\begin{equation}
O_{i1}=\ket{1} =\frac{1}{\sqrt{2}}
\: \left(\begin{array}{c}x \: \Delta m \\1+y\: \Delta m \\1-y \: \Delta m\end{array}\right) +
O[(\Delta m)^{2}] \;, \label{O1icor}
\end{equation}
%%%%%%%%%%%%%%
where 
%%%%%%%%%%%%%
\begin{align}
x &\equiv \frac{1}{(2+ a^{2})} \left [\frac{a^{2}}{m_{\chi_{1}^{0}} - m_{\chi_{2}^{0}}} 
+ \frac{2}{m_{\chi_{1}^{0}} - m_{\chi_{3}^{0}}} \right ]\;, \\[2mm]
y &\equiv \frac{a}{(2+ a^{2})} \left [\frac{1}{m_{\chi_{1}^{0}} - m_{\chi_{2}^{0}}} -
\frac{1}{{m_{\chi_{1}^{0}} - m_{\chi_{3}^{0}}}} \right ] \;.\label{eq:y}
\end{align}
%%%%%%%%%%%%%%%%
Simple substitution of \eq{O1icor} into  \eqs{yhxx}{gZxx} gives
%%%%%%%%%%%%%%%%
\begin{align}
Y^{h\chi_{1}^{0}\chi_{1}^{0}} &=  \frac{(\Delta m)^{2}}{\sqrt{2} v} \: x \: (1+ 2 m y) + 
\mathrm{O}[(\Delta m)^{2}/m^{2}]\;, \label{eq:Yhxx}\\[3mm]
 g^{Z\chi_{1}^{0}\chi_{1}^{0}} &\equiv \frac{g}{c_{W}} O_{11}^{\prime\prime \, L} = 
 -\frac{g}{c_{W}} \: y \: \Delta m + \mathrm{O}[(\Delta m)^{2}/m^{2}]\;.\label{eq:Zxx}
 \end{align}
 %%%%%%%%%%%%%%%%%%%%%%%%%%
  Obviously, for sufficiently small mass difference
  $\Delta m$, the Spin-Independent (SI) coupling 
 $(Y^{h\chi_{1}^{0}\chi_{1}^{0}})$
 is suppressed 
 by $(\Delta m)^{2}/m^{2}$ while the Spin-Dependent (SD) one $(g^{Z\chi_{1}^{0}\chi_{1}^{0}})$
 is suppressed by $\Delta m/m$ relative to their values away from the 
 $SU(2)_{R}$-symmetric limit. This maybe the reason why we have not detected 
 DM-nucleon interactions so far.  
 A question arises immediately about the stability of $\Delta m$  under radiative corrections.
 A quick RGE analysis~\cite{Giudice:2011cg,ArkaniHamed:2012kq} 
 shows that the $\beta$-function for $\Delta m$ at 1-loop is
 %%%%%%%%%%%%%%%%%%%%%%%%%%%%%%%
 \begin{equation}
 \frac{d \Delta m}{d \ln(Q)} \ = \ \frac{\Delta m}{16 \pi^{2}} \,
 \left [ \frac{29}{4} Y^{2} + 3 Y_{t}^{2} - \frac{9}{20} g_{1}^{2} - \frac{33}{4} g_{2}^{2} \right ] \;,
 \label{eq:dmrge}
 \end{equation} 
 %%%%%%%%%%%%%%%%%%%%%%%%%%%%%%%%%%%%%%%%
 where $Y_{t}$ is the top-Yukawa coupling, $Y\equiv Y_{1} \simeq Y_{2}$,  and $g_{1,2}$ the
 hypercharge and weak gauge couplings, respectively. 
 \Eq{eq:dmrge} means that $\Delta m$ is only multiplicatively renormalized. Therefore, setting
 $\Delta m$ to zero at tree level stays zero at 1-loop and possibly 
 at higher orders\footnote{We confirm that   this result remains  unchanged at two-loops.}
 because this is a parameter point protected by the global symmetry.
 From \eqs{eq:Yhxx}{eq:Zxx} we conclude that for $\Delta m=0$, 
 only finite (threshold) and calculable 
 quantum corrections will affect 
 the couplings $Y^{h\chi_{1}^{0}\chi_{1}^{0}}$ and  $g^{Z\chi_{1}^{0}\chi_{1}^{0}}$
 which are relevant to Direct DM searches. 
 We confirm this consequence with a direct calculation of $\delta Y^{h\chi_{1}^{0}\chi_{1}^{0}}$ in  
 section~\ref{sec:direct} and in~\ref{sec:appA}.

 Note that $x$  vanishes in the limit $M_{D} \to 0$ while $(1+ 2 m y)$ vanishes at both
 $M_{D}\to 0$ and $M_{D} \to M_{T}$ limits.
 However, \eq{eq:Yhxx} is not accurate since 
 $(\Delta m)^{2}/m^{2}$-terms are missing in our perturbative expansion.
 It turns out that the $M_{D}\to M_{T}$ limit is violated by those and higher terms,
 but the limit $M_{D}\to 0$
 is protected  because of the $U(1)_{X}$-symmetry 
 that we discussed in section~\ref{sec:symmetry}.
  In contrast, \eq{eq:Zxx} is within 1\% of its exact numerical outcome.
 It is also worth noticing that in the case where the Majorana masses are dominant,
 $M_{D},M_{T}\gg m$, then  $y\to 0$ and therefore $g^{Z\chi_{1}^{0}\chi_{1}^{0}} \to 0$,
 up to higher order terms. 
 
 %%%%%%%%%%%%%% PLOTS %%%%%%%%%%%%%%%%%%%%
\begin{figure*}[t!] %  figure placement: here, top, bottom, or page
   \centering
%   \begin{tabular}{ll}
%      \labellist
%\large \hair 2pt
%\pinlabel {\bf (a)} at 160 260
%\pinlabel {\bf (b)} at 530 255
%\endlabellist
 %  \includegraphics[width=2.5in]{MDMTmx-plot.pdf} ~&
  % \includegraphics[width=2.5in]{MDmmx-plot.pdf} \\
%  \begin{subfigure}[t]{0.5\textwidth}
  \subfloat[]{\includegraphics[width=3.0in]{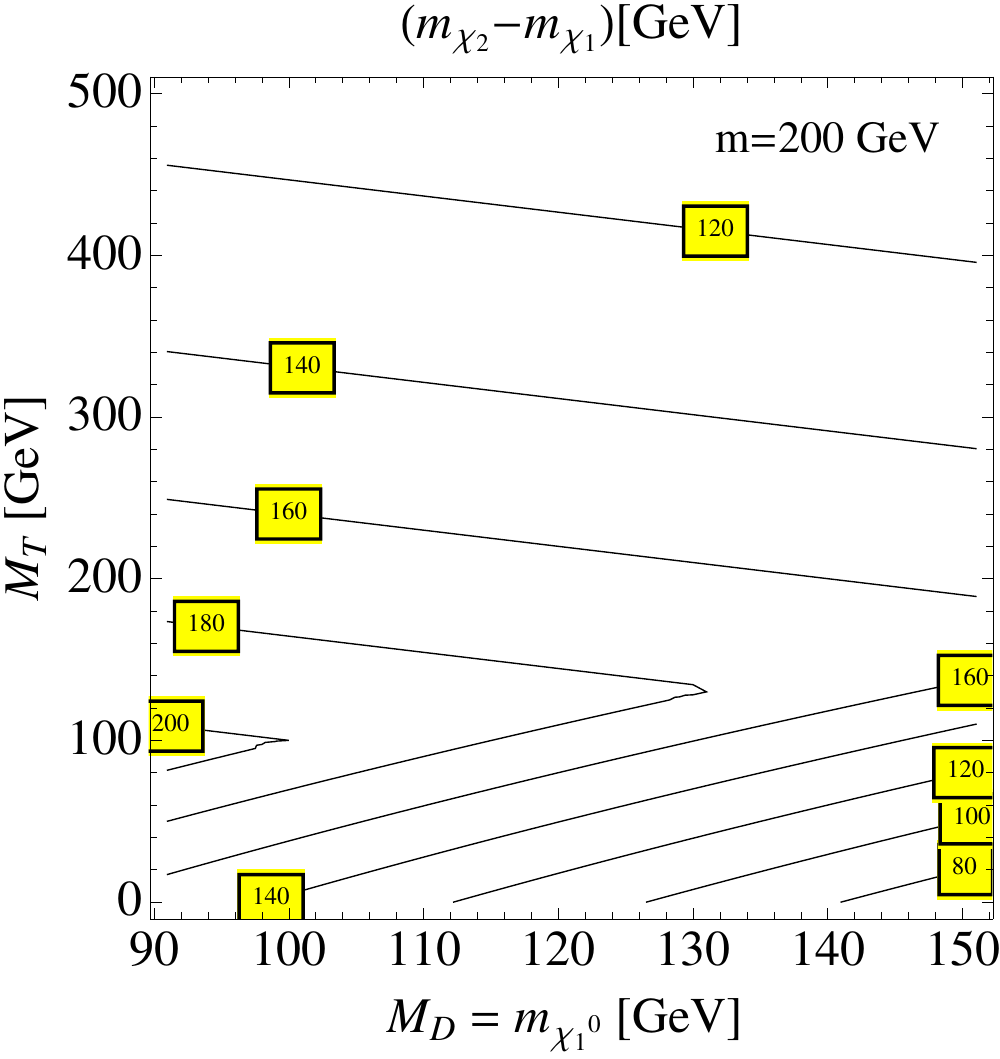}   }
 %  \caption{}
 %  \end{subfigure}%
%   \qquad
%   \begin{subfigure}[t]{0.5\textwidth}
 \subfloat[]{\includegraphics[width=3.0in]{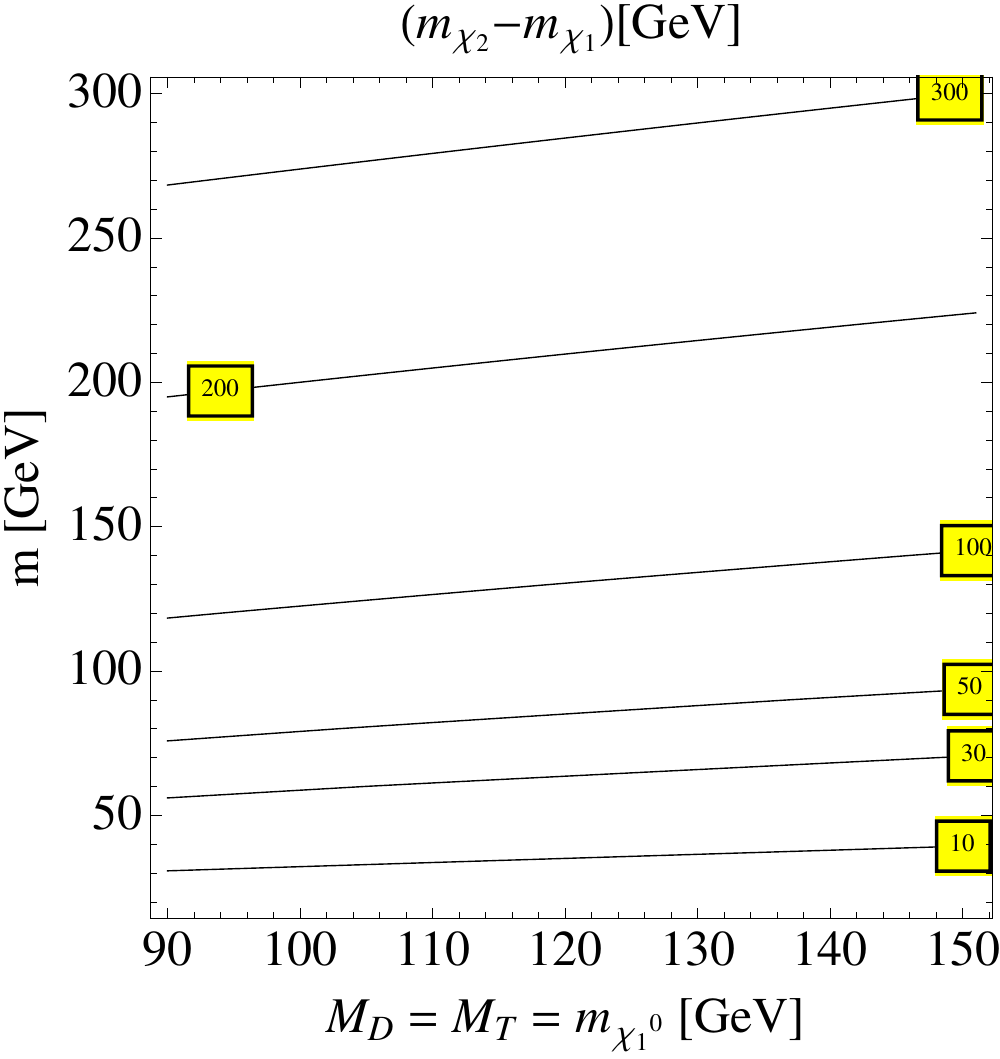} }
%   \caption{}
 %  \end{subfigure}
%   \end{tabular}
   \caption{\sl The mass difference, $|m_{\chi_{2}^{0}}|-|m_{\chi_{1}^{0}}|$,   
   between the next-to-lightest  and the lightest neutral
   particle state in the doublet-triplet fermionic DM model on (a) $M_{D}$ vs. $M_{T}$  
   with $m=200$ GeV,
   and (b) on $M_{D}$ vs. $m$ with $M_{T}=M_{D}$, plane.  
   For both plots and for the rest to come, it is always, $m_{\chi_{1}^{0}}=M_{D}$. }
   \label{fig:masses}
\end{figure*}
%%%%%%%%%%%%%%%%%%%%%%%%%%%%%

It will be useful for the discussion, especially on the relic density,
to show the mass difference 
 between the next-to-lightest ($|m_{\chi_{2}^{0}}|$) and the 
 lightest ($|m_{\chi_{1}^{0}}|$)  neutral fermion states.
 This is depicted as contour lines in Fig.~\ref{fig:masses}(a,b)
 on the $M_{D} - M_{T}$  plane (left plot) 
 and on the $M_{D} - m$ plane with 
 $M_{T} = M_{D}$   (right plot). Note that  $M_{D}$ coincides with the LP mass \ie
  $M_{D} = m_{\chi_{1}^{0}}$, everywhere in these graphs. For $m=200$ GeV, the
  mass difference is nowhere smaller than approximately 80 GeV, and typically, it is as large
  as the parameter $m$ with the maximum value at $M_{D}=M_{T}$. Subsequently, 
  in Fig.~\ref{fig:masses}b,
  we plot the maximum values of the mass difference on $M_{D}-m$ plane. 
   Alternatively,   it is easy to read from   Fig.~\ref{fig:masses}, the 
 parameter $a$ defined in \eq{eq:a}, because for the $M_{D}$ values taken throughout, 
  it is  $a = - (|m_{\chi_{2}^{0}}| - |m_{\chi_{1}^{0}}|)/m$. For instance, in the  plots shown, 
  this parameter varies,
  approximately, in the region, 
  $a \in [-1,-0.3]$.

 %%%%%%%%%%%%%%%%%%%%%%%%%%%%%%%%% 
 \subsection{Analytical expressions for the new interactions under the symmetry} 
 \label{sec:a}
%%%%%%%%%%%%%%%%%%%%%%%%%%%%%%%%%

As we have already discussed in section~\ref{spec}, 
in the symmetric $SU(2)_{R}$ limit of  (\ref{ass}),
 two of the eigenvalues from the charged fermion mass matrix are  degenerate
 respectively with those of the neutral fermion masses given in \eqs{eig2}{eig3},
 %%%%%%%%%%%%%%
 \begin{equation}
 m_{\chi_{1}^{\pm}} = m_{\chi_{2}^{0}}\;, \quad m_{\chi_{2}^{\pm}} = m_{\chi_{3}^{0}}\;.
 \label{cusmass}
 \end{equation}
 %%%%%%%%%%%%%%%%%%
%This is easily understood directly from the look of the - now symmetric - charged and neutral
%fermion mass matrices in \eq{mcmn}, since after the $\Sigma$ similarity transformation,
%$\mathcal{M}_{C}$ becomes principal submatrix of $\mathcal{M}_{N}$. They share 
%two common eigenvalues. 
%For reasons we discussed in the introduction In this article 
%we are particularly interested in this $SU(2)_{R}$-limit. It is therefore 
In addition, it is useful for further reference to present analytical 
expressions for all new interactions appear in the model. All these new interactions 
can be simply 
written in matrix forms containing 
(at most) one parameter, the real parameter $a$ of \eq{eq:a}. 
For example,  rotation matrices defined in \eq{omat}  read
%%%%%%%%%%%%%%
\begin{eqnarray}
U = U_{L} = U_{R}= \frac{1}{\sqrt{2+ a^{2}}} \: \left(\begin{array}{cc}a & -\sqrt{2} \\\sqrt{2} & a\end{array}\right) \;, \quad 
O = \left(
\begin{array}{ccc}
 0 & -\frac{a}{\sqrt{2+a^2}} & \frac{\sqrt{2}}{\sqrt{2+a^2}} \\
 \frac{1}{\sqrt{2}} & -\frac{1}{\sqrt{2+a^2}} & -\frac{a}{\sqrt{2} \sqrt{2+a^2}} \\
 \frac{1}{\sqrt{2}} & \frac{1}{\sqrt{2+a^2}} & \frac{a}{\sqrt{2} \sqrt{2+a^2}}
\end{array}
\right) \;.
\end{eqnarray}
%%%%%%%%%%%%%%%
 The couplings between $\chi_{1}^{0}$, $W$ and $\chi^{\pm}$
given in \eq{eq:OLOR} become explicitly:
%%%%%%%%%%%%%
%\begin{eqnarray}
%O_{1j}^{L} = - O_{1j}^{R\, *}\;, \qquad O_{11}^{L} = - \frac{\sqrt{2}}{2}\: \frac{1}{\sqrt{2+a^{2}}} \;,
%\quad  O_{12}^{L} = - \frac{a}{2}\: \frac{1}{\sqrt{2+a^{2}}} \;.
%\end{eqnarray}
%%%%%%%%%%%%%%%%%%%%%%%%%%%
%More explicitly we obtain:
%%%%%%%%%%% MATRICES in the Custodial Limit %%%%%%%%%%%%%%%%
\begin{eqnarray}
O_{1j}^{L} = - O_{1j}^{R\, *}\;, \quad
O^{L} = \left(
\begin{array}{cc}
 -\frac{1}{\sqrt{2} \sqrt{2+a^2}} & -\frac{a}{2 \sqrt{2+a^2}} \\
 -\frac{1+a^2}{2+a^2} & \frac{a}{\sqrt{2} \left(2+a^2\right)} \\
 \frac{a}{\sqrt{2} \left(2+a^2\right)} & -\frac{4+a^2}{4+2 a^2}
\end{array}
\right) \;, \quad
O^{R} = \left(
\begin{array}{cc}
 \frac{1}{\sqrt{2} \sqrt{2+a^2}} & \frac{a}{2 \sqrt{2+a^2}} \\
 -\frac{1+a^2}{2+a^2} & \frac{a}{\sqrt{2} \left(2+a^2\right)} \\
 \frac{a}{\sqrt{2} \left(2+a^2\right)} & -\frac{4+a^2}{4+2 a^2}
\end{array}
\right) \;,
\end{eqnarray}
%%%%%%%%%%%%%%%%%%%%%%%%%%%%%%%%%%%
while those in \eqss{eq:OLp}{eq:ORp}{gZxx},
%%%%%%%%%%%%%%%%%%%%%%%%%%%%%%%
\begin{eqnarray}
O^{\prime\, L(R)}  =\left(
\begin{array}{cc}
 \frac{-1-a^2+\left(2+a^2\right) s_W^2}{2+a^2} & \frac{a}{\sqrt{2} \left(2+a^2\right)} \\
 \frac{a}{\sqrt{2} \left(2+a^2\right)} & -\frac{4+a^2-2 \left(2+a^2\right) s_W^2}{2 \left(2+a^2\right)}
\end{array}
\right)\;, \quad
O^{\prime\prime\, L}  = \left(
\begin{array}{ccc}
 0 & \frac{1}{\sqrt{2} \sqrt{2+a^2}} & \frac{a}{2 \sqrt{2+a^2}} \\
 \frac{1}{\sqrt{2} \sqrt{2+a^2}} & 0 & 0 \\
 \frac{a}{2 \sqrt{2+a^2}} & 0 & 0
\end{array}
\right)\;. \label{eq:anOLpp}
\end{eqnarray}
%%%%%%%%%%%%%%%%%%%%%%%%%%%%%%%%%%%%%%%
Finally, the Higgs couplings to neutral and charged fermions 
in \eqs{yhxx}{yhpp} are respectively:
%%%%%%%%%%%%%%%%%%%%%%
\begin{eqnarray}
Y^{h\chi^{0}\chi^{0}} = \frac{m}{v} \left(
\begin{array}{ccc}
 0 & 0 & 0 \\
 0 & \frac{2 \sqrt{2} a }{\left(2+a^2\right) } & \frac{\left(-2+a^2\right) }{\left(2+a^2\right) } \\
 0 & \frac{\left(-2+a^2\right) }{\left(2+a^2\right) } & -\frac{2 \sqrt{2} a }{\left(2+a^2\right) }
\end{array}
\right)\;, \qquad
Y^{h\chi^{-}\chi^{+}} = \frac{m}{v} \left(
\begin{array}{cc}
 \frac{2 \sqrt{2} a }{\left(2+a^2\right) } & \frac{\left(-2+a^2\right) }{\left(2+a^2\right) } \\
 \frac{\left(-2+a^2\right) }{\left(2+a^2\right) } & -\frac{2 \sqrt{2} a }{\left(2+a^2\right) }
\end{array}
\right)\;, \label{eq:hx0x0}
\end{eqnarray}
%%%%%%%%%%%%%%%%%%%%%%%%
while those to Goldstone bosons given in \eq{LGB}, can now be simply written as
%%%%%%%%%%%%%%%%%%%%%%%%%
\begin{eqnarray}
Y^{G^{0}\chi^{0}\chi^{0}} = \frac{i m}{v} \left(
\begin{array}{ccc}
 0 &- \frac{a}{\sqrt{2+a^2}} & \frac{\sqrt{2}}{\sqrt{2+a^2}} \\
 -\frac{a}{\sqrt{2+a^2}} & 0 & 0 \\
 \frac{\sqrt{2}}{\sqrt{2+a^2}} & 0 & 0
\end{array}
\right)\;, \qquad
Y^{G^{0}\chi^{-}\chi^{+}}= \frac{ i m}{v} \left(
\begin{array}{cc}
 0 & -1 \\
 1 & 0
\end{array}
\right)\, \forall{a}\;,
\end{eqnarray}
%%%%%%%%%%%%%%%%%%%
and 
%%%%%%%%%%%%%%%%%%%%%%%
\begin{eqnarray}
Y^{G^{+}\chi^{-}\chi^{0}} = \frac{m}{v}\left(
\begin{array}{ccc}
 \frac{a}{\sqrt{2+a^2}} & 0 & -1 \\
 -\frac{\sqrt{2}}{\sqrt{2+a^2}} & 1 & 0
\end{array}
\right)\;, \qquad
Y^{G^{-}\chi^{+}\chi^{0}} = \frac{m}{v}
\left(
\begin{array}{ccc}
 -\frac{a}{\sqrt{2+a^2}} & 0 & -1 \\
 \frac{\sqrt{2}}{\sqrt{2+a^2}} & 1 & 0
\end{array}
\right)\;.
\end{eqnarray}
%%%%%%%%%%%%%%%%%%%

Depending on  whether the chiral mass $m$ or the vectorial 
masses $M_{D}$ and $M_{T}$ are dominant, and for  $M_{D}>0$, 
there are two extreme limits for the model at hand
%%%%%%%%%%%%%%%%%%%%%
\begin{align}
\mathrm{``Majorana~ dominance'' } &: M_{T} \approx M_{D} \gg m \Rightarrow a \approx 0 \;, \quad
m_{\chi_{1}^{0}}^{2} \approx m_{\chi_{2}^{0}}^{2} \approx M_{D}^{2}\;, 
\quad m^{2}_{\chi_{3}^{0}} \approx M^{2}_{T}\;.\\[3mm]
\mathrm{``Dirac~ dominance '' } &: M_{T} \approx M_{D} \ll m \Rightarrow a \approx -\sqrt{2} \;, \quad
m_{\chi_{2}^{0}}^{2} \approx m_{\chi^{0}_{3}}^{2} \approx M_{D}^{2}+2 m^{2}  \;.
\label{eq:dd}
\end{align}
%%%%%%%%%%%%%%%%%%%%
The ``Majorana dominance'' limit corresponds more or less 
to the ``higgsino-wino'' scenario of the MSSM
where the first two neutral particle masses are degenerate, while the ``Dirac dominance'' limit is 
the imprint of a  large Yuakawa coupling in \eq{LDM}. It is  the latter case that in addition
to $SU(2)_{R}$-symmetry, it is protected by the 
global $U(1)_{X}$ symmetry. For example, plugging in $a=-\sqrt{2}$ into \eq{eq:hx0x0}, 
we immediately see that the Higgs couplings to new fermions 
become diagonal resulting in a vanishing, 
as long as $M_{D}\to 0$,
one loop corrections to the $h-\chi_{1}^{0}-\chi_{1}^{0}$ vertex, 
as we qualitatively confirmed in section~\ref{sec:LP} below  \eq{eq:Yhxx},
and as we shall see below in section~\ref{sec:direct}.

%%%%%%%%%%%%%%%%%%%%%%%%%
\subsection{Composition of  the lightest Neutral Fermion}
%%%%%%%%%%%%%%%%%%%%%%%%

As we showed in \eqs{eigen}{vecs}, in the symmetric limit $m_{1}=m_{2}$,
the neutral fermion mass matrix $\mathcal{M}_{N}$,
can be diagonalized analytically into three mass eigenstates 
%%%%%%%%%%%%%%%%
\begin{equation}
\ket{\chi_{i}^{0}} \ = \  O_{i1}\,  \ket{1} + O_{i2}\, \ket{2} + O_{i3}\, \ket{3} \;.
\end{equation}
%%%%%%%%%%%%%%%%
Following  conventional MSSM nomenclature~\cite{Jungman:1995df},
lets define the ``Doublet'' composition  of the $\chi_{i}^{0}$ as 
%%%%%%%%%%%%%%%
\begin{equation}
F_{D}^{i} = |O_{i2}|^{2} + |O_{i3}|^{2}\;.\label{eq:fd}
\end{equation}
%%%%%%%%%%%%%%%%%
Then we say that a state of $\chi_{i}^{0}$
 is  (D)oublet-like  if $F_{D}^{i} > 0.99$, it is (T)riplet like if $F_{D}^{i} < 0.01$
 and it is (M)ixed state if $0.01<F_{D}^{i} <0.99$. 
 %From time to time we shall refer to the 
 %special case with $F_{D}^{i} \approx 0.66$, as a well-tempered composition.

%%%%%%%%%%%%%%%%%%%%%%%%%
 \begin{figure}[htbp] %  figure placement: here, top, bottom, or page
   \centering
   \includegraphics[width=3in]{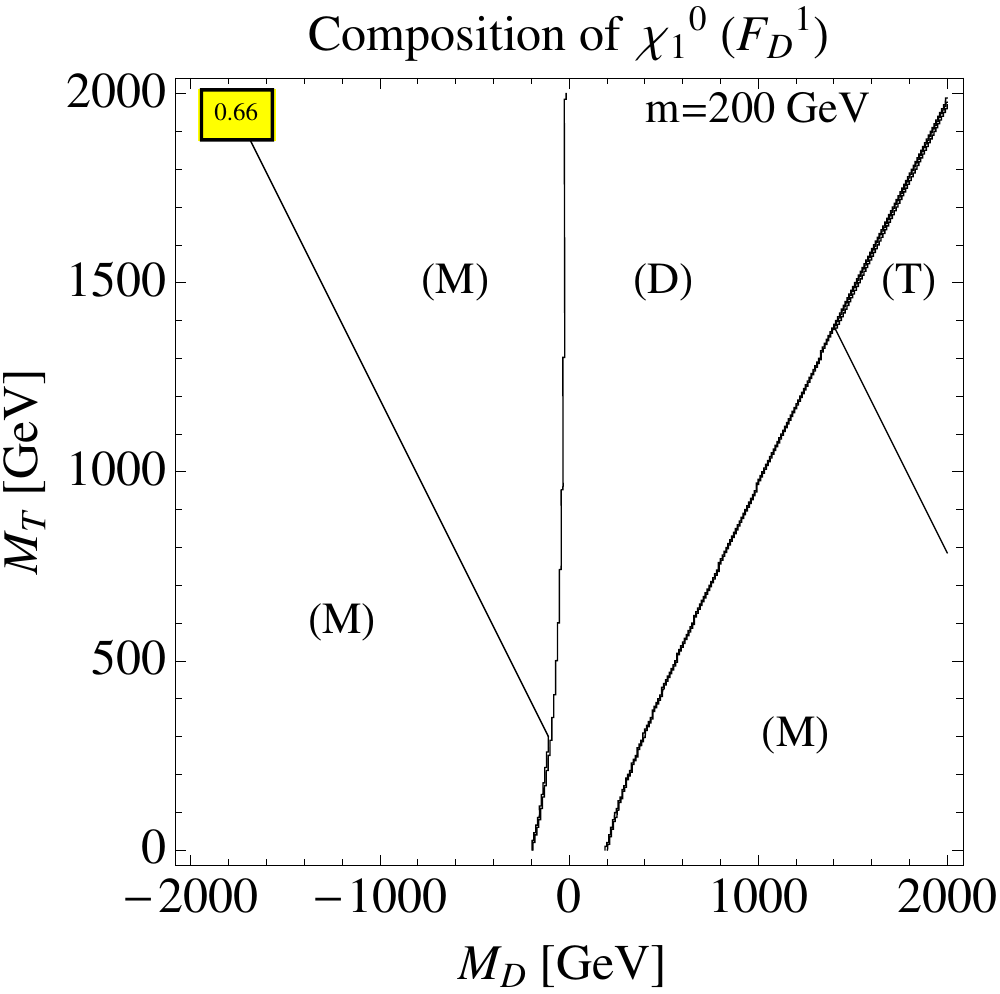} 
   \caption{\sl The composition of the WIMP in terms of (D)oublet, (T)riplet and (M)ixed states
   following the definition given in the paragraph below \eq{eq:fd}, on a $M_{D}$ vs. $M_{T}$ plane
   and for fixed (common) Yukawa coupling, $Y=m/v\simeq 200/174\simeq 1.15$. }
   \label{fig:comp}
\end{figure}
%%%%%%%%%%%%%%%%%%%%%%%%%%%

In Fig.~\ref{fig:comp} we present the  composition of the DM candidate particle $\chi_{1}^{0}$
on a $M_{D}$ vs. $M_{T}$ plane for fixed mass, $m=200$ GeV.
Both $Z$ and Higgs-boson couplings to pairs of $\chi_{1}^{0}$'s
vanish at tree level only  in the region denoted by (D)
(for Doublet) where $M_{D}$ is (most of the time) 
positive and equal to or less than $M_{T}$. 
It is mostly  in this region we are focusing  on in this article, because in this region 
the model evades, without further tweaks,
 direct DM detection experimental bounds. Note also that  for
light $M_{D} = \chi_{1}^{0}\lesssim 150~\mathrm{GeV} << m$,
the WIMP  composition satisfies (D) condition for every value of $M_{T}$. 
For negative values of $M_{D}$, $\chi_{1}^{0}$ is  a pure doublet only
in the region $|M_{D}| \le m$ but shrinks down to unacceptably small 
$M_{D}$ for large values of $M_{T}$; otherwise it is a mixed state 
everywhere in Fig.~\ref{fig:comp}.
For large $M_{D} \gg M_{T}$, the $\chi_{1}^{0}$-composition consists of mainly 
a triplet.

Note that when the lightest state is pure (D)oublet the heavier states are exactly 
an equal admixture of doublets and the triplet \ie $F_{D}^{2,3} = 0.5$.

%%%%%%%%%%%%%%%%%%%%%%%%%%
\section{Estimate of Electroweak Corrections}
\label{sec:STU}
%%%%%%%%%%%%%%%%%%%%%%%%%%%%
In the limit of large Yukawa couplings, $Y = Y_{1} = Y_{2} \simeq 1$, 
we generally expect large contributions from the new 
fermions, $\chi^{0},\chi^{\pm}$,
to $(Z,W)$-gauge boson self-energy one-loop
diagrams. In this section we investigate  constraints on the doublet-triplet fermion
model parameter space, $\{M_{D}, M_{T}, m\}$, from the 
oblique  electroweak parameters $S,T$ and $U$~\cite{Peskin:1991sw}.

Due to ${Z}_{2}$-parity symmetry, at one-loop level, there is no mixing
between the extra fermions, $\chi^{0},\chi^{\pm}$, and the SM leptons. 
Therefore corrections
to electroweak precision observables involving light fermions 
 arise only from gauge bosons vacuum polarisation 
 Feynman diagrams \ie there are only oblique electroweak corrections.   
In order to estimate these corrections it is convenient to calculate 
the  $S, T$ and  $U$ parameters, in the limit where 
$m_{\chi^{0}}, m_{\chi^{\pm}} \gtrsim m_{Z}$. This is true when the doublet mass 
$M_{D},$ is greater than $m_{Z}$ and $m$ is much greater than $m_{Z}$ 
(see Fig.~\ref{fig:masses}). We shall not consider the case of a  light dark matter
particle, $m_{\chi_{1}^{0}} \lesssim m_{Z}$.

Following closely the notation by Peskin and Takeuchi in \Ref{Peskin:1991sw}, 
we write,
%%%%%%%%%%%%%%
\begin{subequations}
\begin{align}
\alpha \, S &\equiv 4 e^{2} \: \frac{d}{dp^{2}} \left [\Pi_{33}(p^{2}) - \Pi_{3Q}(p^{2})
\right ] \biggl |_{p^{2}=0} \;, \\
\alpha \, T &\equiv \frac{e^{2}}{s_{W}^{2} c_{W}^{2} m_{Z}^{2}}\: 
\left [\Pi_{11}(0) - \Pi_{33}(0) \right ]\;, \\
\alpha \, U &\equiv 4 e^{2} \: \frac{d}{dp^{2}} \left [\Pi_{11}(p^{2}) - \Pi_{33}(p^{2})
\right ]\biggl |_{p^{2}=0} \;,  
\end{align}
\label{eq:STU}
\end{subequations}
%%%%%%%%%%%%%%%%
where $\alpha = e^{2}/4\pi$. In numerics we use input parameters 
from \Ref{PDG}, the bare value at lowest order 
$s_{W}^{2} = g^{'2}/(g^{'2}+g^{2}) \simeq 0.2312$ and the $Z$-pole mass $m_{Z}=91.1874$ GeV.  
We calculate corrections arising only from the extra fermions, $\chi_{i=1..3}^{0}, \chi_{i=1..2}^{\pm}$, 
to the $g^{\mu\nu}$ part of the gauge boson self energy amplitudes, 
$\Pi_{IJ}\equiv \Pi_{IJ}(p^{2})$,
where $I$ and $J$ may be photon ($\gamma)$, $W$ or $Z$, 
%%%%%%%%%%%%%%%%%%
\begin{subequations}
\begin{align}
\Pi_{\gamma\gamma} &= e^{2}\: \Pi_{QQ} \;, \\
\Pi_{Z\gamma} &= \frac{e^{2}}{c_{W}s_{W}} \: \left ( \Pi_{3Q} - s^{2} \Pi_{QQ} \right )\;,\\
\Pi_{ZZ} &= \frac{e^{2}}{c_{W}^{2} s_{W}^{2}} \: 
\left (\Pi_{33} - 2 s^{2} \Pi_{3Q} + s^{4} \Pi_{QQ}\right )\;,\\
\Pi_{WW} &= \frac{e^{2}}{s_{W}^{2}} \: \Pi_{11} \;,
\end{align}
\end{subequations}
%%%%%%%%%%%%%%%%% 
where $s_{W}=\sin\theta_{W}, c_{W}=\cos\theta_{W}$. We find,
%%%%%%%%%%%%%%%%%%
\begin{subequations}
\begin{align}
\Pi_{QQ} &= - \frac{p^{2}}{8 \pi^{2}} \sum_{i=1}^{2} \: \left [ \frac{2}{3}\, E - 
4 \,b_{2}(p^{2}, m_{\chi_{i}^{\pm}}^{2},m_{\chi^{\pm}_{i}}^{2}) \right ]\;, \\
%%%%%%%%%%%%%%%%%%%%%%%%%%
\Pi_{3Q} &= \frac{p^{2}}{16\pi^{2}}\: \sum_{i=1}^{2} (Z_{ii}^{L} + Z_{ii}^{R}) \:
\left [ \frac{2}{3}\, E - 4\, b_{2}(p^{2}, m_{\chi_{i}^{\pm}}^{2},m_{\chi^{\pm}_{i}}^{2}) \right ]\;, \\
%%%%%%%%%%%%%%%%%%%%%%%%%%%%%%%%
\Pi_{33} &= \frac{1}{16\pi^{2}} \: \sum_{i,j=1}^{2} 
\left [ (Z_{ij}^{L} Z_{ji}^{L} + Z_{ij}^{R} Z_{ji}^{R}) \:
G(p^{2}, m_{\chi_{i}^{\pm}}^{2},m_{\chi^{\pm}_{j}}^{2}) -2 \, Z_{ij}^{L} Z_{ji}^{R}\:  
m_{\chi^{\pm}_{i}}m_{\chi^{\pm}_{j}} \: I(p^{2}, m_{\chi_{i}^{\pm}}^{2},m_{\chi^{\pm}_{j}}^{2}) \right ]
\nonumber \\
&+ \frac{1}{16\pi^{2}}\: \sum_{i,j=1}^{3}\left [ O_{ij}^{\prime\prime \, L} O_{ji}^{\prime\prime \, L} \:
G(p^{2}, m_{\chi_{i}^{0}}^{2},m_{\chi^{0}_{j}}^{2}) + (O_{ij}^{\prime\prime \, L})^{2} \: 
m_{\chi^{0}_{i}}m_{\chi^{0}_{j}} \: I(p^{2}, m_{\chi_{i}^{0}}^{2},m_{\chi^{0}_{j}}^{2}) \right ]\;, \\
%%%%%%%%%%%%%%%%%%%%%%%%%%%%%
\Pi_{11} &= \frac{1}{16\pi^{2}} \: \sum_{i=1}^{3} \sum_{j=1}^{2} \left [ (|O_{ij}^{L}|^{2} + 
|O_{ij}^{R}|^{2} ) \: G(p^{2}, m_{\chi_{i}^{0}}^{2},m_{\chi^{\pm}_{j}}^{2}) - 
2 \Re e(O_{ij}^{L\, *} O_{ij}^{R}) \, m_{\chi_{i}^{0}}  m_{\chi^{\pm}_{j}}\:  I(p^{2}, m_{\chi_{i}^{0}}^{2},m_{\chi^{\pm}_{j}}^{2}) \right ] \;, 
%%%%%%%%%%%%%%%%%%%%%%%%%%%%%
\end{align}
\label{eq:pij}
\end{subequations}
%%%%%%%%%%%%%%%%%%%%%%%%%
where $Z_{ij}^{L(R)} \equiv O_{ij}^{\prime \, L(R)} -s_{W}^{2} \delta_{ij}$. In addition,
$E\equiv \frac{2}{\epsilon} -\gamma+ \log 4\pi - \log Q^{2}$ is the infinite part of 
loop diagrams. The various one-loop functions in eqs.~(\ref{eq:pij}) are given by  
%%%%%%%%%%
\begin{align}
G(p^{2},x,y) &= -\frac{2}{3} p^{2} \: E + (x+y) \: E + 4 \: p^{2}\: b_{2}(p^{2},x,y)
- 2 \left [ y\: b_{1}(p^{2},x,y) + x\: b_{1}(p^{2},y,x) \right ] \;, \\
I(p^{2},x,y) &= 2 \: E - 2\: b_{0}(p^{2},x,y) \;, \\
b_{0}(p^{2},x,y) &= \int_{0}^{1} dt \log \frac{\Delta}{Q^{2}} \;, \quad
b_{1}(p^{2},x,y) = \int_{0}^{1} dt \, t \,\log \frac{\Delta}{Q^{2}} \;, \\
b_{2}(p^{2},x,y) &= \int_{0}^{1} dt \, t\, (1-t)\,\log \frac{\Delta}{Q^{2}} \;, \quad
\Delta = t y + (1-t) x - t(1-t)\, p^{2} - i\epsilon\;.
\end{align}
%%%%%%%%%%%%%%%%%%%%%%%%%%
There are numerous useful identities, 
%%%%%%%%%%
\begin{align}
b_{0}(p^{2},x,y) &= b_{0}(p^{2},y,x)\;, \quad b_{2}(p^{2},x,y) = b_{2}(p^{2},y,x)\;, \\
G(p^{2},x,y) &= G(p^{2},y,x) \;, \quad I(p^{2},x,y) = I(p^{2},y,x) \;, \\
b_{1}(p^{2}, x,y) &= b_{0}(p^{2},y,x) - b_{1}(p^{2},y,x)\;, \quad 
b_{1}(p^{2},x,x) =\frac{b_{0}(p^{2},x,x)}{2}\;,
\end{align}
%%%%%%%%%%%%%%%%
that will help us to simplify our expressions below.
%
%%%%%%%%%%%%%%%%%%%%%%%%%%%%%%%%%%%%%
\begin{figure}[t]
  \centering
%
%   \begin{tabular}{ll}
%      \labellist
%\large \hair 2pt
%\pinlabel {\bf (a)} at 130 300
%\pinlabel {\bf (b)} at 545 300
%\endlabellist
%    \includegraphics[width=90mm]{contours-x1-M1=M2=200.eps}&
 %   \includegraphics[width=90mm]{contours-x1-x2-M1=M2=200.eps}\\
\subfloat[]{\includegraphics[width=85mm]{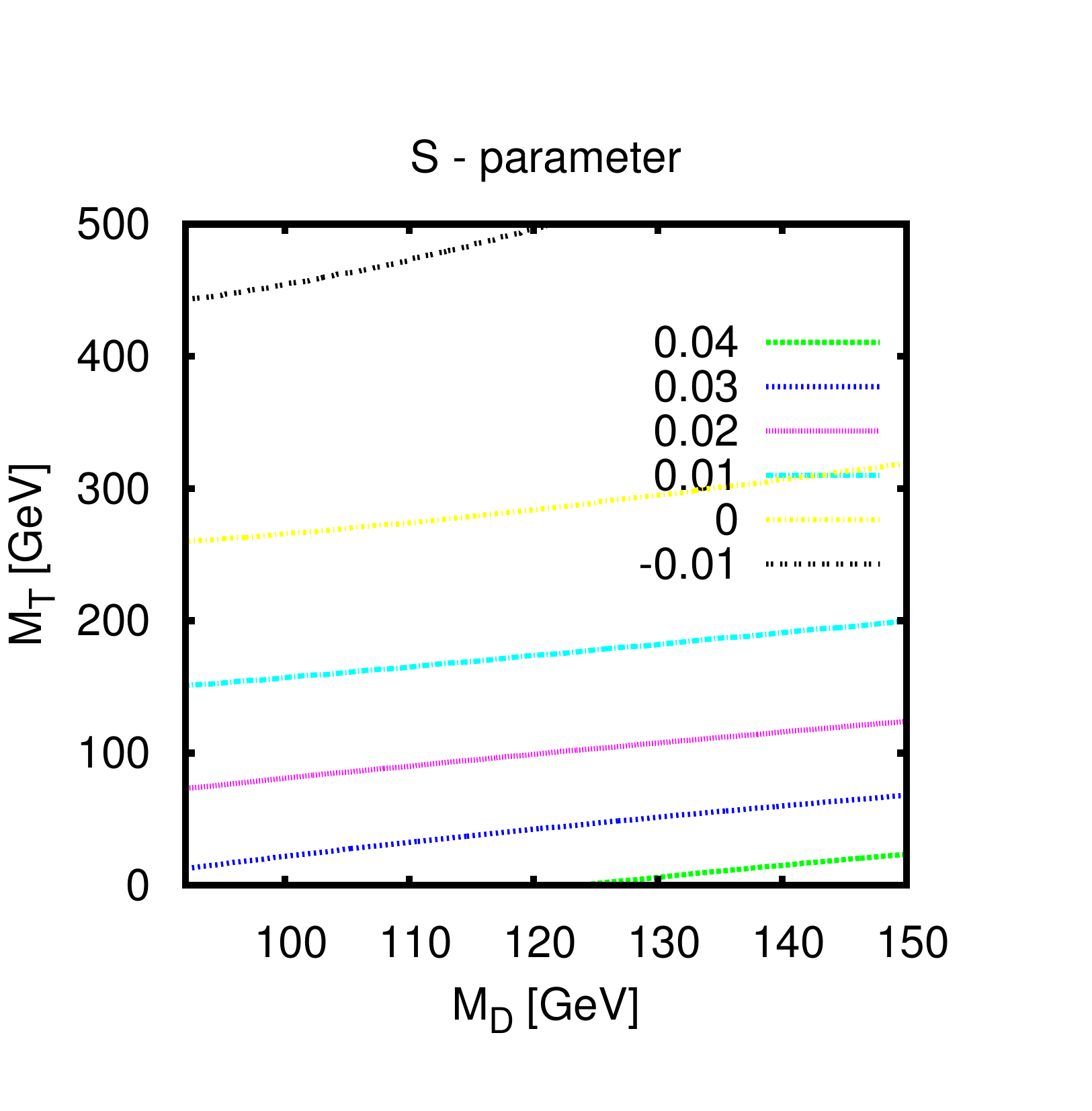}}%&\hspace*{-1.5cm}
\subfloat[]{\includegraphics[width=85mm]{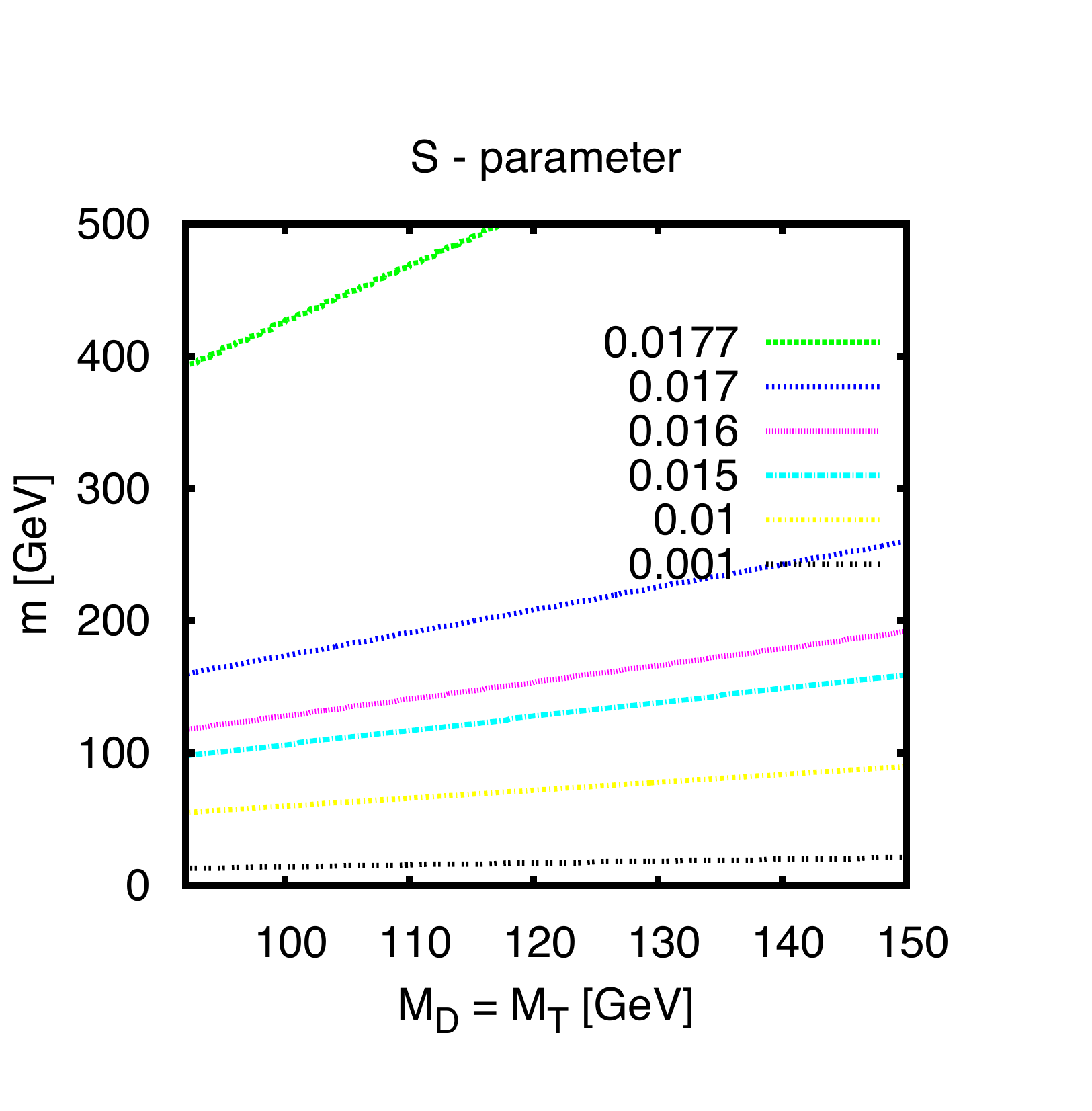}}
%  \end{tabular}
\caption{\sl Contour plots of the $S$-parameter on the  $M_{D}$ vs. $M_{T}$ plane (left)
for $m=200$ GeV  and on $M_{D}$ vs. $m$ plane (right) for $M_{T}=M_{D}$.}
\label{fig:STU}
\end{figure}
%%%%%%%%%%%%%%%%%%%%%%%%%%%%%%%%%%%%
%
Furthermore, in the exact $SU(2)_{R}$ limit  where $m_{1} = m_{2}$, there 
is no isospin breaking in $\bar{D}$-components  and therefore $T=0$, while the $S$-parameter
receives non-zero, non-decoupled, contributions due to the enlarged particle number 
 of the $SU(2)$-sector.  
Specifically, 
in the limit where $M_{D} = M_{T} \ll m=m_{1}=m_{2}$, there is a light
neutral fermion $(m_{\chi_{1}^{0}})$ 
and heavy degenerate other four (two neutral and two charged) fermions, 
with squared mass $x$, resulting in
%%%%%%%%%%%%%%%%%%%%%%%%%
\begin{subequations}
\begin{align}
\Pi_{3Q}^{\prime} (0) &\approx \frac{1}{16\pi^{2}} \: \biggl [ - 2 E + 2 \ln(\frac{x}{Q^{2}} )\biggr ] \;, \\
\Pi_{33}^{\prime} (0) &= \Pi_{11}^{\prime}(0)\approx \frac{1}{16\pi^{2}} \: \biggl [ - 2 E + 
2 \ln(\frac{x}{Q^{2}} )+ 
\frac{1}{18} \biggr ] \;, \label{p33p}\\
\Pi_{33}(0) &= \Pi_{11}(0) \approx \frac{1}{16\pi^{2}}\: \biggl [ \frac{3 x}{2} E -
\frac{3 x}{2} \ln (\frac{x}{Q^{2}}) + 
\frac{x}{4} \biggr ]\;.
\end{align}
\label{eq:appSTU}
\end{subequations}
%%%%%%%%%%%%%%%%%%%%%%%%
Plugging in eqs.~(\ref{eq:appSTU}) into eqs.~(\ref{eq:STU}) we arrive at the approximate values
expressions
%%%%%%%%%%%%%%%%%%%%%%
\begin{equation}
S \ \approx \ \frac{1}{18\pi}\;, \quad T \approx U \approx 0 \;.
\label{appresSTU}
\end{equation}
%%%%%%%%%%%%%%%%%%%%%%%%%%%%%%%%% 
This result is also confirmed numerically in Fig.~\ref{fig:STU}
where we draw contours of the  $S$-parameter
on  $M_{D}$ vs. $M_{T}$ plane (left plot) and on $M_{D}$ vs. $m$ plane (right plot). 
As it is shown, for large $m$ we obtain $S \to 1/18\pi \simeq 0.0177$
while for $m\to 0$ we obtain $S\to 0$,
as expected because in this case only
vector-like  masses will exist in $\mathscr{L}_{\rm Yuk}^{\rm DM}$ of \eq{LDM},
that make no contribution to parameter $S$.
Experimentally, we know~\cite{PDG} that when the $U$-parameter is zero, 
%PDG~\cite{PDG} quotes the 
%following result for 
the parameters $S$ and $T$ which fit the electroweak data are constrained to be
%%%%%%%%%%%%%%%%%%%
\begin{subequations}
\begin{align}
S &= 0.04 \pm 0.09\;, \label{PDG-S}\\
T &= 0.07 \pm 0.08 \;.
\end{align}  
\end{subequations}
%%%%%%%%%%%%%%%%%%
Predictions for the $S$-parameter  shown in  Fig.~\ref{fig:STU}a,b 
comfortably  fall within the bound of (\ref{PDG-S}). 
 In addition, even though it is not shown, the $T,U$-parameters are always negligibly small. 
%

%%%%%%%%%%%%%%%%%%%%%%%%%%%%%%
\section{The Thermal Relic Dark Matter Abundance}
\label{sec:relic}
%%%%%%%%%%%%%%%%%%%%%%%%%%%%%%

%%%%%%%%%%%%%%%%%%%%%%%%%%%%%%%%%
\begin{figure}[t]
   \centering
   \includegraphics[width=3in]{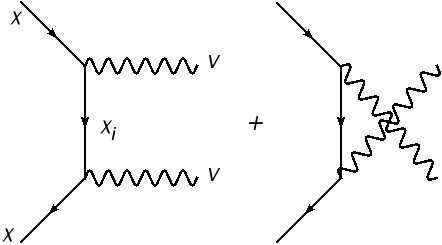} % requires the graphicx package
   \caption{\sl Lower level 
   Feynman diagrams contributing to annihilation cross section for the process $\chi+\chi \to V +V$
   for $V=W,Z$.}
   \label{fig:xxtoVV}
\end{figure}
%%%%%%%%%%%%%%%%%%%%%%%%%%%%%%%%

As we have seen, $V\chi_{1}^{0}\chi_{1}^{0}$ with $V=W,Z$ and $h\chi_{1}^{0}\chi_{1}^{0}$ 
are forbidden at tree level if $\chi_{1}^{0}$ is a pure doublet \ie 
$m_{\chi_{1}^{0}} = M_{D}$, in the exact $SU(2)_{R}$-limit. 
Therefore,
the annihilation cross section for the lightest  neutral fermion
results solely from the following $t-$ and $u-$channel tree level Feynman diagrams,
shown in Fig.~\ref{fig:xxtoVV},
with neutral or charged  fermion exchange,  collectively shown 
as $\chi_{i}$, with axial-vector interactions
%%%%%%%%%%%%%%%%%%%
\begin{subequations}
\begin{align}
\chi_{1}^{0} \ +\ \chi_{1}^{0} & \rightarrow  W^{+} \ + \ W^{-} \;, \\
\chi_{1}^{0} \ +\ \chi_{1}^{0} & \rightarrow  Z \ + \  Z \;.
%\chi_{1}^{0} \ +\ \chi_{1}^{0} &\rightarrow  Z  \ + \  h \;.
\end{align}
\end{subequations}
%%%%%%%%%%%%%%%%%%%%%%%%%%%
All other processes vanish at tree level. This can easily be understood  by looking at  the matrix
forms of $O^{\prime\prime\,L}$ and 
$Y^{h\chi^{0}\chi^{0}}$ in \eqs{eq:anOLpp}{eq:hx0x0}.
Before presenting our results for the annihilation cross section
it is helpful to (order of magnitude) estimate the thermal 
dark matter relic density for $\chi_{1}^{0}$s.
Consequently, by expanding the total cross section 
as $\sigma_{Ann} v = a_{V} + b_{V} v^{2} +\, ...$~\cite{Drees:1992am,Jungman:1995df} and keeping 
only the zero-relative-velocity $a$-terms we find
%\footnote{Our $a_{V}$ and $b_{V}$'s 
%are by a factor of two smaller than the ones given in 
%\Ref{Drees:1992am,Jungman:1995df}.} 
(for $M_{D} = M_{T}$):
%%%%%%%%%%%%%%%%%%%%%%%
\begin{subequations}
\begin{align}
a_{W} &=  \frac{g^{4}\, \beta_{W}^{3}}{32 \pi} \: 
\frac{m_{\chi}^{2}}{(m_{\chi}^{2} + m_{\chi_{j}}^{2} - m_{W}^{2})^{2}} 
\ \xrightarrow[m \gg M_{D}]{m_{\chi_{j}} \gg m_{\chi}} \ \frac{g^{4}\, 
\beta_{W}^{3}}{32 \pi} \: \left(\frac{m_{\chi}}{m_{\chi_{j}}} \right )^{4} \: \frac{1}{m_{\chi}^{2}}
\;, \label{eq:aW} \\[2mm]
a_{Z} &=  \frac{g^{4}\, \beta_{Z}^{3}}{64 \pi \, c_{W}^{4}} \:
\frac{m_{\chi}^{2}}{(m_{\chi}^{2} + m_{\chi_{j}}^{2} - m_{W}^{2})^{2}} \ 
\xrightarrow[m \gg M_{D}]{m_{\chi_{j}} \gg m_{\chi}} \
\frac{g^{4}\, \beta_{Z}^{3}}{64 \pi \, c_{W}^{4}} \: 
\left(\frac{m_{\chi}}{m_{\chi_{j}}} \right )^{4} \: \frac{1}{m_{\chi}^{2}}
\;,  
\label{eq:aZ}
\end{align}
\label{eq:as}
\end{subequations}
%%%%%%%%%%%%%%%%%%%%
where $g\approx 0.65$ is the electroweak coupling, $\beta_{V}= \sqrt{1-m_{V}^{2}/m_{\chi}^{2}}$
for $V=W,Z$, and in order to simplify notation, we take
$m_{\chi}\equiv m_{\chi_{1}^{0}}$ to denote  the DM particle 
mass and $m_{\chi_{j}} \equiv m_{\chi_{j}^{0}} = m_{\chi_{j-1}^{\pm}} \ge m_{\chi}$ 
for $j=2,3$ [see \eq{cusmass}]  the heavier neutral and charged fermions of the DM-sector.
In the case where $M_{D}=M_{T}$, the heavier fermions are degenerate with mass,
$m^{2}_{\chi_{j}} = 2 m^{2} +  M_{D}^{2}$, and the mass spectrum pattern is similar to
the one shown in Fig.~\ref{fig:spec}.  
Following this pattern in \eq{eq:as} we have  taken the limit of $m\gg M_{D}$ 
or alternatively, $m_{\chi_{j}} \gg m_{\chi}$. 

\begin{figure}[t] %  figure placement: here, top, bottom, or page
   \centering
   \includegraphics[width=3in]{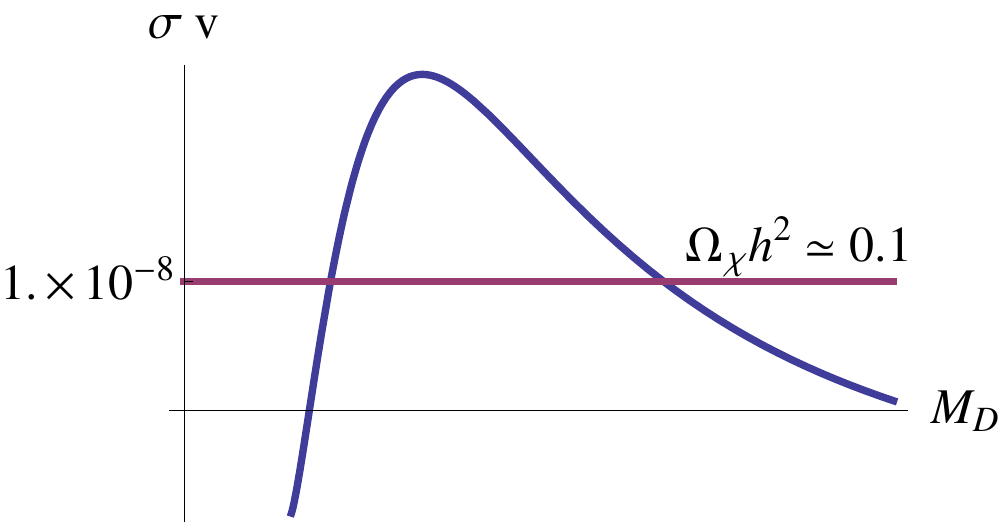} 
   \caption{Sketch of the resulting annihilation cross section. }
   %The maximum value of the curve is obtained at $M_{D}\approx \sqrt{2} m$.}
   \label{fig:xsec}
\end{figure}

Obviously, \eqs{eq:aW}{eq:aZ} viewed as functions of $M_{D}$, 
exhibit a maximum extremum since both $a$'s vanish in 
the limits of $M_{D} \to 0$ and $M_{D} \to \infty$ and, in addition,  they are positive definite. The
maximum cross section is obtained approximately at $M_{D} \approx \sqrt{2} m$. The situation 
is clearly sketched in Fig.~\ref{fig:xsec}. 
%%%%%%%%%%%%%%%%%%%%%
%%%%%%%%%%%%%%%%%%%%%%%
Once again,  we assume  that  particle-$\chi$
is a cold thermal relic, and that  its mass is about  few  tens
bigger than its freeze-out temperature.  Then, universe's critical  density 
times the Hubble constant squared (in units of 100 km/s/Mpc, $h^{2}\simeq 0.5$)
for $\chi$'s is 
%%%%%%%%%%%%%%%%%%%% 
\begin{equation}
\Omega_{\chi}\, h^{2} \sim 0.1\; \frac{10^{-8} \; \mathrm{GeV}^{-2}}{\sigma v} \;.
\label{ome}
\end{equation}
%%%%%%%%%%%%%%%%%%%%%%
Therefore, if the correct cross section, $\sigma v \approx 10^{-8}~\mathrm{GeV}^{-2}$,
 that produces the right relic density, $\Omega_{\chi}\, h^{2} \sim 0.1$,
happens to be below the maximum of $\sigma v$ in Fig.~\ref{fig:xsec} then there are
two of its points crossing the observed relic density: one for low $M_{D}$ and one
for high $M_{D}$ with the single crossing point being at  $M_{D} \approx \sqrt{2} m$.
The mass spectrum of  new fermions with high $M_{D}$ exhibits nearly
degeneracy in the first two states \ie $m_{\chi} = m_{\chi_{2}} \simeq M_{D}$. 
This shares similarities with the MSSM (or more precisely with the 
Split SUSY with $\tan\beta =1$ ``wino-higgsino''
scenario) for higgsino Dark Matter 
which is well studied and we are not going to pursue further. 
The other case, on the other hand, with 
low $M_{D} \lesssim  m$, exhibits a mass hierarchy between the DM candidate
particle ($\chi$) and all the rest ($\chi_{j}$) particles. 
It is the suppression factor $(m_{\chi}/m_{\chi_{j}})$ to the fourth power
in \eqs{eq:aW}{eq:aZ} that prohibits the cross section from taking on very large values. 
It is therefore evident  that this low $M_{D}$ scenario can provide the SM with a DM candidate
particle with mass $M_{D}$ that lies ``naturally'' at the EW scale as this suggested 
by the observation  $\sigma \approx 10^{-8}~\mathrm{GeV}^{-2}$, and is accompanied by 
heavy fermions few to several times heavier (depending on the value of $m$) than $M_{D}$.

Before proceeding further, it is worth looking back at Fig.~\ref{fig:masses},
the mass difference between the first two neutral states. For $m \gtrsim 100$ GeV 
the mass difference is always more than 50\% than the lightest mass $m_{\chi}$. 
This  in turn suggests that
\emph{no significant} contributions to $\Omega_{\chi} h^{2}$
are anticipated from co-annihilation effects~\cite{Griest:1990kh}.  
 %

%%%%%%%%%%%%%%%%%%%%%%%%%%%%%%%%%%%%%
\begin{figure}[t]
  \centering
%
%\begin{tabular}{ll}
 %  \labellist
%\large \hair 2pt
%\pinlabel {\bf (a)} at 130 350
%\pinlabel {\bf (b)} at 545 350
%\endlabellist
 \subfloat[]{\includegraphics[height=85mm]{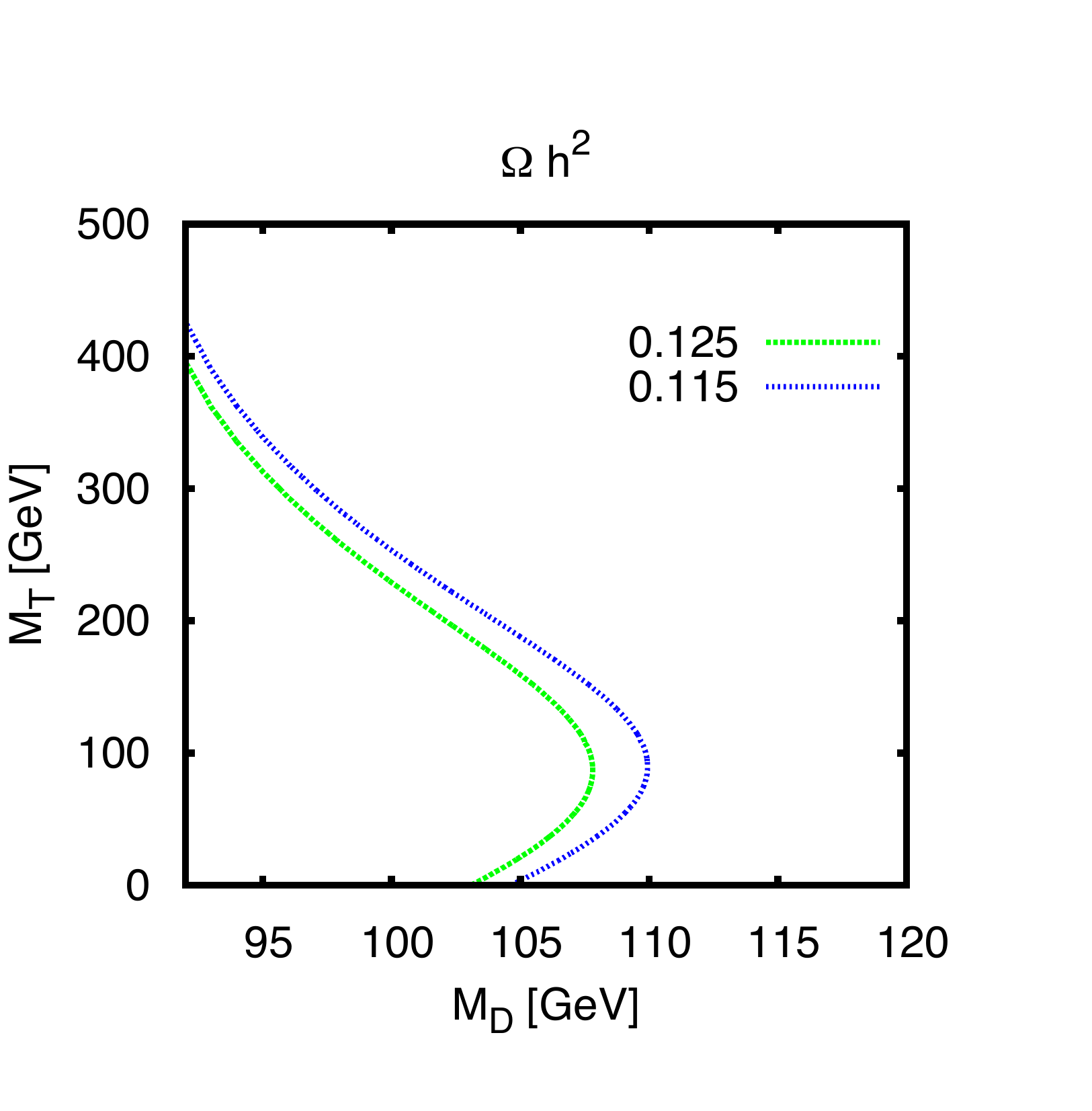}}% &
 % \hspace*{-1.5cm}  
  \subfloat[]{\includegraphics[height=85mm]{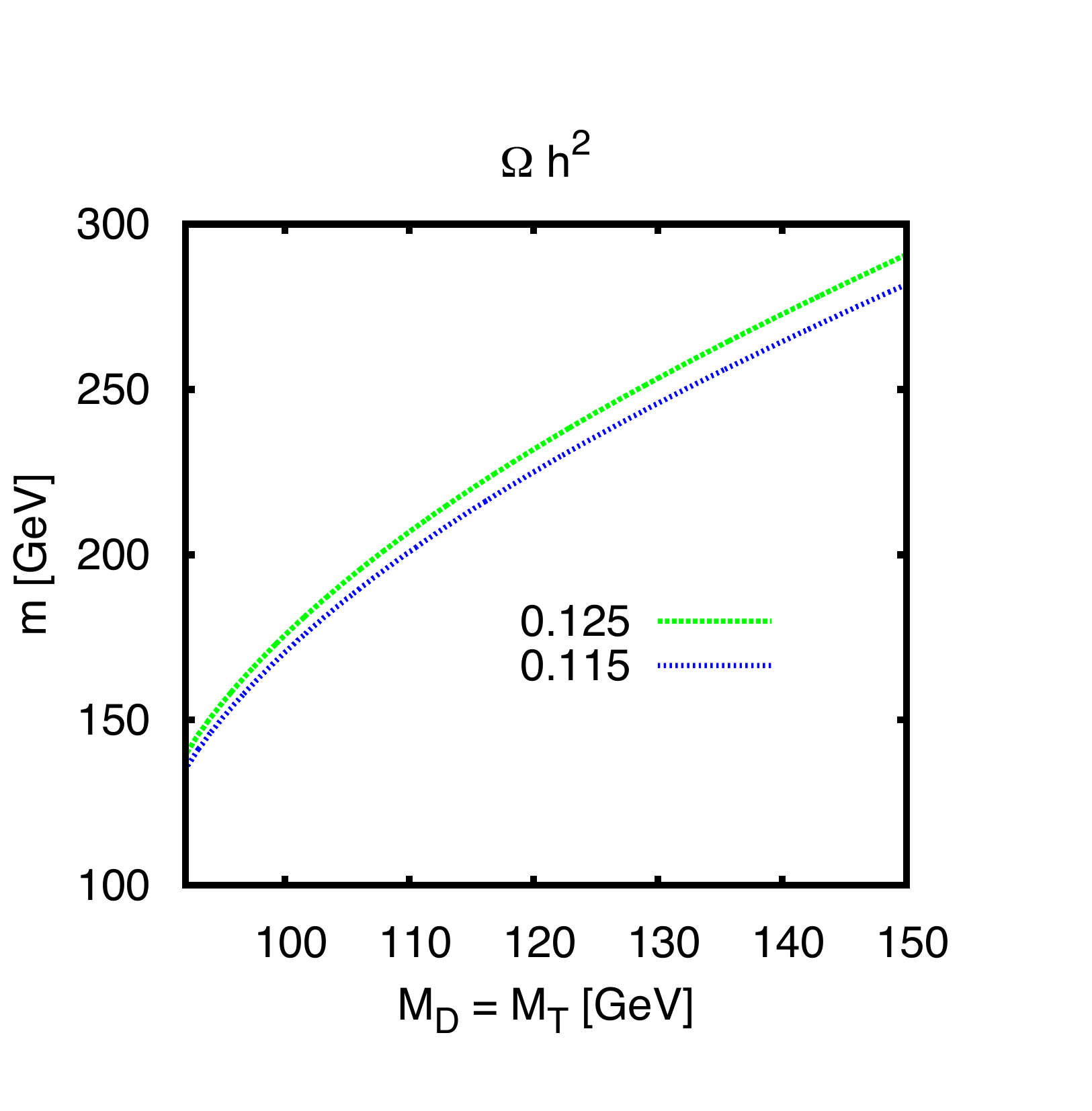} }
%    \end{tabular}
\caption{\sl (Left) Contour plots on the plane $M_{D}$ vs $M_{T}$ for the observed 
relic density $\Omega_{\chi} h^{2}$ [see \eq{eq:omobs}]. 
of the lightest neutral fermion with $m=200$ GeV. 
(Right) The same on  $M_{D}$ vs. $m$ plane for $M_{D}=M_{T}$. Recall that for both  plots
it is $m_{\chi} = M_{D}$.}
\label{figos}
\end{figure}
%%%%%%%%%%%%%%%%%%%%%%%%%%%%%%%%%%%%

In the end, we have calculated the today's relic density of the neutral, stable, 
and therefore, DM-candidate particle $\chi$. Our calculation
is a tree level one; see however comments below. 
Within the context of the (spatially flat) six-parameters
standard cosmological model, {\em Planck} experiment~\cite{Ade:2013zuv}  
reports a  density for cold, non-baryonic, dark matter, that is
%%%%%%%%%%%%%%%%%%%
\begin{eqnarray}
\Omega \, h^{2} = 0.1199 \pm 0.0027 \;. \label{eq:omobs}
\end{eqnarray}
%%%%%%%%%%%%%%%%%% 
The 2-$\sigma$ value is satisfied only in the area between the two lines 
in  both plots in Fig.~\ref{figos}. This happens for rather low $m_{\chi}=M_{D}$ in the region
$92 \lesssim m_{\chi_{1}^{0}} \lesssim 110$ GeV and for $M_{T} \lesssim 420$ GeV 
on the $M_{D}-M_{T}$ plane with fixed $m=200$ GeV, 
in  Fig.~\ref{figos}a.\footnote{We have not considered the 
case $M_{D} < M_{Z}$ as this would require further three body
decay analysis which is beyond the scope of this paper.} 
We also observe that the result for $\Omega_{\chi} h^{2}$ is not very sensitive to the 
triplet mass, $M_{T}$. Even vanishing $M_{T}$-values are in accordance with
the observed $\Omega_{\chi} h^{2}$, with  mass values
$m_{\chi}$  laying nearby the EW scale.  
(If $M_{D}$ is in the region $m_{W} < M_{D} < m_{Z}$, and if we neglect three body decays,
then the cross section becomes 
about  half the one for $M_{D}>m_{Z}$. This means that $\Omega h^{2}$ is
doubled and therefore larger $M_{T}$ (about twice as large) 
masses may be consistent with the observed  
$\Omega h^{2}$ values in \eq{eq:omobs}.)

We also consider the effect on $\Omega_{\chi} h^{2}$ from varying $m$ and  $M_{D}$, with
  $M_{D}=M_{T}$, in Fig.~\ref{figos}b. Obviously,
 the lower the $m$ is the lower the $M_{D}$ should be. For
   $m_{\chi} \simeq 91$ GeV the correct density is obtained for $m \simeq 140$ GeV. 
 As we move to heavier values \ie $m\approx 300$ GeV,  
 $M_{D}$ (which is equal to $m_{\chi}$), is required to be
 heavier, but not much heavier, than $M_{Z}$. However,   
 as we shall discuss in section~\ref{sec:Landau}, those heavy values
 of $m$ are not accepted by the vacuum stability constraint without modifying the model. 

%%%%%%%%%%%%%%%%%%%%%%%%%%
\begin{figure}[t]
   \centering
   \includegraphics[width=90mm]{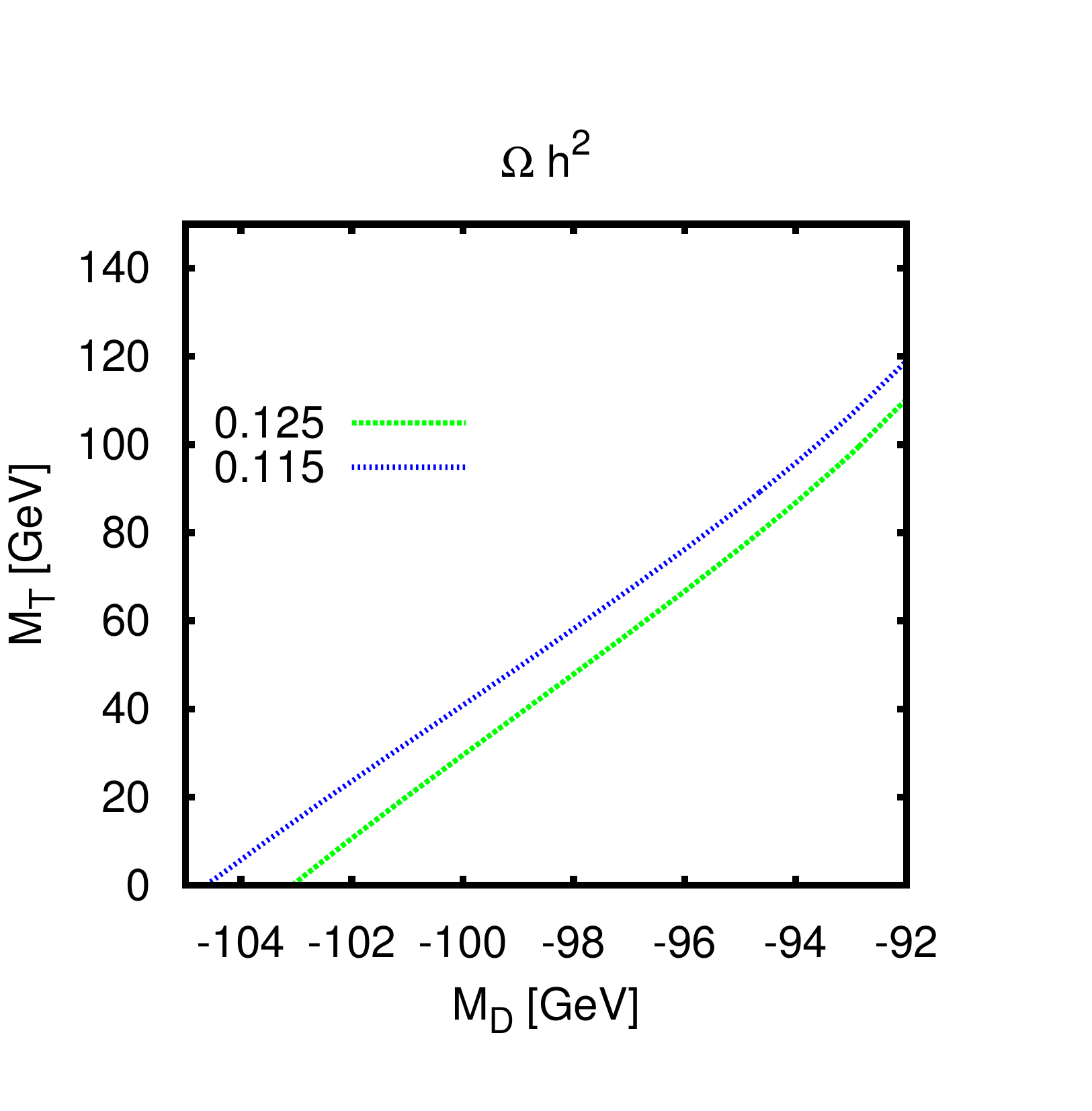} 
   \caption{Same as Fig.~\ref{figos}a  but for negative values of $M_{D}$.}
   \label{figosn}
\end{figure}
%%%%%%%%%%%%%%%%%%%%%%

Consistent $\Omega_{\chi} h^{2}$ with observation is also achieved 
for negative values of $M_{D}$ in the same region as for positive $M_{D}$
as it is shown  in Fig.~\ref{figosn}. (This is the small area for negative $M_{D}$ shown in 
Fig.~\ref{fig:comp} where $\chi_{1}^{0}$ is Doublet).  The $M_{T}$ values  where this 
happens are limited in the mass region smaller than  about 120 GeV. The
EW $S$ parameter in this region is slightly moved upwards  but is still
consistent with  \eq{PDG-S}.  However, as we shall see below,
the  $M_{D}<0$ region suffers from huge suppression relative to  SM
in the $h\to \gamma\gamma$ decay rate. 
 
 One loop corrections to the annihilation cross section contribute only to
 the $b_{V}$-parameter \ie they are $p$-wave suppressed, if
 $m_{\chi} \lesssim (m_{Z} + m_{h})/2$. Our estimate, using the crude formula of \eq{eq:LET} below,
 shows that one loop induced $b_{V}$-terms are, numerically, about 10 times smaller than the tree
 level ones. However, if the above limit is not hold, then  ($s$-wave) $a$-terms are 
 coming into the final $\sigma_{Ann} v$. These terms could be of the same order
 as for the tree level $b$-terms and, in principle, for a precise $\Omega_{\chi} h^{2}$ prediction,
 they have to be included in the calculation.

We therefore conclude that, DM particle mass around the EW scale is possible
and this requires large couplings of the heavy fermions to the Higgs boson
{\it i.e.,} large $m=Y v$ with $Y\approx 1$.
and secondarily, relatively low values of triplet mass {\it i.e.,} $M_{T} \simeq M_{D}$.  
This scenario can be hinted  or completely excluded at the LHC because the couplings 
of the heavy new fermions (both neutral and charged) to the Higgs and gauge bosons
are, in general, not suppressed in the symmetry limit [see discussion in Section~\ref{sec:LHC}].

%%%%%%%%%%%%%%%%%%%%%
\section{Direct DM Detection}
\label{sec:direct} 
%%%%%%%%%%%%%%%%%%%%

\begin{figure}[t] %  figure placement: here, top, bottom, or page
   \centering
   \includegraphics[width=5in]{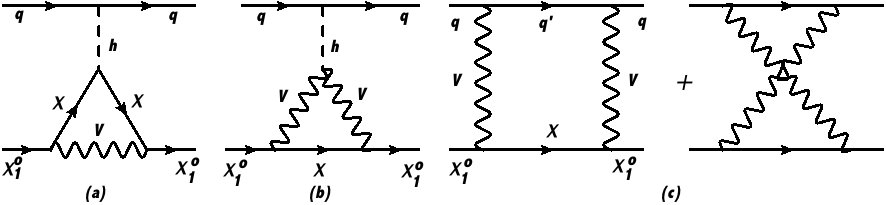} 
   \caption{\sl  Feynman diagrams (in unitary gauge) related to spin independent (SI)  elastic 
   cross section $\chi_{1}^{0} + q \rightarrow \chi_{1}^{0} + q$ where $q=u,d,s$ --
   light quarks. 
   Particle $V$ represents $W$ or $Z$ and $\chi$ represents $\chi_{i=1..2}^{\pm}$ or 
   $\chi_{i=1..3}^{0}$, respectively. 
   One loop self energy corrections are absent
   in the particular scenario we have chosen. }
   \label{fig:1loophiggs}
\end{figure}

Following the notation of Drees and Nojiri in \Ref{Drees:1993bu}, the Higgs boson mediated
part of the effective Lagrangian 
for light quark $(u,d,s)$ - WIMP (\ie the neutral fermion $\chi_{1}^{0}$) interaction is given by 
%%%%%%%%%%%%%%%%%
\begin{equation}
\mathcal{L}_{\mathrm{scalar}} = f^{(h)}_{q}\, \bar{\chi}_{1}^{0} \chi_{1}^{0}\, \bar{q} q \;.
\label{effxxqq}
\end{equation} 
%%%%%%%%%%%%%%%%%%%%%%%%%%
Note that in this model there are no tensor contributions (at 1-loop level) 
since $\chi_{1}^{0}$ does not interact directly with coloured particles (as opposed
to supersymmetric neutralino for example).  The next step is to form the nucleonic
matrix elements for the $\bar{q}q$ operator in \eq{effxxqq} and we write
%%%%%%%%%%%%%%%%%
\begin{equation}
\bra{n}{m_{q} \bar{q} q}\ket{n} = m_{n} f_{Tq}^{(n)} \;,
\end{equation}
%%%%%%%%%%%%%%%%%%%%%%%%%%%
where $m_{n}=0.94$ GeV, is the nucleon mass.
The form factors $f_{Tq}^{(n)}$ are obtained within chiral perturbation theory 
and the experimental measurements of pion-nucleon interaction term, and they
are subject to significant uncertainties. $f_{Tq}^{(n)}$ for $q=u,d$~\cite{Crivellin:2013ipa} 
are generically small
by, say, a factor of $O(10)$ compared to $f_{Ts}=0.14$ obtained from  \Ref{Gasser:1990ce} 
value which we adopt into our numerical findings  here. 
%
%\textcolor{red}{
However, bear in mind that $f_{Ts}$  
is  subject to large theoretical errors~\cite{Jungman:1995df,Crivellin:2013ipa}. 
For instance, the average value quoted from lattice calculations~\cite{Junnarkar:2013ac} is
$0.043\pm 0.011$, which is smaller by a factor of three  from the one obtained 
from chiral perturbation theory. This will result in, at least, a factor of $\mathcal{O}(10)$ 
reduction in the WIMP-nucleon cross section results,
 presented in Fig.~\ref{fig:1loophiggs}, below.
%}

The Higgs boson  couples to quarks and then to  
gluons through the one-loop triangle diagram. Subsequently, the gluons ($G$)
couple to the heavy quark current through the heavy quarks ($Q=c,b,t$)  in loop.
The analogous ($q\to Q$) matrix element in \eq{effxxqq} for $m_{Q} \bar{Q} Q$
can be replaced by the trace anomaly operator $-(\alpha_{s}/12\pi) G\cdot G$ to obtain 
%%%%%%%%%%%%%%%%%
\begin{equation}
\bra{n}{m_{Q} \bar{Q} Q}\ket{n} = \frac{2}{27} m_{n} \biggl [
1 - \sum_{q=u,d,s} f^{n}_{Tq} \biggr ] \equiv \frac{2}{27} m_{n} f_{TG}   \;.
\end{equation}
%%%%%%%%%%%%%%%%%%%%%%%%%%%
We are ready now to write down the effective couplings of $\chi_{1}^{0}$ to 
nucleons ($n=p,n$):
%%%%%%%%%%%%%%%%%%%%%%%%%%%    
\begin{equation}
\frac{f_{n}}{m_{n}} = \sum_{q} \frac{f_{q}^{(h)}}{m_{q}} f^{(n)}_{Tq} \ + \ \frac{2}{27} 
 \sum_{Q} \frac{f_{Q}^{(h)}}{m_{Q}} f_{TG} \;. \label{fnfacs}
\end{equation}
%%%%%%%%%%%%%%%%%%%%%%%%%%%%% 
Note that the bigger the $f_{Ts}$ is, the bigger the $f_{n}$ becomes. Also
note that $f_{q}^{(h)} \propto m_{q}$. Furthermore, for $f_{Ts} \simeq 0.14$ 
 the second term in \eq{fnfacs}, 
 which is formally a two loop contribution to $f_{n}$, is about a factor of two smaller
 than the first one. 
 Under the above assumption for the $f_{Ts}$ dominance we obtain
 $f_{p} = f_{n}$. In this case, the Spin Independent (SI) elastic scattering 
 cross section at zero momentum transfer,
 of the WIMP $\chi_{1}^{0}$ scattering off a given target nucleus with 
 mass $m_{N}$ in terms of the coupling $f_{p}$ is  
 %%%%%%%%%%%%%%%%%%
 \begin{equation}
 \sigma_{0 \mathrm{(scalar)}} = \frac{4}{\pi} \,  
 \frac{ m_{\chi_{1}^{0}}^{2} m_{N}^{4} }{(m_{\chi_{1}^{0}} + m_{N} )^{2} } \, 
 \left (\frac{f_{p}}{m_{n}} \right )^{2}   \;.
 \end{equation}
%%%%%%%%%%%%%%%%%%%
The perturbative dynamics of the model is contained in the factor $f_{p}$ and 
therefore, from \eq{fnfacs}, in $f_{q}^{(h)}$ and $f_{Q}^{(h)}$. 
%
%%%%%%%%%%%%%%%%%%%%%%%%%%%%%%%%%%%%%%
%\begin{figure}[t]
 %       \centering
%        \begin{subfigure}[b]{0.3\textwidth}
%                \includegraphics[width=\textwidth]{x1q->x1q1.png}
 %               \caption{$M_{1} ^{\chi}$}
%                \label{M1}
%        \end{subfigure}
 %       \begin{subfigure}[b]{0.3\textwidth}
 %               \includegraphics[width=\textwidth]{x1q->x1q2.png}
%                \caption{$M_{2} ^{\chi} $}
 %               \label{M2}
 %       \end{subfigure}
 %       \caption{The one loop correction to the Higgs vertex, where $\chi_i = \chi_{1,2} ^0 $ for  $V = Z$ and $\chi_i = \chi_{1,2} ^{\pm} $ for $V = W$}\label{1loop_higgs}
%\end{figure}
 %%%%%%%%%%%%%%%%%%%%%%%%%%%%%%%%%%%%%%%%%%%
 In this particular model the form factor $f_{q}^{(h)}$ reads, 
 \begin{equation}
 \frac{f_{q}^{(h)}}{m_{q}} \ = \ \frac{g\: [\Re e (Y^{h \chi^{0}_1 \chi^{0}_1 }) - 
 \delta Y^{h \chi^{0}_1 \chi^{0}_1 }]}{4 \, m_{W} \, m_{h}^{2}} \;.
 \end{equation}
 %%%%%%%%%%%%%%%%%%% 
The Higgs coupling to lightest neutral fermions is given in \eq{yhxx}. In particular, 
under the custodial symmetry consideration we adopt here, it is obvious from \eq{eq:hx0x0},
 that    $Y^{h \chi^{0}_1 \chi^{0}_1 }=0$, at tree level. 
Generic one-loop corrections will be proportional to,  $g^{2} Y/4\pi \approx 0.03$, which
can easily fall in the experimental exclusion region
from current direct experimental DM searches
 for large $Y\sim 1$ coupling
(see for instance eq.(3) in \Ref{Cheung:2012qy}).
We therefore need to calculate the one loop
corrections,  $\delta Y^{h \chi^{0}_1 \chi^{0}_1 } \equiv \delta Y$ to the 
$h \chi^{0}_1 \chi^{0}_1$-vertex.  
 
There is a fairly quick way to get an order of magnitude reliable calculation
of $\delta Y$ through the Low Energy Higgs 
Theorem (LEHT)~\cite{Ellis:1975ap,Vainshtein:1980ea,Kniehl:1995tn,Pilaftsis:1997fe}.  
Application of LEHT in  the region of our interest
i.e.,  $m_{\chi_{1}^{0}} \approx m_{W} \approx m_{h} \ll m_{\chi_{i}^{\pm}}$ 
or $M_{D} \approx M_{T} \approx m_{W} \ll m$,  and considering only Goldstone
boson contributions to $\chi_{1}^{0}$ one-loop self energy diagrams,
results in 
%%%%%%%%%%%%%%%%%%%%%%%
\begin{equation}
\delta Y \ =\  \frac{\partial}{\partial v}\delta M_{D}(v) \ \approx \ \frac{Y^{3}}{4 \pi^{2}} \,\frac{  M_{D} \, m}{M_{D}^{2} + 2 m^{2}} \;, \quad  M_{D} \approx M_{T} \approx m_{W} \approx m_{h} \ll m \;.\label{eq:LET}
\end{equation}
%%%%%%%%%%%%%%%%%%%%%%%%% 
Let's inspect  \eq{eq:LET}. First, the middle term explains trivially  why the Higgs coupling is
zero at tree level: the lightest eigenvalue of the neutral mass matrix is $M_{D}$ which
is independent on any vacuum expectation value (v.e.v.). Then because at one loop, the $\chi_{1}^{0}$ self-energies
involve only the heavy fermion masses (both charged and neutral) which depend on 
the v.e.v through $m=Y v$ or through $m_{W}, m_{Z}$ in the propagators of 
$\chi^{\pm}_{i}, \chi^{0}_{i=2,3}$ and $W,Z$ respectively, the one loop correction $\delta Y$ 
\emph{does not} in general vanish. Second, the third term of the equality \eq{eq:LET} shows that 
the effect increases by the third power of the Yukawa coupling $Y$ [recall  \eq{LDMsym}]
and vanishes when $M_{D} \rightarrow 0$ [the $U(1)_{X}$ symmetry limit]. 
As for the numerical approximation,
\eq{eq:LET} is always less than  20\% of the exact calculation (see below) even though
we have completely neglected the non-Goldstone diagrams that are proportional
to gauge couplings.  It is however a crude approximation which is only relevant 
when the new heavy fermions are far heavier than the $Z,W,h$-bosons as well as
from the lightest neutral fermion. 
 
%%%%%%%%%%%%%%%%%%%%%%%%%%%%%%%%
%%%%%%%%%%%%%%%%%%%%%%%%
%

In \ref{sec:appA}, 
we calculate the exact one-loop amplitude for the vertex $h-\chi_{1}^{0}-\chi_{1}^{0}$
with physical external  $\chi_{1}^{0}$ particles  at a zero Higgs-boson momentum transfer.
A similar calculation has been carried out in \Ref{Hisano:2004pv} for the MSSM  and 
 in \Ref{Hisano:2011cs} for minimal DM
models. However, due to peculiarities of this model
that have been stressed out in the introduction with 
respect to the aforementioned models,
 a general calculation is needed. 
 The one-loop corrected vertex amplitude arises from 
(a) and (b) diagrams\footnote{Note that, 
\Eq{eq:hx0x0} implies that
there are no self energy contributions to - $i \,\delta Y$ - at one-loop. }
depicted in Fig.~\ref{fig:1loophiggs} involving
vector bosons ($W$ or $Z$) and new charged ($\chi_{i=1,2}^{\pm}$) or neutral ($\chi_{i=1..3}^{0}$) 
fermions, as
%%%%%%%%%%%%%%%
\begin{equation}
i \,\delta Y = \sum_{j=(a),(b)}  (i \,\delta Y_{j}^{\chi^{\pm}} + i\, \delta Y_{j}^{\chi^{0}})\;.
\end{equation}
%%%%%%%%%%%%%%%%%%%%%%%%
Detailed forms, not resorting to $CP$-conservation, for $\delta Y$'s are given  in ~\ref{sec:appA}. 
 We have proven both analytically and numerically that when the external particles 
$\chi_{1}^{0}$ are
on-shell,  infinities cancel  in the sum of
the two  vertex diagrams in Fig.~\ref{fig:1loophiggs}a,b 
without the need of any renormalization prescription, and the resulting 
amplitude - $i \,\delta Y$ - is finite and renormalization scale invariant.

We have also carried out the one-loop 
calculation of the box diagrams in Fig.~\ref{fig:1loophiggs}c. 
The effective operators for box diagrams consist of scalar, $f_{q}^{\mathrm{(box)}}$ 
[like the $f_{q}$ in \eq{effxxqq}] and twist operators, $g_{q}^{(1)}$ and $g_{q}^{(2)}$
written explicitly for example in   \Ref{Drees:1993bu}.  
In the parameter space of our interest where $M_{D} \ll m$, 
the $f_{q}^{\mathrm{(box)}}$ contributions to  $f_{q}^{(h)}$ in \eq{fnfacs},
are in general two orders of magnitude smaller
than the vertex ones arising from Fig.~\ref{fig:1loophiggs}(a,b), 
and they are only important in the case where  the latter 
cancel out among each other. 
% 2-loop
%\textcolor{red}{
Moreover, it has recently been shown in 
\Refs{Hill:2013hoa,Hill:2014yka,Hisano:2012wm}
that, the full two-loop
gluonic contributions are relevant for a correct order of magnitude 
estimate of the cross section in the heavy WIMP mass limit, 
especially when adopting the  ``lattice'' value for $f_{T_{s}}$.  
We are not aware, however, of any study dealing with those corrections 
and WIMP mass around the electroweak scale which is the case  of our  interest. 
Such a calculation is quite involved and is beyond the scope 
of the present article.  
%}

\begin{figure}[t] %  figure placement: here, top, bottom, or page
   \centering
 %  \begin{tabular}{ll}
%   \labellist
%\large \hair 2pt
%\pinlabel {\bf (a)} at 160 250
%\pinlabel {\bf (b)} at 530 250
%\endlabellist
\subfloat[]{   \includegraphics[width=3in]{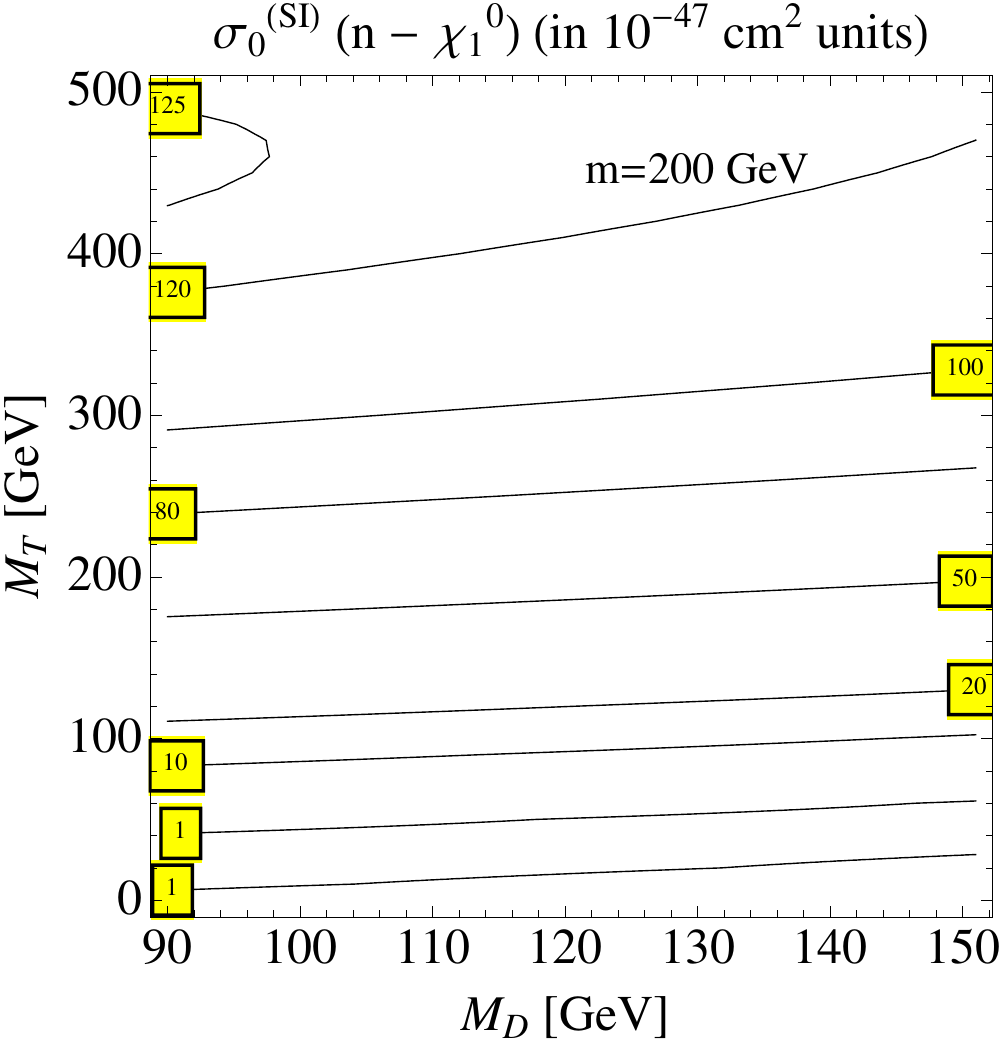}  }%&
 \subfloat[]{  \includegraphics[width=3in]{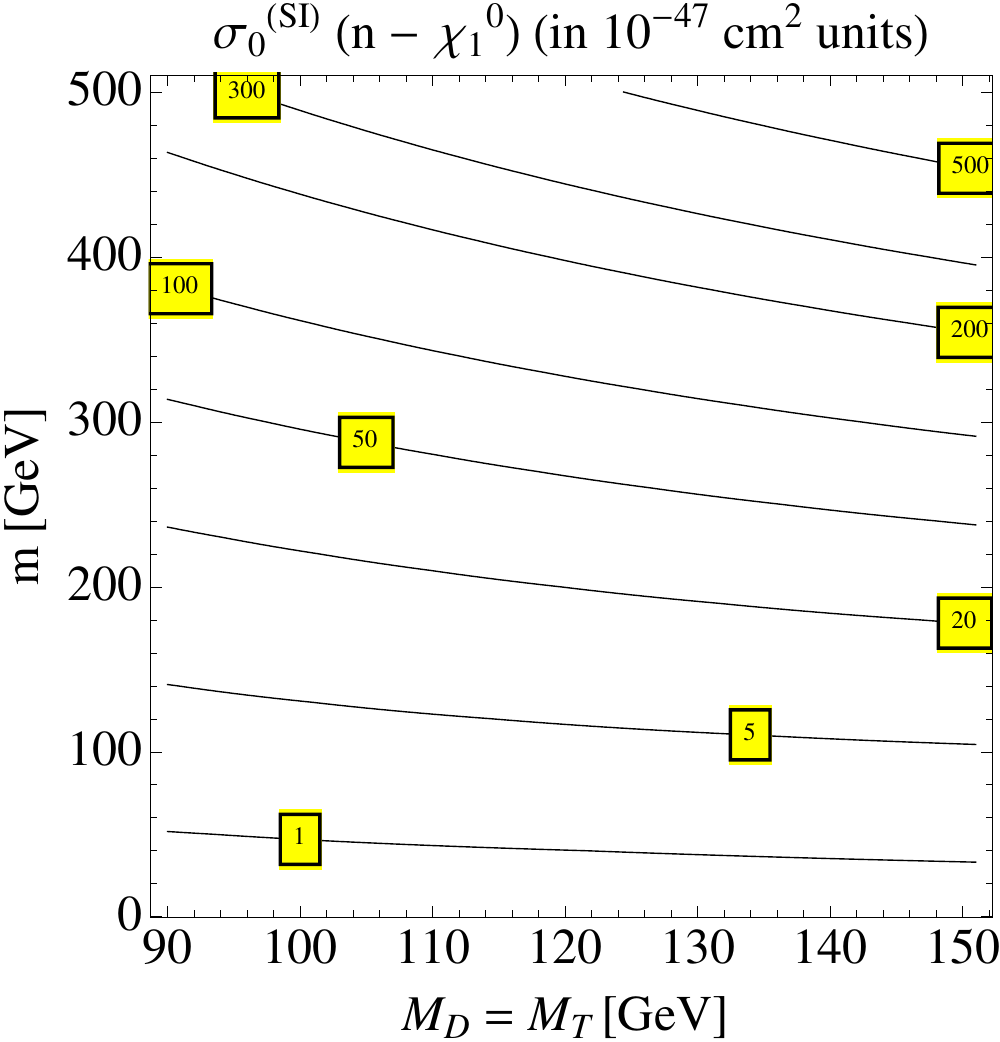}}
 %  \end{tabular}
   \caption{\sl Results (in boxed labels)
   for the Spin-Independent (SI) scattering cross section for the nucleon - WIMP 
   (n-$\chi_{1}^{0}$) in units of $10^{-47} cm^{2}$ on a $M_{D}$ vs. $M_{T}$ plane 
   for fixed parameter $m=Y v=200$ GeV (left) and on $M_{D}$ vs. $m$ plane  
   for fixed $M_{T}=M_{D}$ GeV (right).  }
   \label{fig:directX}
\end{figure}

In Fig.~\ref{fig:directX} we present our numerical results for the SI nucleon-WIMP 
cross section. The current LUX~\cite{LUX} (XENON100~\cite{XENON100})
experimental bounds for a 100 GeV WIMP  mass 
is $\sigma_{0}^{(SI)} \lesssim 1 (2)\times 10^{-45}~\mathrm{cm}^{2}$ at 90\% C.L. 
%%%%%%%%%%%%%%%%%%%%%%%%%%%%%%%%%
%%%%%%%%%%%%%%%%%%%%%%%%%%%%%%%%%%%%

From the left panel of Fig.~\ref{fig:directX} 
we observe that in the region where $M_{T}\ll M_{D} \ll m$ the cross section 
is by one to two orders of magnitude smaller than the current experimental bound.
More specifically, in the region where we obtain the right relic density 
[see Fig.~\ref{figos}a]
the prediction for the $\sigma_{0}^{\mathrm(SI)}$ is about to be observed only
for large values of $M_{T}$ ($M_{T}\approx 500$ GeV),
while it is by an order of magnitude smaller for
low values of $M_{T}$ ($M_{T} \lesssim 100$ GeV). 
There is  a region, around $M_{T} \approx 25$ GeV,
where  box corrections, that arise from the diagram in Fig.~\ref{fig:1loophiggs}c,
on scalar and twist-2 operators  become important because the vertex corrections
mutually cancel out. However, in this region the cross-section becomes two to four orders 
of magnitude smaller than the current experimental sensitivity.
%
%\textcolor{red}{
We also remark that 
$\sigma_{0}^{\mathrm(SI)}$ reaches a maximum value, 
indicated by the closed contour line in the upper left corner of Fig.~\ref{fig:directX}a,
and then starts decreasing for larger  $M_{T}$ and $M_{D}$ values,
a situation that looks like following the Appelquist-Carazzone
decoupling theorem~\cite{Appelquist:1974tg}. 
However, even at very large masses, $M_{D}$ and $M_{T}$, not shown in 
 Fig.~\ref{fig:directX}, there is a 
constant piece of $\delta Y$, and hence of $\sigma_{0}^{\mathrm(SI)}$, 
that does not decouple. This can be traced respectively in the second and the first 
terms of integrals $I_{4}^{V}$, and $I_{5}^{V}$  of \eq{I5f}, in the limit $M_{D}=M_{T}\to \infty$.
 This non-decoupling can 
also been seen in the heavy particle, effective field theory analysis of
\Ref{Hill:2013hoa} and also in \Refs{Cirelli:2005uq,Hisano:2011cs}.
%}
%
We have also checked numerically that $\sigma_{0}^{\mathrm(SI)}$ vanishes
at $M_{D}\to 0$ as expected from \eq{eq:LET} and from
 the $U(1)_{X}$-symmetry.\footnote{Because only
$\bar{D}_{1,2}$ are charged under $U(1)_{X}$ (not the Higgs boson), and 
 $\chi_{1}^{0}$ is a linear combination of only $\bar{D}$'s.}

In Fig.~\ref{fig:directX}b, we also plot 
predictions for the doublet-triplet fermionic model 
on SI cross section $\sigma_{0}^{\mathrm(SI)}$
 on $M_{D}$ vs. $m$ plane for $M_{T}=M_{D}$.  As we recall from \eq{eq:LET},
 the cross section increases with $m$ (or $Y$)
 as $m^{2}\propto Y^{2}$. 
 It becomes within current  experimental sensitivity 
 reach for $m \gtrsim 400$ GeV while for
  low  $m\approx 100$ GeV,   $\sigma_{0}^{\mathrm(SI)}$ 
  is about 100 times smaller.  
  Besides, for heavy $M_{D}$ and $m$ (upper right corner), $\sigma_{0}^{\mathrm(SI)}$
  becomes excluded by current searches although vacuum stability  bounds  hit first.
  If we compare with the corresponding plot for the relic density 
  in Fig.~\ref{figos}b, we see that the observed $\Omega_{\chi} h^{2}$ is allowed by 
  current experimental searches on  $\sigma_{0}^{\mathrm(SI)}$ but it will certainly
  be under scrutiny in the forthcoming experiments~\cite{Panci:2014gga}.
  
 Finally, for negative values of $M_{D}$ consistent with the observed  density depicted
 in Fig.~\ref{figosn}, it turns out that  $\sigma_{0}^{\mathrm(SI)}$ is by a factor of about $\sim 10$
 bigger than the corresponding parameter space for $M_{D}>0$ given in Fig.~\ref{fig:directX}a.
 In fact, the region of 1-loop cancellations happened for $M_{T} \approx 20$ GeV, do not
 take place for $M_{D}<0$.
  However, within errors discussed at the beginning of this 
 section, this is still consistent with current experimental bounds.

%%%%%%%%%%%%%%%%%%%%%%%%%
\section{Higgs boson decays to two photons} 
\label{sec:h2gg}
%%%%%%%%%%%%%%%%%%%%%%%%%

In the doublet-triplet fermionic model there are two pairs of electromagnetically 
charged fermions and antifermions, namely, $\chi_{1}^{\pm},\chi_{2}^{\pm}$. 
They have electromagnetic interactions with charge $Q=\pm1$ and interactions
with the Higgs boson, $Y^{h\chi^{-}\chi^{+}}$, given in general by   \eqs{lyint}{yhpp},
or in particular, in the symmetry limit, by \eq{eq:hx0x0}. These latter interactions 
are of similar size as of the top-quark-antiquark pairs with the Higgs boson \ie $Y\sim 1$.
Hence, we expect a substantial modification of the decay rate, 
$\Gamma(h \to \gamma\gamma)$ relative to the SM one\footnote{The Higgs boson 
production cross section is the same with the SM because the new fermions are uncoloured.}  
$\Gamma(h \to \gamma\gamma)_{SM}$, through the famous triangle graph~\cite{Ellis:1975ap},
%
%
%%%%%%%%%%%%%%%%%%%%%%%%%
\begin{figure}[t]
%\begin{tabular}{ll}
%\labellist
%\large \hair 2pt
%\pinlabel {\bf (a)} at 120 350
%\pinlabel {\bf (b)} at 540 350
%\endlabellist
   \centering
 \subfloat[]{  \includegraphics[width=90mm]{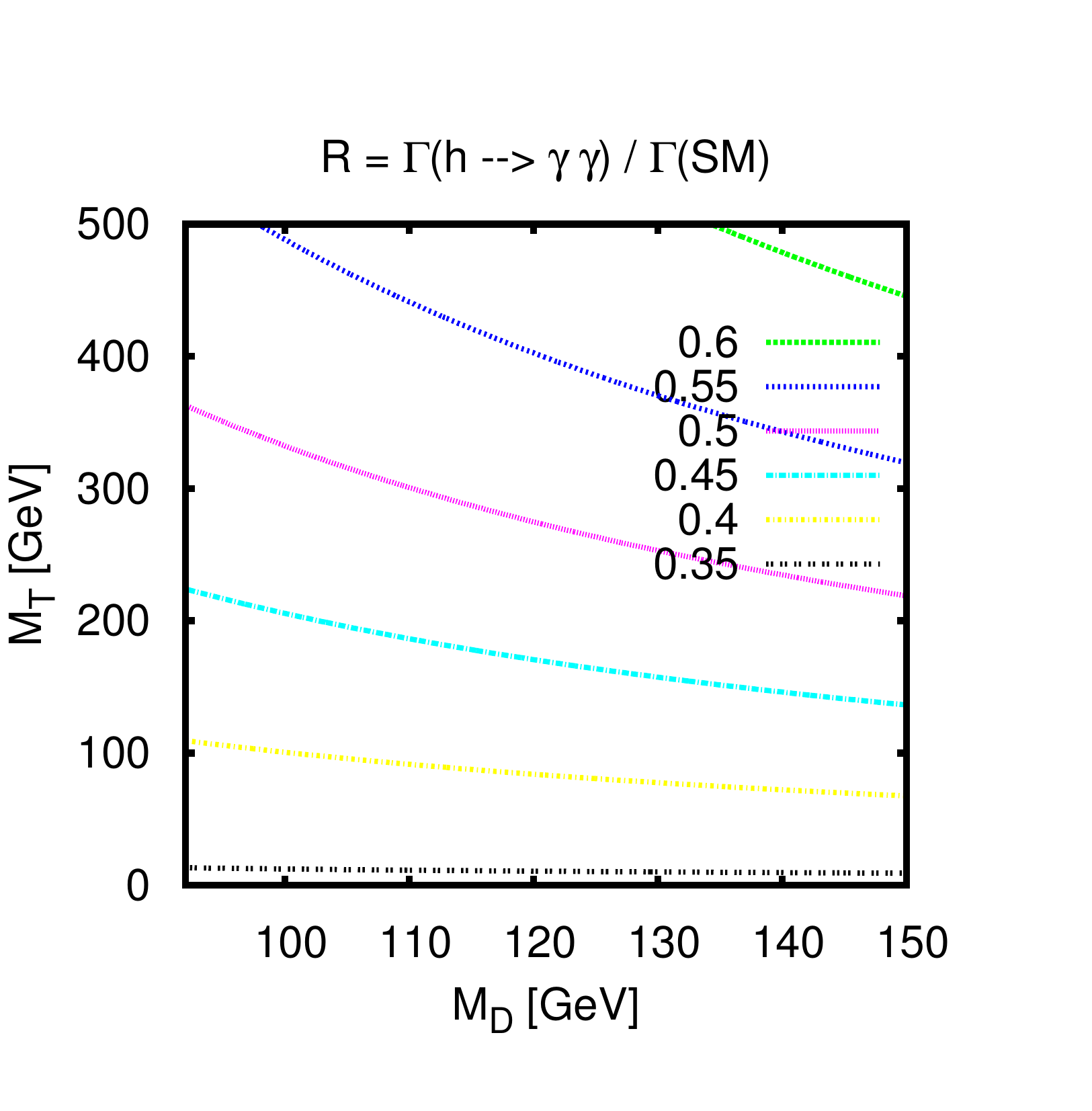} } \hspace*{-1.5cm} 
\subfloat[]{   \includegraphics[width=90mm]{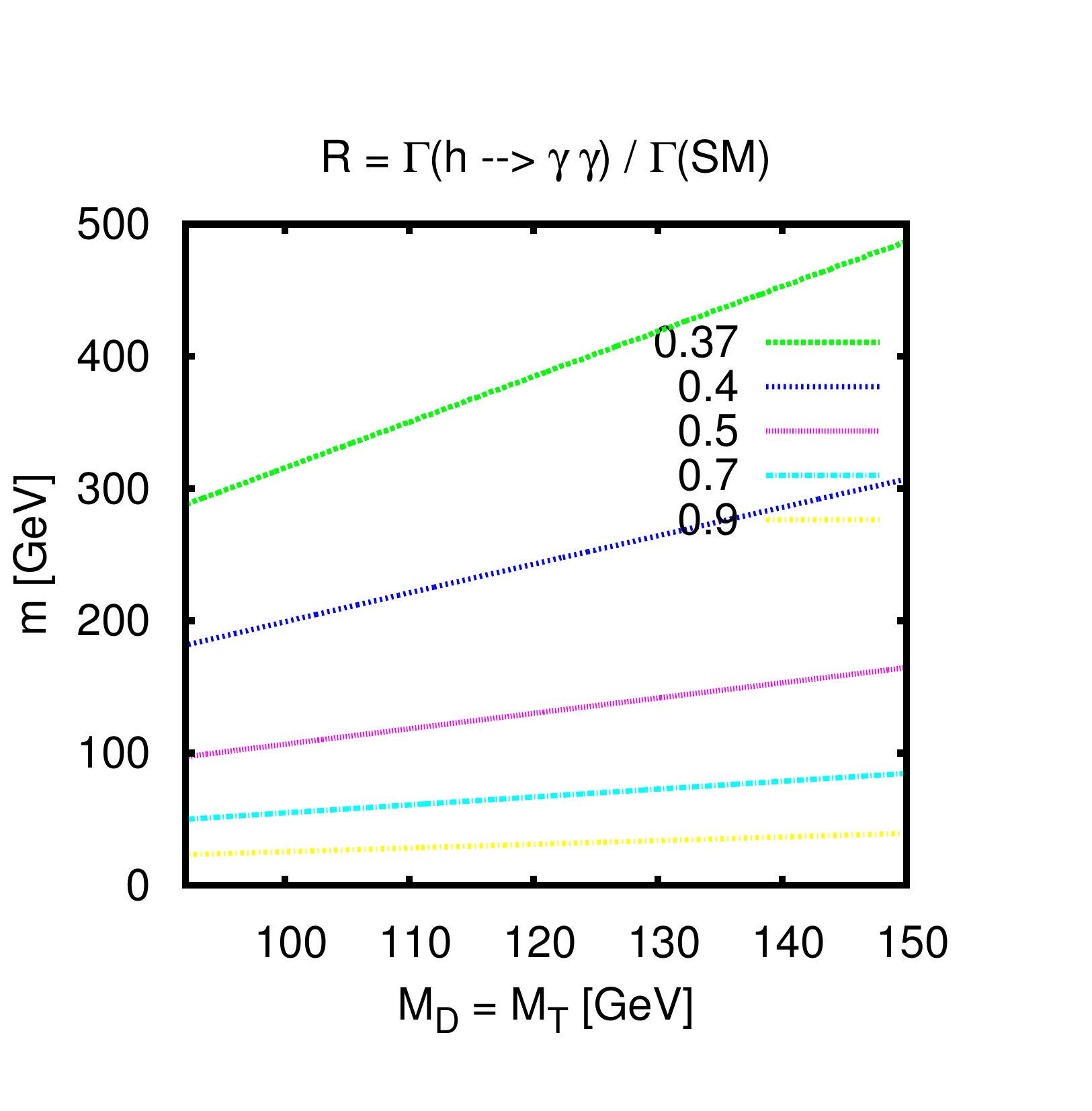}}
%   \end{tabular}
   \caption{\sl Contour lines for the ratio,
   $R=\Gamma(h\to \gamma\gamma)/\Gamma(h\to \gamma\gamma)_\mathrm{SM}$,
    for the decay rate of Higgs boson decays into two photons over the SM
   prediction on (a) $M_{D}$ vs. $M_{T}$ 
   plane with   $m=200$ GeV and (b) on $M_{D}$ vs. $m$ plane with 
   $M_{T}=M_{D}$.}
   \label{fig:h2gg}
\end{figure}
%%%%%%%%%%%%%%%%%%%%%%%%%%%%%%%%%%
%
involving $W$-gauge bosons, the top-quark ($t$) and the new fermions $\chi_{i}^{\pm}$. 
Under the assumption of real $M_{D}$,  $Y^{h\chi_{i}^{-}\chi_{i}^{+}}$ is also real, 
and we obtain:
%%%%%%%%%%%%%
\begin{equation}
R\equiv \frac{\Gamma(h\to \gamma\gamma)}{\Gamma(h\to \gamma\gamma)_{\mathrm{(SM)}}} \ = \  
\: \biggl |\: 1 \ + \  
\frac{1}{A_{\mathrm{SM}}} \: \sum_{i= \chi_{1}^{\pm}, \chi_{2}^{\pm}} \,  
\sqrt{2} \: \frac{Y^{h\chi_{i}^{-}\chi_{i}^{+}}\, v}{m_{\chi_{i}^{+}}} \: A_{1/2}(\tau_{i})\: \biggr |^{2}\;,
\label{hi2g}
\end{equation}
%%%%%%%%%%%%%%%%%%
%
where $A_{\mathrm{SM}} \simeq -6.5$ for $m_{h}=125$ GeV,
 is the SM result dominated by the $W$-loop~\cite{Dedes:2012hf},
with $\tau_{i} = m_{h}^{2}/4 m_{i}^{2}$ and $A_{1/2}$ is the well known function
given for example in \Ref{Djouadi:2005gi}\footnote{The Higgs-fermion vertex is parametrized here
as $\mathcal{L} \supset - Y\, h\: \bar{f}\: f +\mathrm{h.c.}$ and therefore   
for the top-quark Yukawa we obtain
$Y_{i} \to Y_{t}/\sqrt{2}$ from \eq{LSM} while for the new charged fermions 
$Y_{i} \to Y^{h\chi_{i}^{-}\chi_{i}^{+}}$ from \eqs{lyint}{yhpp}.}. 
%Note that $A_{1}(\tau \to 0)=-7$ and $A_{1/2}(\tau\to 0)=4/3$ come with opposite
%signs.  The top-quark contribution ($Q=2/3,N_{c}=3$) inside the modulo brackets is: 
%$\frac{4}{3} A_{1/2}(\tau_{t}) = +1.8$. 
The $\chi_{i}^{\pm}$-fermion contribution ($Q=1,N_{c}=1$), 
is also positive because the ratio, 
${Y^{h\chi_{i}^{-}\chi_{i}^{+}}}/{m_{\chi_{i}^{+}}}$,  is always positive when 
$m_{\chi_{1}^{0}} = M_{D}$,
as can be seen by inspecting \eqss{eq:hx0x0}{eq:dd}{eigen}.
After using the simplified (by symmetry) \eq{eq:hx0x0} with $a \approx -\sqrt{2}$,
we approximately obtain 
%%%%%%%%%%%%%%%%%%%%%%%%%%%
\begin{equation}
\sum_{i} \frac{\sqrt{2}\, m}{m_{\chi_{i}^{+}}} \: A_{1/2}(\tau_{\chi_{i}^{+}}) \approx + \frac{8}{3}\;,
\end{equation}
%%%%%%%%%%%%%%%%%%%%%
%
which means that $\Gamma(h\to \gamma\gamma)$  is  smaller than the SM expectation.
But how much smaller?
In Fig.~\ref{fig:h2gg}  
we plot contours of the ratio
$R\equiv \Gamma(h\to \gamma\gamma)/\Gamma(h\to \gamma\gamma)_{(\mathrm{SM})}$
 on $(M_{D}$ vs. $M_{T})$-plane  for $m=200$ GeV (Fig~\ref{fig:h2gg}a) 
 and $M_{D}$ vs. $m$-plane for $(M_{T}=M_{D})$ (Fig~\ref{fig:h2gg}b). 
 Our numerical results plotted in Fig.~\ref{fig:h2gg} are  exact at one-loop.
  We observe that the new charged fermions render  the ratio less than unity
%%%%%%%%%%%%%%%%%%%
\begin{equation}
R \lesssim 1 \;,
\end{equation}
%%%%%%%%%%%%%%%%%%
everywhere
 in the parameter space considered. 
 Let's look at this in a more detail. 
 The contribution
  of fermions $\chi_{i}^{\pm}$ in \eq{hi2g},
depends on the quantity\footnote{This quantity is obtained also by using the low energy Higgs theorem 
as in \Ref{Joglekar:2012vc} for the singlet-doublet DM-case.}
%%%%%%%%%%%%%%
\begin{eqnarray}
\sim  \:\frac{2\, m^{2}}{2\, m^{2} + M_{D}\, M_{T}}\;,
\label{eq:supap}
\end{eqnarray}
%%%%%%%%%%%%%%
which is always positive for $M_{D}, M_{T}>0$ 
\ie it adds to the top-quark contribution and subtracts 
from  the large and negative $W$-boson one
resulting in a suppressed $R$-ratio.  If instead we choose $M_{D}<0$,  
then for  $|M_{D} M_{T}| > \sqrt{2} |m|$, one can obtain
$R\gtrsim 1$, a situation which is explored in \Ref{ArkaniHamed:2012kq}.
As can be seen from Fig.~\ref{fig:comp} however, in this case 
the DM candidate particle $(\chi_{1}^{0})$ is not a pure doublet. It is instead a mixed state.
(In fact the states $\ket{1}$ and $\ket{2}$ are interchanged in \eq{vecs}).
As a consequence, 
there is a non-zero (and generically large) 
$h\chi_{1}^{0}\chi_{1}^{0}$-coupling already  present at tree level, and,
bear in mind fine tuning,  it is excluded by direct DM search bounds. 

By comparing areas with the observed relic density in  Figs.\ref{figos}(a,b) we see that,
the results for $0.35\lesssim  R \lesssim 0.5$ 
shown in  Figs.~\ref{fig:h2gg}(a,b) are within $1\sigma$-error compatible with  
current central values of CMS measurements~\cite{Chatrchyan:2012ufa} $(0.78 \pm 0.27)$
but are highly ``disfavoured''  by those from ATLAS~\cite{Aad:2012tfa} ones, $1.65 \pm 0.24(stat)^{+0.25}_{-0.18}(syst)$. 
The forthcoming second LHC run will be decidable in favour or against this outcomes here. 

Fig.~\ref{fig:h2gg}(a) or \eq{eq:supap}, 
shows also that when  $M_{T}$ becomes heavy
 the ratio $R$ approaches the current CMS central value.  
 This happens because one of the two charged fermion eigenvalues 
 becomes very heavy, $m_{\chi_{2}^{+}} \approx M_{T}$, and therefore it  is decoupled
 from the ratio. 
As we discussed in section~\ref{sec:relic}, large $M_{T}\sim 1$ TeV 
values, may be consistent with the observed $\Omega_{\chi} h^{2}$ for $m_{W} < M_{D} < m_{Z}$.
We have found that even in this case, $R$ is always smaller than $0.65$.

 If we assume that $M_{D}<0$ and $\chi_{1}^{0}$ pure doublet as shown 
 in Fig.~\ref{fig:comp}, then it is always $R<1$. In fact, using the input values from 
 Fig.~\ref{figosn} for the correct relic density, the suppression of $R$
 is even higher, $0.25 \lesssim R \lesssim 0.35$.   
 Alternatively, if we assume that $M_{D}$ is a general complex parameter, then the coupling,
 $Y^{h\chi^{-}_{1}\chi^{+}_{1}}$, is complex too. In this case
  one has to add the CP-odd Higgs contribution
 into \eq{hi2g} which is always positive definite. For large phases relatively large $M_{T}$ the ratio
 $R$ may be greater than one, however, again the direct detection bounds are violated 
 by a factor of more than 10-1000.   
 
Of course, if we increase $M_{D}$, the parameter space may be compatible with the observed
relic density seen in the right side of ``heavy'' $M_{D}$-branch in Fig.~\ref{fig:xsec}.
However, following our motivation for ``\emph{only} EW scale DM'' we do not 
discuss this region further which is anyhow very well known from MSSM studies. 

We therefore conclude that in the doublet triplet fermionic model
\emph{thermal DM relic abundance for low DM particle mass
$m_{\chi_{1}^{0}} \approx M_{Z}$, consistent with observation~\cite{Ade:2013zuv} and with 
 direct DM searches~\cite{LUX,XENON100} leads to a 
substantial suppression (45-75\%)
for the rate $\Gamma(h\to \gamma\gamma)$ relative to the SM
expectation.}  

We have also calculated the ratio $R$ for the Higgs boson decay into $Z\gamma$.
The results are similar to the case of $R(h \to \gamma\gamma)$. In particular, in the
parameter space explored in  Fig.~\ref{fig:h2gg}(a), we observe exactly the same shape of 
lines with a ratio slightly shifted upwards in the region, 
$0.4 \lesssim R(h\to Z\gamma) \lesssim 0.7$. This suppression is due to the same 
reason discussed  in the paragraph below \eq{eq:supap}.

%%%%%%%%%%%%%%%%%%%%%
\section{Vacuum Stability} 
\label{sec:Landau}
%%%%%%%%%%%%%%%%%%%%

The stability of the Standard Model vacuum is an important issue, so we need to find an energy scale ($\Lambda_{UV}$) where new physics is needed, in order to 
make the vacuum stable or a metastable (unstable with lifetime larger than the age of the universe). 
To make an estimate about the $\Lambda_{UV}$ of the theory,
one needs to calculate the tunnelling rate between the false and the true vacuum 
and impose that the SM vacuum has survived until 
today\footnote{The probability of the tunnelling has been calculated at tree level 
in \Ref{Coleman:1977py}.}.
%
%%%%%%%%%%%%%%%%%%%%%%%%%%%%%%%%%%%%  
 \begin{figure}[t!]
 \centering
  \includegraphics[width=0.55\linewidth]{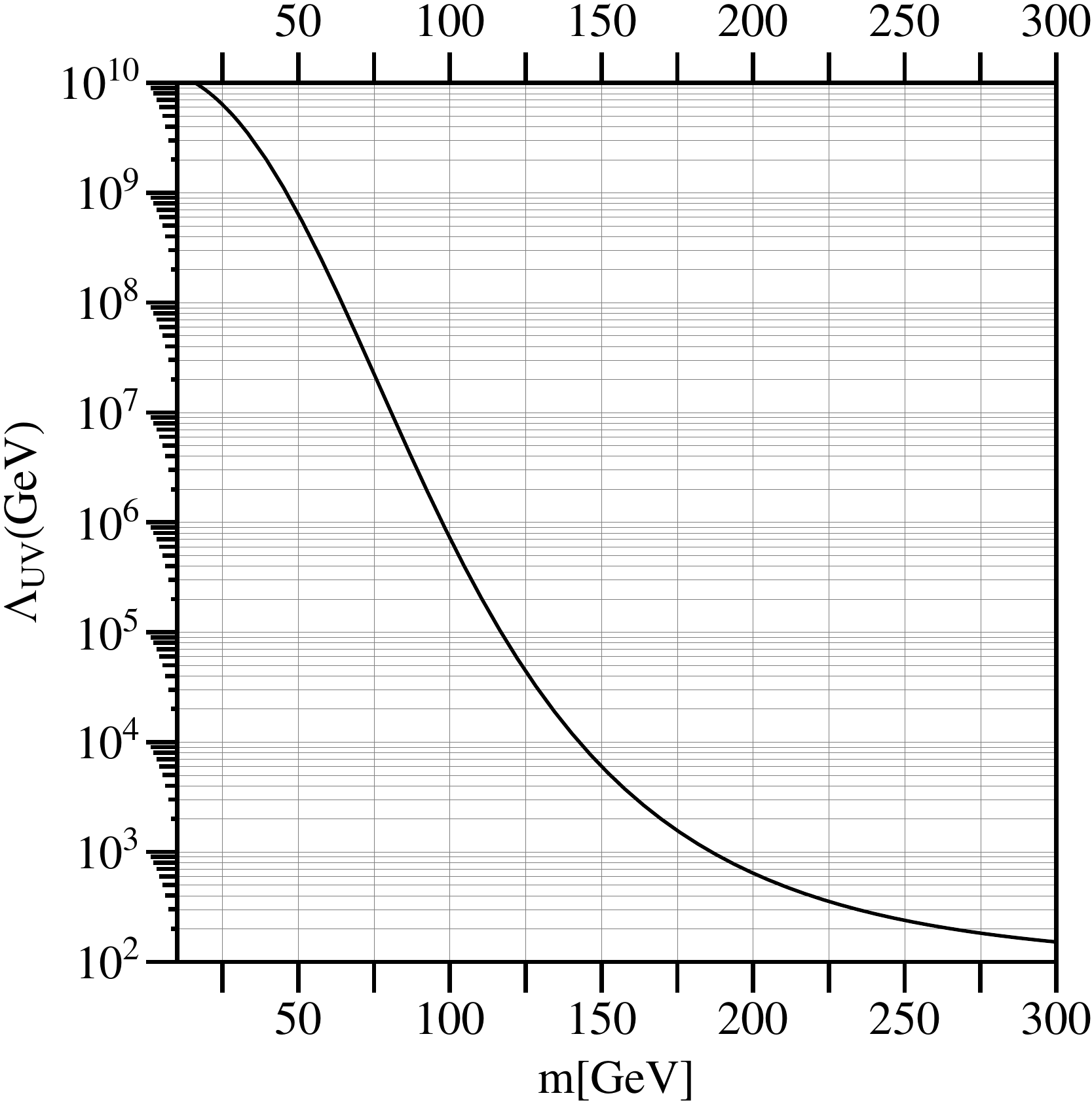}
  \caption{\sl The vacuum stability plot: $\Lambda_{UV}$ against $m=Y \upsilon$.}
     \label{LambdaUV}
  \end{figure}
%%%%%%%%%%%%%%%%%%%%%%%%%
%
Following \Ref{Isidori:2001bm}, we can see that the bound for the Higgs self coupling, $\lambda$,  becomes\footnote{This bound can also be found in \Ref{ArkaniHamed:2012kq}.}: 
%
%%%%%%%%%%%%%%%%%%%%%%%
\begin{equation}\label{luv}
 \lambda(\Lambda_{UV}) = \frac{4 \pi ^2}{ 3 \; \ln{\left( \dfrac{H}{\Lambda_{UV}}\right) }},
\end{equation}
%%%%%%%%%%%%%%%%%%%%%%%%%%
where $\Lambda_{UV}$ is the cut off scale and $H$ 
is the Hubble constant $H=1.5 \times 10^{-42}$ GeV.
In order to impose the contstraint  (\ref{luv}),  we also need to find the running parameter  
$\lambda$  by solving the renormalization group equations. 
The one-loop beta functions for the model at hand  are given in 
\Ref{Giudice:2011cg,Cheung:2012nb,ArkaniHamed:2012kq}\footnote{We need to make the substitutions 
$\tilde{g}_{2d} \rightarrow -Y_{1}$ 
and $\tilde{g}_{2u} \rightarrow - Y_{2}$ because of  different conventions with 
\Ref{Giudice:2011cg}.}, and we  solve this set of differential equations using as initial input
parameters:
%%%%%%%%%%%%%%%%%%%%%%% 
\begin{eqnarray}\label{in_con}
\alpha_3( M_Z )  &=  0.1184 \;, \quad
\alpha_2 ( M_Z )  = 0.0337\;, \quad
\alpha_1 ( M_Z )  = 0.0168 \;, \\
\lambda ( M_Z ) & =  0.1303\;, \quad
y_t ( M_Z )  =      0.9948 \;, \quad
M_Z  = 91.1876 \;\; \mathrm{GeV}.
\end{eqnarray}
%%%%%%%%%%%%%%%%%%%%%%%%%%%%%%%%%%%%%
%
The result for the cut off scale as a function of $m=Y\upsilon$ is given in Fig.~\ref{LambdaUV}. 
As we can see, $\Lambda_{UV} \approx 600$  GeV  for $m \approx 200$ GeV which is quite 
small while  $\Lambda \approx 20$ TeV for $m\approx 130$ GeV. 
The result for $\Lambda_{UV}$ in Fig.~\ref{LambdaUV} is only approximate. Threshold effects,
from the physical masses of the
doublet, triplet and even the top-quark,  together with comparable 
two-loop corrections to $\beta$-functions, which can be found for example in 
\Refs{Giudice:2011cg,Cheung:2012nb}, are
missing in Fig.~\ref{LambdaUV}. 
These effects may change the outcome for $\Lambda_{UV}$ by a factor of two or
so but they will not change the conclusion, that extra new physics is required already nearby
the TeV-scale. The form of new physics will probably be in terms of new scalar 
fields since  extra new fermions will make $\Lambda_{UV}$ even smaller. These scalars may be
well within reach at the second run  of the LHC~\cite{ArkaniHamed:2012kq} but
it is our assumption here that they do not intervene with the  DM sector.
%in sections~\ref{sec:relic} and \ref{sec:direct}. 
%

As far as the (1-loop) perturbativity of the Yukawa couplings $Y\sim 1.2$ (for $m=200$)
 and $Y_{t}$, is concerned,
these exceed the value $4\pi$ at around 
the respective scales, $10^{9}$ and $10^{10}$ GeV.
Given the modifications of the model that must be performed 
at  $\Lambda_{UV} \sim \mathrm{TeV}$
scale, the perturbativity  bound  is of secondary importance here. 
%
%For the same reason,
%unification of gauge couplings in this model is  of 
%peripheral importance. For perturbative $Y$, for example $Y\sim g$,
%the fermion content added is not capable to unify the gauge couplings
%and new matter has to be added. 

%%%%%%%%%%%%%%%%%%%%%%%%%%%%%%
\section{Heavy fermion production and decays}
\label{sec:LHC}
%%%%%%%%%%%%%%%%%%%%%%%%%%%%%%

The unknown new fermions that have been introduced 
into this model to accompany the DM mechanism can be searched for
at the LHC in a similar fashion as for charginos and neutralinos of the MSSM. Multilepton 
final states associated with missing energy may arise in three different ways from 
the decays of new fermion pairs:  $\chi^{+}_{i}\chi^{-}_{j}$,
$\chi^{\pm}_{i} \chi^{0}_{j}$, and $\chi_{i}^{0} \chi_{j}^{0}$.

%%%%%%%%%%%%%%%%%%%%%%%%
\subsection{Production}
%%%%%%%%%%%%%%%%%%%%%%

%%%%%%%%%%%%%%%%%%%%%%%%%%%%%%
\begin{figure}[t!]
   \centering
   \includegraphics[width=80mm]{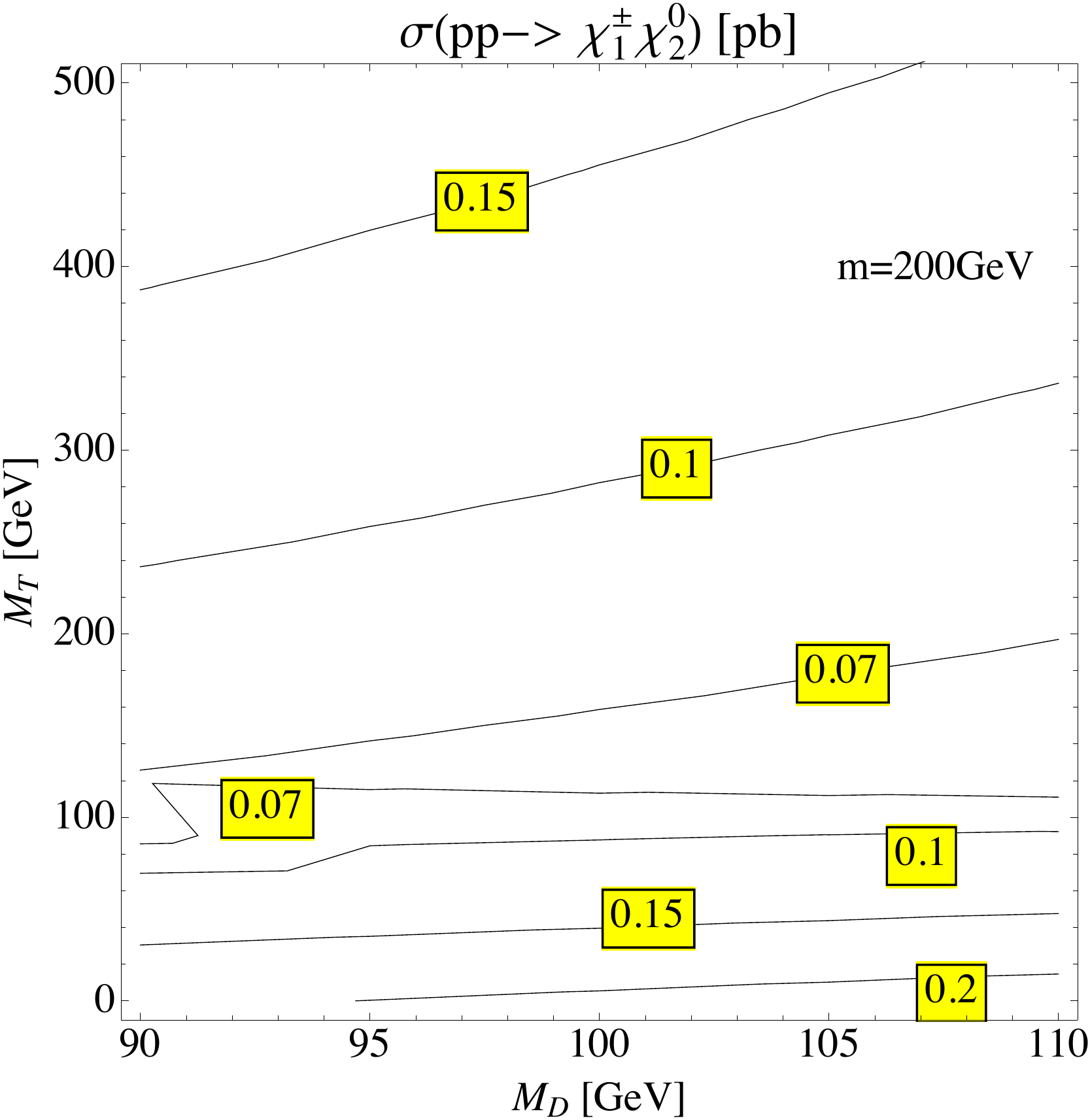} % requires the graphicx package
   \caption{\sl Contours of the production cross section for the new fermions, 
   $\sigma(pp\to \chi_{1}^{\pm} \chi_{2}^{0})$ [in $pb$], 
   on $M_{D}$ vs. $M_{T}$ plane, at LHC with $\sqrt{s} = 8$ TeV.}
   \label{fig:ppxx}
\end{figure}
%%%%%%%%%%%%%%%%%%%%%%%%%%%%%%%%

A  recent study at LHC~\cite{ATLAS:2013rla,CMS:2013dea}
has presented upper limits in the signal 
production cross sections for  charginos and neutralinos, in  the process
%$\chi^{+}_{1} \chi_{2}^{0}$
%%%%%%%%%%%%%%
\begin{equation}
p \ + \ p \ \rightarrow \ W^{*} \ \rightarrow \ \chi^{+}_{1} \ +  \ \chi_{2}^{0} \;,
\end{equation}
%%%%%%%%%%%%%%%%%%%%
which is mediated by the $W$-gauge boson. One can  use Fig.~9b from \Ref{ATLAS:2013rla} 
to set limits to the cross section and therefore to constrain the parameter space. 
This figure fits perfectly into our study since  
it assumes a) 100\% branching ratio for the $\chi^{+}_{1}$ and $\chi_{2}^{0}$ decays
as it is the case here [see section~\ref{dec} below] 
b) degenerate  masses for $\chi^{+}_{1}$ and $\chi_{2}^{0}$ as it is exactly the case
here as shown in \eq{cusmass}. The production cross section has been calculated 
in \Ref{Beenakker:1999xh} also including next to leading order QCD corrections.
The parton-level, tree level, result is
%%%%%%%%%%%%%%%%%
\begin{equation}
\frac{d \hat{\sigma}}{d\hat{t}}(u  +  d^{\dagger}  \rightarrow  W^{*}  
\rightarrow  \chi^{+}_{i} + \chi_{j}^{0} )  \ = \ \frac{1}{16 \pi \hat{s}^{2}} \, \left ( \frac{1}{3\cdot 4} 
\sum_{\rm spins} |\mathcal{M}|^{2} \right ) \;, \label{xsec}
\end{equation}
%%%%%%%%%%%%%%%%
where the factors $1/3$ and $1/4$ arise from colour and spin average
of initial states,   $\hat{s},\hat{t},\hat{u}$ are the Mandelstam variables at the parton level,  
and
%%%%%%%%%%%%%%%%
\begin{eqnarray}
\sum_{\rm spins} |\mathcal{M}|^{2}  =  |c_{1}|^{2} (\hat{u} - m_{\chi_{i}^{+}}^{2}) 
(\hat{u} - m_{\chi_{j}^{0}}^{2}) + |c_{2}|^{2}  (\hat{t} - m_{\chi_{i}^{+}}^{2}) 
(\hat{t} - m_{\chi_{j}^{0}}^{2})
  +  2 \Re e [c_{1} c_{2}^{*} ]          m_{\chi_{i}^{+}}      m_{\chi_{j}^{0}} \,  \hat{s} \;,
  \label{cs} 
 \end{eqnarray}
%%%%%%%%%%%%%%%%%%%
with the coefficients $c_{i}$ being
%%%%%%%%%%%%%%%%
\begin{equation}
c_{1} = -\frac{\sqrt{2} \, g^{2} }{\hat{s}-m_{W}^{2}} \, O^{L\,*}_{ji}\;, \qquad
c_{2} = - \frac{\sqrt{2} \, g^{2} }{\hat{s}-m_{W}^{2}} \, O^{R\,*}_{ji} \;.
\nonumber
\end{equation}
%%%%%%%%%%%%%%%%%
%
We let the indices $i=1,2$ and $j=1,2,3$ free as
there is a situation of a complete mass degeneracy between
the heavy neutral and charged fermions when $M_{D} = M_{T}$.  
Our result in \eqs{xsec}{cs} are in agreement with \Refs{Beenakker:1999xh,Dreiner:2008tw}.

By convoluting \eq{xsec} with the proton's pdfs 
and integrating over phase space we obtain 
in Fig.~\ref{fig:ppxx},
 the production cross section for $\sigma(pp\to \chi_{1}^{\pm} \chi_{2}^{0})$ [in $pb$]. 
In the region with correct DM relic density, 
we obtain typical values varying in the interval $(0.07 - 0.2) pb$  for $\sqrt{s}=8$ TeV.
This is about 1400-4000  events at LHC before any experimental cuts assuming 
$20 fb^{-1}$ of accumulated luminosity. 
This is  within current  sensitivity
search and  analysis has been performed 
by ATLAS~\cite{ATLAS:2013rla} and CMS~\cite{CMS:2013dea}
for simplified supersymmetric models. 
Looking for example in Fig.~9b in  ATLAS~\cite{ATLAS:2013rla}, for the same
parameter space as in our Fig.~\ref{fig:ppxx},
 the observed upper limit on the signal cross section varies in the
interval  (0.14-1.2) $pb$. 
In the region where $M_{D}=M_{T}$, all heavy fermions are
mass degenerate. In this case the total cross section is the sum of all possible
production modes $\chi_{1,2}^{\pm} \chi_{2,3}^{0}$, and the total cross 
section is about 0.15 $pb$ which is on the spot of  current LHC sensitivity 
($0.14$ pb)~\cite{ATLAS:2013rla}.

%Bear in mind however that, as mentioned in the introduction,  there are certain differences between
%(Split-)SUSY and the present model and therefore a dedicated Monte-Carlo analysis is needed. 

%%%%%%%%%%%%%%%%%%%%%%%%
\subsection{Decays}
\label{dec}
%%%%%%%%%%%%%%%%%%%%%%

Just by looking at a typical spectrum of the model in Fig.~\ref{fig:spec}, 
we see that the heavy fermions can decay on-shell to two final
states with a gauge boson and the lightest neutral stable particle.
Therefore, the lightest charged and the next to lightest neutral fermions
decay like
%%%%%%%%%%%%%%%%%%
\begin{subequations}
\begin{align}
\chi^{\pm}_{1} \ &\rightarrow \ \chi^{0}_{1}\ + \ W^{\pm} \;, \label{rw}\\
\chi^{0}_{2} \ &\rightarrow \ \chi^{0}_{1} \ + \ Z \label{rz}\;.
\end{align}
\end{subequations}
%%%%%%%%%%%%%%%%%%
 In our case where $\chi_{1}^{0}$ is a ``well tempered doublet'' there 
are no-off diagonal couplings to the Higgs boson,  like for example $h\chi_{1}^{0}\chi_{2}^{0}$. 
Therefore, particles $\chi_{1}^{\pm}$ and $\chi_{2}^{0}$ decay purely to final states following
(\ref{rw})  and (\ref{rz}) with 100\% branching fractions. 
The signature at hadron colliders is the well know from SUSY searches,
trileptons plus missing energy.

\begin{figure}[t!]
   \centering
%   \begin{tabular}{ll}
%   \labellist
%\large \hair 2pt
%\pinlabel {\bf (a)} at 130 350
%\pinlabel {\bf (b)} at 540 350
%\endlabellist
 \subfloat[]{  \includegraphics[width=80mm]{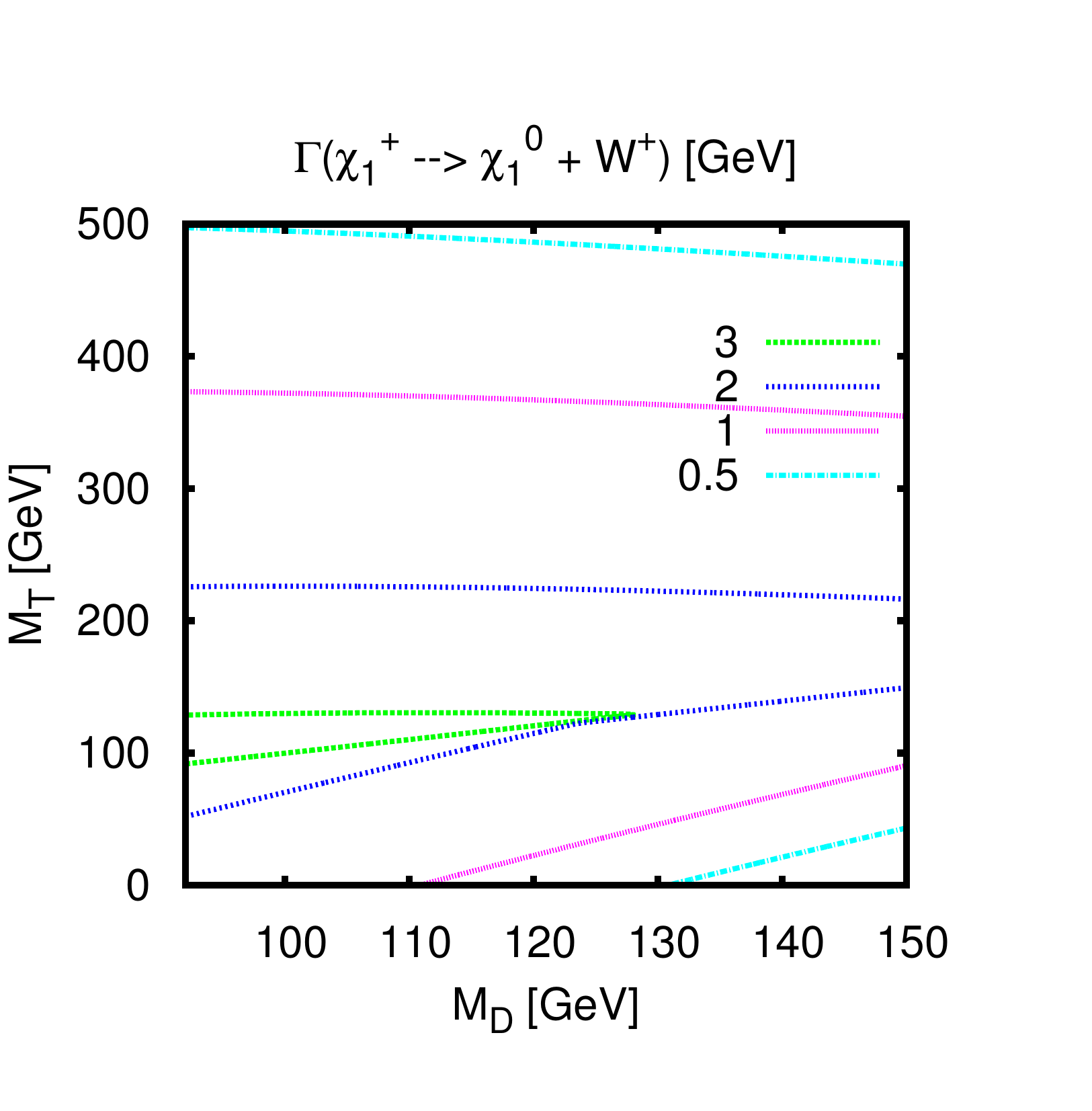} } \hspace*{-1.5cm}
\subfloat[]{  \includegraphics[width=80mm]{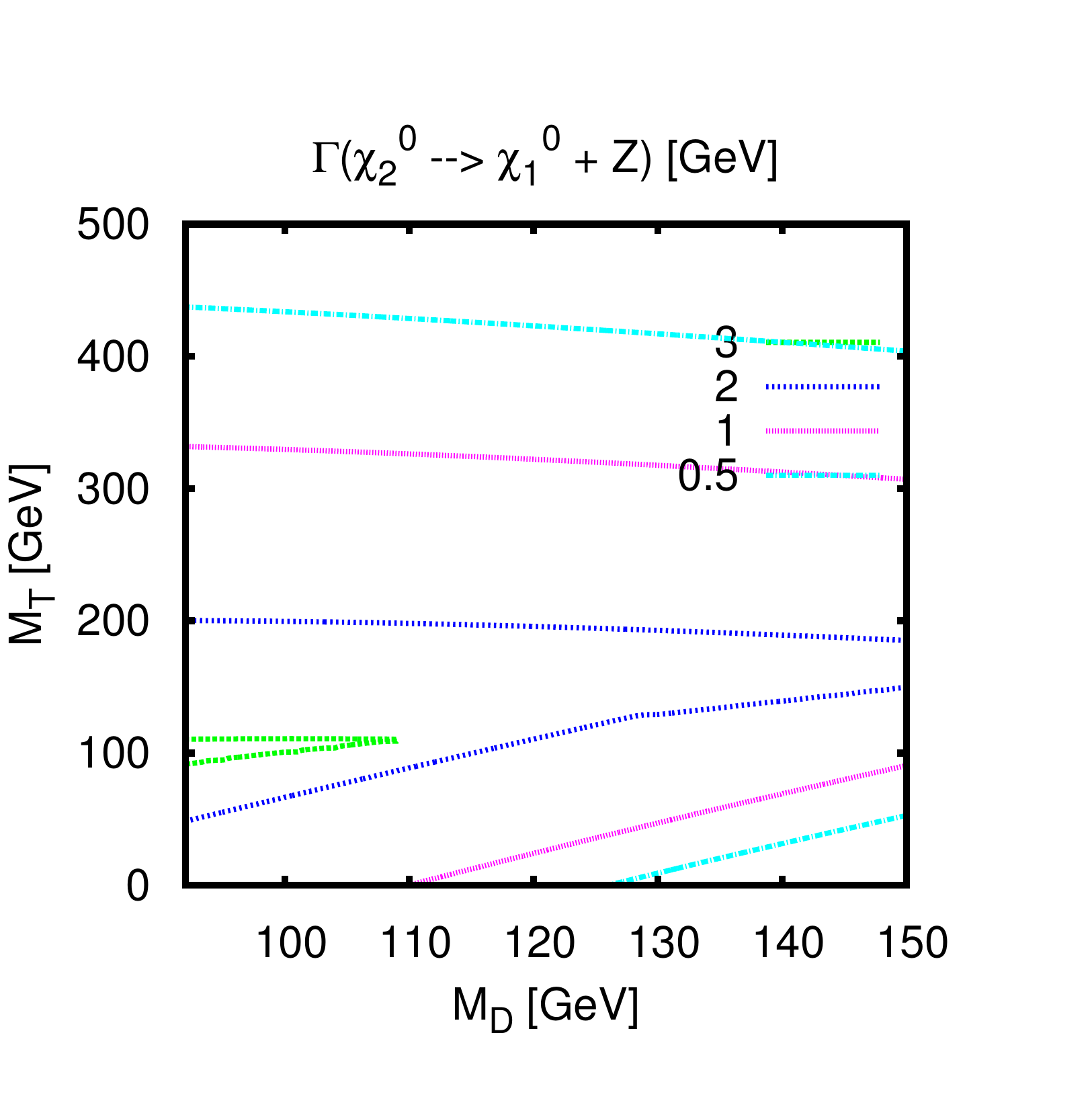}}
   %\\
   %\includegraphics[width=80mm]{contours-chi2tochi0W-M1=M2=200.eps} &
   %\includegraphics[width=80mm]{contours-chi2tochi01W-M1=M2=200.eps}
%   \end{tabular}
   \caption{\sl Contour plots for the 
   decay rates [in GeV] for the processes $\chi_{1}^{+}\to \chi^{0}_{1} + W^{+}$ (left)
   and $\chi_{2}^{0}\to \chi_{1}^{0} + Z$ (right). 
   We assume $m=200$ GeV.}
   \label{fig:chidecays}
\end{figure}

Analytically we find  the decay widths~\cite{Gunion:1987yh,Dreiner:2008tw}:
%%%%%%%%%%%%%%%%%
\begin{align}
%\Gamma(\chi^{+}_{i} \to \chi^{0}_{j} + W^{+}) &= \frac{g^{2} \sqrt{\lambda}}{32\pi m_{\chi^{+}_{i}}^{3} m_{W}^{2}} \, \biggl \{ (|O^{L}_{ji}|^{2} + |O^{R}_{ji}|^{2}) \,
%\biggl [\lambda + 3\, m_{W}^{2} 
%(m_{\chi^{+}_{i}}^{2} + m_{\chi^{0}_{j}}^{2} - m_{W}^{2}) \biggr]  \\ \nonumber 
%&- 12\, \Re e(O^{L}_{ji} O^{R\,*}_{ji})\, m_{\chi^{+}_{i}} m_{\chi^{0}_{j}} \, m_{W}^{2} \biggr \} \;,
\Gamma(\chi^{+}_{i} \to \chi^{0}_{j} + W^{+}) &= \frac{g^{2} \, m_{\chi^{+}_{i}}}{32 \pi}\, 
\lambda^{1/2}(1,r_{W}, r_{j}) \, \biggl \{  (|O^{L}_{ji}|^{2} + |O^{R}_{ji}|^{2})\,
  \biggl [1 + r_{j} - 2 r_{W} + (1-r_{j})^{2}/r_{W}  \biggr ]
\nonumber \\[2mm] &- 12\, \sqrt{r_{j}} \, \Re e(O^{L\,*}_{ji} O^{R}_{ji}) \biggr \}\;,
\nonumber \\[2mm] 
\Gamma(\chi^{0}_{i} \to \chi^{0}_{j} + Z) &= \frac{g^{2} \, m_{\chi^{0}_{i}}}{16 \pi c_{W}^{2}}\, 
\lambda^{1/2}(1,r_{Z}, r_{j}) \, \biggl \{  |O^{\prime\prime \, L}_{ij}|^{2}\,
  \biggl [1 + r_{j}^{0} - 2 r_{Z} + (1-r_{j}^{0})^{2}/r_{Z}  \biggr ]
\nonumber \\[2mm] &+ 6\, \sqrt{r_{j}^{0}} \, \Re e[(O^{\prime\prime\, L}_{ij})^{2}] \biggr \}\;,
\end{align}
%%%%%%%%%%%%%%%%%%%%
where 
%%%%%%%%%%%%%%%
\begin{subequations}
\begin{align}
r_{W} \equiv m_{W}^{2}/m_{\chi_{i}^{+}}^{2}\;, & \quad  r_{Z} \equiv m_{Z}^{2}/m_{\chi_{i}^{0}}^{2}\;, 
 \quad r_{j} \equiv m_{\chi_{j}^{0}}^{2}/m_{\chi_{i}^{+}}^{2} \;, 
 \quad  r_{j}^{0} \equiv m_{\chi_{j}^{0}}^{2}/m_{\chi_{i}^{0}}^{2}  \\[1mm]
& \lambda(x,y,z) \equiv x^{2} + y^{2}  + z^{2} -2 x y - 2 x z - 2 y z \;.
\end{align}
\end{subequations}
%%%%%%%%%%%%%%%%   
Numerical results for the decay widths for the processes (\ref{rw}) and (\ref{rz}) 
in the area of interest 
are depicted in Fig.~\ref{fig:chidecays}(a) and (b), respectively.
%
%%%%%%%%%%%%%%%%%%%%%%%%%
%%%%%%%%%%%%%%%%%%%%%%%%%%%%%%%%%%
%
Both decay widths behave similarly. In the area $M_{D} \approx M_{T} \approx 100$ GeV
we observe maximum values 
$\Gamma \approx 3$ GeV.  As $M_{T}$ is increases or decreases,
the widths get smaller than $1$ GeV. 
This is easily   understood  if we look back at the mass difference 
$|m_{\chi_{2}^{0}}|-|m_{\chi_{1}^{0}}|$ in Fig.~\ref{fig:masses}(a)
and recall that for the parameter considered  in Fig.~\ref{fig:chidecays}, it is 
 $m_{\chi_{1}^{0}} = M_{D}$ and $m_{\chi_{2}^{0}} = m_{\chi_{1}^{\pm}}$.

For heavier charged fermions, new decay channels  include
%%%%%%%%%%%%%%%%%%
\begin{subequations}
\begin{align}
\chi_{2}^{+} &\rightarrow \chi_{1}^{+} + Z\;, \\
\chi_{2}^{+} &\rightarrow \chi_{1}^{+} + h \;,
\end{align}
\end{subequations}
%%%%%%%%%%%%%%%%%%%%%%%%%%%
that are mostly kinematically allowed in the low $M_{D} \approx 100$ GeV but high 
$M_{T} \gtrsim 220$ GeV  regime.
For the heavier neutral particles, if kinematically allowed they would decay
to $W,Z$-gauge bosons and/or the Higgs boson,  
%%%%%%%%%%%%%%%%%%
\begin{subequations}
\begin{align}
\chi_{3}^{0} &\rightarrow \chi^{\pm}_{1} + W^{\mp} \;, \\
\chi_{3}^{0} &\rightarrow \chi_{2}^{0} + Z \;, \\
\chi_{3}^{0} &\rightarrow \chi_{2}^{0} + h \;.
\end{align}
\end{subequations}
%%%%%%%%%%%%%%%%%%%%% 

%%%%%%%%%%%%%%%%%%%%%%%%%%%%%%%%%%%%%
\section{Conclusions and Future Directions}
\label{sec:conclusions}
%%%%%%%%%%%%%%%%%%%%%%%%%%%%%%%%%%%%%

Our  motivation for writing this paper is to import a simple DM sector in the 
SM with particles in the vicinity of the electroweak scale responsible for
the observed DM relic abundance, preferably not relying on co-annihilations
or resonant effects, and capable of escaping current detection 
from nucleon-recoil   experiments. Meanwhile, we study consequences of this model 
in EW observables and Higgs boson decays ($h\to \gamma\gamma, Z\gamma$)
and other possible signatures at LHC.  

This SM extension 
consists of two fermionic $SU(2)_{W}$-doublets with opposite hypercharges and
a  fermionic $SU(2)_{W}$-triplet with zero hypercharge.
The new interaction  Lagrangian is given in \eq{LDM}, and contains both Yukawa 
trilinear terms together with explicit mass terms for the doublets and triplet fields.
Under the assumption of 
a certain global $SU(2)_{R}$-symmetry, discussed in section~\ref{sec:symmetry},
 that rotates $H$ to $H^{\dagger}$ and 
$\bar{D}_{1}$ to $\bar{D}_{2}$,  the two 
Yukawa couplings become equal with certain consequences that 
capture our interest throughout  this work. 
After electroweak symmetry breaking this sector widens the SM with
two charged  Dirac fermions and
 three  neutral Majorana fermions, the lightest ($\chi_{1}^{0}$) 
of which plays the role of the DM particle.
Under the symmetry assumption and for Yukawa couplings comparable to top-quark,
the lightest neutral particle ($\chi_{1}^{0}$) may 
have mass equal to the vector-like mass of the doublets, $M_{D}$, 
and its field composition contains only
an equal amount of the two doublets [see Fig.~\ref{fig:comp}]. As a result, 
the couplings of the Higgs and the $Z$ bosons to the lightest neutral fermion
pair vanish at tree level.  

Within this framework  we observe in Fig.~\ref{figos}, that $\Omega_{\chi} h^{2}$,
is in accordance  with observation [\eq{eq:omobs}] provided that the parameters of the model, 
$M_{D}, M_{T}$ and $m$, lie  naturally   at the EW scale \ie without the need for
resonant or co-annihilation effects.
Moreover, the $\chi_{1}^{0}$-nucleon SI cross section appears at one-loop,
turns out to be around 1-100 times smaller than the current experimental sensitivity
from LUX and XENON1T as it is shown in Fig.~\ref{fig:directX}.
In addition, we find that the oblique electroweak parameters $S,T$ and $U$ are 
all compatible with EW data fits as it is shown in Fig.~\ref{fig:STU},
a result which is partly a consequence of the global symmetry exploited.

We  also look for direct implications at the LHC.  We find that the existence 
of the extra charged fermions reduces substantially  the ratios
of the Higgs decay to di-photon (see Fig.~\ref{fig:h2gg}) and to $Z\gamma$ w.r.t the SM.
This is a certain prediction of this scenario that cannot be avoided by 
changing the parameter space. For very large Yukawa coupling, 
this reduction maybe of up to 65\% relative  to the SM expectation as we
obtain from Fig.~\ref{fig:h2gg}.  
Furthermore, the production and decays of those new charged/neutral fermion states, 
is within current and forthcoming LHC  reach.  
Decay rates for some of these states are shown in Fig.~\ref{fig:chidecays}. 

We should notice here that the minimality of the  Higgs sector
together with the $Z_{2}$-parity symmetry  preserves the appearance 
of new flavour changing or CP-violating effects beyond those of the SM,
for up to two-loop order (for a nice discussion of effects on EDMs from the 
charged fermions, see \Ref{Fan:2013qn}). 

On top of  collider/astrophysical constraints,  we  made an estimate of 
the consequences of the new states to vacuum stability of the model. 
The 1-loop result for the UV cutt off scale, above  which the model
needs some completion, is given in Fig.~\ref{LambdaUV}. We see that for the parameter
space of interest,  new physics, probably in the form of new, supersymmetric, scalars is needed 
already nearby the  TeV or multi-TeV scale to cancel fermionic 
contributions in the quartic Higgs coupling.
For example,  this solution may take the form of an MSSM extension with
 $\bar{D}_{1,2}$ and $T$ superfield (extensions with   
 a triplet superfield have been explored in  \Ref{Delgado:2012sm}).

%{\bf 
% Future work} may include
%LHC monojet signatures,  baryogenesis, indirect DM searches,  neutrino masses etc
%CP-violations; relax $M_{D}$ reality

In summary, in this work we basically studied the synergy between three observables:
$\Omega_{\chi} h^{2}, \sigma^{SI}_{0}$, and  $R(h\to \gamma\gamma)$, in a simple 
fermionic DM model.
If charged fermion states are discovered at the second run of LHC
and are compatible with $\Omega_{\chi} h^{2}$ with $m_{\chi}\sim m_{Z}$, 
then $R(h\to \gamma\gamma)$
has to be suppressed \ie $R$ will turn towards the CMS central value. 
If instead $R(h\to \gamma\gamma)\gtrsim 1$ is enhanced, then the DM particle is heavy, 
$m_{\chi}\sim 1$ TeV, or otherwise  
excluded by direct DM detection bounds. If $R\sim 1$, then one has to go to large $M_{T}$ values
where, however,  $\Omega_{\chi} h^{2}$ is only barely compatible with $m_{\chi} \simeq m_{Z}$. 
In this latter case,  the mass of the DM particle may be  below the EW gauge boson masses.
 However, in this case an entire new analysis  is required.

Apart from studying the regime with  mass $m_{\chi}$  lower than $M_{Z}$, 
this work can be extended in several ways as for example,
to investigate the role of $CP$-violating phases of $M_{D}$ on baryogenesis.
Indirect DM searches could be also an interesting avenue together with 
extensions of the Higgs sector.   We postpone all these interesting phenomena for future study.

%%%%%%%%%%%
\section*{Acknowledgements}

{\it We are grateful to Susanne Westhoff for helping us fiinding a mistake 
in  our formulae in the Appendix A. Our results for nucleon-WIMP cross section, depicted
in Fig.~10, are now  in good agreement with \Ref{Freitas:2015hsa}.}
A. D. would like to thank, M. Drees for useful comments,
 A. Barucha for discussions on the ongoing 
LHC chargino searches, C. Wagner for drawing our attention to \Ref{Carena:2004ha},
and F. Goertz for discussions on custodians. We are grateful to J. Rosiek 
for letting us using his code for squaring matrix elements and to compare with 
our analytical results. Our numerical routines for matrix diagonalization and for loop
functions follow  closely those in \Refs{Hahn:2006hr,Hahn:2006qw}, respectively.

This research Project  is co-financed by the European Union -  
European Social Fund (ESF) and National Sources, in the framework of the  
program ``THALIS" of the ``Operational Program Education and Lifelong  
Learning" of the National Strategic Reference Framework (NSRF)  
2007-2013.
D.K. acknowledges full financial support from the research program ``THALIS".
%%%%%%%%%%%%%%%%%%%%%%%%%%%%%%%%

%%%%%%%%%%%%%%%%%%%%%%%%%%%%%%%%%%%%%%%%
% APPENDICES
%\newpage
%%%%%%%%%%%%%%%%%%%%%%%%%%%%%%%%%%%%%%%%
\renewcommand{\thesection}{Appendix~\Alph{section}}
\renewcommand{\theequation}{\Alph{section}.\arabic{equation}}

\setcounter{equation}{0}  % reset counter
\setcounter{section}{0}
\bigskip

%%%%%%%%%%%%%%%%%%%%%%%
\section{}
\label{sec:appA}

The 1-loop corrected vertex amplitude arises from 
(a) and (b) diagrams  depicted in Fig.~\ref{fig:1loophiggs} involving
vector bosons ($W$ or $Z$) and new charged ($\chi_{i=1,2}^{\pm}$) or neutral ($\chi_{i=1..3}^{0}$) 
fermions. It can be written as,
%%%%%%%%%%%%%%%
\begin{equation}
i \,\delta Y = \sum_{j=(a),(b)}  (i \,\delta Y_{j}^{\chi^{\pm}} + i\, \delta Y_{j}^{\chi^{0}})\;,
\end{equation}
%%%%%%%%%%%%%%%%%%%%%%%%
where
%%%%%%%%%%%%%%%%%%%%%%%%%%%%%%%%%%%%%%%%
\begin{subequations}
\begin{eqnarray}
i\, \delta Y_{(a)}^{\chi^{\pm}} &=&
- {g^2} \sum_{i,j=1}^{2} \biggr \{ \left( O_{1j} ^R \, O_{1i} ^{L*}\, 
Y^{h \chi^{-}_j \chi^{+}_i } \ + \ O_{1i}^R \, O_{1j}^{L*}\,  Y^{h  \chi^{-}_i \chi^{+}_j}  \right)\, 
I^{Wij} _1 \label{M1xc} \nonumber \\[1mm] 
&  + &   m_{\chi_i ^{+}}\: m_{\chi_j ^{+}}\, \left( O_{1j}^R\, O_{1i}^{L*}\, 
Y^{h \chi^{-}_i \chi^{+}_j  *} \ + \ 
O_{1i}^R\, O_{1j}^{L*}\, Y^{h \chi^{-}_j \chi^{+}_i * }  \right)\, I^{Wij} _2  \\[1mm]
&+&  \left[ O_{1j}^L\, O_{1i}^{L*} \left( m_{\chi_i ^{+}}\,  Y^{h \chi^{-}_i \chi^{+}_j *} + 
m_{\chi_j ^{+}}\,Y^{h \chi^{-}_j \chi^{+}_i } \right ) \right.  \nonumber \\[1mm]
& + & \left. O_{1i}^R \, O_{1j}^{R*} \, 
\left( m_{\chi_i^{+}} Y^{h \chi^{-}_j \chi^{+}_i *} 
+  m_{\chi_j^{+}}Y^{h \chi^{-}_i \chi^{+}_j } \right) \right ]\, I^{Wij} _3    \biggr \},  
\nonumber \\[3mm]
%%%%%%%%%%%%%%%%%%%%%%%%%%%%%%%%%%%%%%
i\, \delta Y_{(a)}^{\chi^{0}} 
& = & \dfrac{g^2}{c_W ^2 }\, \sum_{i,j=1}^{3} \biggr \{ 
 O_{j1}^{\prime \prime L}\, O_{i1}^{\prime \prime L } \,Y^{h \chi^{0}_i \chi^{0}_j }
%\ + \ O_{1j}^{\prime \prime L} \, O_{1i}^{\prime \prime L }\, Y^{h \chi^{0}_i \chi^{0}_j * } \right) 
\, I^{Zij} _1 
%\nonumber \\[1mm]
 +  m_{\chi_i ^{0}}\, m_{\chi_j ^{0}}\, \ O_{i1}^{\prime \prime L}\, O_{j1}^{\prime \prime L} 
\, Y^{h \chi^{0}_i \chi^{0}_j *} %\ + \ O_{1j}^{\prime \prime L} \, O_{1i}^{\prime \prime L } \, 
%Y^{h \chi^{0}_i\, \chi^{0}_j }  \right)
\, I^{Zij} _2 
 \nonumber  \\[1mm]
&-&  O_{1j}^{\prime \prime L}\,  O_{i1}^{\prime \prime L} \left( m_{\chi_i ^{0}} 
Y^{h \chi^{0}_i \chi^{0}_j *} + m_{\chi_j^{0}} Y^{h \chi^{0}_i \chi^{0}_j } \right) 
%\right. 
%\nonumber \\[1mm]
% & + & \left.
% O_{j1}^{\prime \prime L}\, O_{1i}^{\prime \prime L} \left( m_{\chi_j ^{0}} Y^{h \chi^{0}_i \chi^{0}_j * }
 % + m_{\chi_i ^{0}} Y^{h \chi^{0}_i \chi^{0}_j } \right) \right ] 
 \, I^{Zij} _3    \biggr \},
\label{M1x0}  \\[3mm]
%%%%%%%%%%%%%%%%%%%%%%%%%
i\, \delta Y_{(b)}^{\chi^{\pm}} &=& 
- \dfrac{\sqrt{2}\, g^2 m_W ^2 }{v} \sum_{i=1}^{2} 
\biggr[ \left( |O_{1i}^L|^{2}  \ + \ |O_{1i}^R|^{2} \right)\,   I_4 ^{Wi} 
\ + \   2 \, m_{\chi_i ^{+}} \, O_{1i}^{L*}\, O_{1i}^{R}  
\,  I_5 ^{Wi} \biggr],   \label{M2xc} 
 \\[3mm]
 %%%%%%%%%%%%%%%%%%%%%%%%%%%%%% 
i\, \delta Y_{(b)} ^{\chi^{0}} &=& - \dfrac{\sqrt{2} g^2 m_Z^2 }{c_W^2   \,v } 
\sum_{i=1}^{3} \biggr \{  O_{i1}^{\prime \prime L} \,  O_{1i}^{\prime \prime L} \, I_4^{Zi}  \ - \ 
m_{\chi_i ^{0}} \, (O_{i1}^{\prime \prime L})^{2}  \,   I_5 ^{Zi} \biggr \} \;, \label{M2x0}
\end{eqnarray}
\end{subequations}
%%%%%%%%%%%%%%%%%%%%%%%%%%%%%%%
where the  integrals, $I_{1...5}^{V}$, are defined in terms of Passarino-Veltman (PV) 
functions~\cite{Passarino:1978jh} as, 
%
 %    
%where $\mu$ is the energy scale. The final form of the above integrals is: 
 %%%%%%%%%%%%%%%%%%%%%%%%%%%%%%%%%%%%%%%    
%\begin{subequations}
\begin{eqnarray}
I_1 ^{Vij} &=&  
(D-1)\, m_i^2 \, C_{0}(-p,p,m_i,m_V,m_j) - \dfrac{m_i ^2}{m_V^2} \, B_0 (0,m_i,m_j) \label{I1f} 
\nonumber \\[2mm] 
&+& (D-1) \, B_0 (p,m_V,m_j) - \dfrac{1}{m_V^2} \, A_0 (m_j) \;, 
 \end{eqnarray}
%%%%%%%%%%%%%%%%%%%%%%%%%%%%%%
\begin{eqnarray}
I_2 ^{Vij} =  (D-1)\,  C_{0}(-p,p,m_i,m_V,m_j) - \dfrac{1}{m_V^2}\, B_0 (0,m_i,m_j),\label{I2f} 
\end{eqnarray}
%%%%%%%%%%%%%%%%%%%%%%%%%%%%%%%
\begin{eqnarray}
I_3 ^{Vij} &=& 
\left( D - 2 +\dfrac{m_i^2}{m_{V}^2} - \dfrac{m_{\chi_1 ^0}^2}{m_V^2}  \right)\,
m_{\chi_1^0} \, [C_{11}(-p,p,m_i,m_V,m_j) - C_{12}(-p,p,m_i,m_V,m_j)]
\nonumber  \\[1mm]
 &+&
  \left ( 1+\dfrac{m_i ^2}{m_{V}^2} - \dfrac{m_{\chi_1 ^0}^2}{m_V^2}  \right)\,m_{\chi_1 ^0}\,
   C_{0}(-p,p,m_i,m_V,m_j) 
   -\dfrac{m_{\chi_1 ^0}}{m_V^2}\, 
  B_1 (p,m_V,m_j) 
  \nonumber \\[1mm]
  & + & \dfrac{m_{\chi_1 ^0}}{m_V^2}\, B_0 (0,m_i,m_j)\;,
  \end{eqnarray}
%%%%%%%%%%%%%%%%%%%%%%%%%%%%%%%%%%%%%%%%%%  
\begin{eqnarray}
I_4 ^{Vi} &=& \left( 2-D -\dfrac{m_i ^2}{m_{V}^2} + \dfrac{m_{\chi_1 ^0}^2}{m_V^2}\,  
\right) m_{\chi_1 ^0} \left [C_{11}(p,-p,m_V,m_i,m_V) - C_{12}(p,-p,m_V,m_i,m_V) \right ]
 \label{I4f} \nonumber \\
& - & (D-3) \, m_{\chi_1 ^0}\, C_{0}(p,-p,m_V,m_i,m_V) \ + \ \dfrac{m_{\chi_1 ^0}}{m_V ^4}\, (m_i ^2 - m_{\chi_1 ^0}^2)\,  B_1 (p,m_V,m_i) \nonumber \\ &-&  \dfrac{m_{\chi_1 ^0}}{m_V^4} \, A_0 (m_i)\;,
\end{eqnarray}
%%%%%%%%%%%%%%%%%%%%%%%%%%%%%%%%%%%%%%%%%
\begin{eqnarray}
I_5 ^{Vi} &=& (D-1) \, C_{0}(p,-p,m_V, m_i, m_V) \ + \  \dfrac{1}{m_V^4}A_0 (m_i) \;, \label{I5f}
\end{eqnarray}
%\end{subequations}
%%%%%%%%%%%%%%%%%%%%%%%%%%%%%%%%%%%%%%%%%%%
where $D\equiv 4- 2 \,\epsilon \, \delta_{\overline{\mathrm{MS}}}$ and 
$\delta_{\overline{\mathrm{MS}}} =1$ for $\overline{\mathrm{MS}}$ and 
$\delta_{\overline{\mathrm{MS}}}=0$ for $\overline{\mathrm{DR}}$ scheme.
All external particles (i.e., $\chi_{1}^{0}$) 
are taken on-shell and $m_i = m_{\chi_i^0} $ for $V = Z$ and 
$m_i = m_{\chi_i^{\pm}} $ for $V = W$. Our notation for PV-functions  $A,B,C$,
 follows closely the one  defined in the Appendix of \Ref{Axelrod:1982yc}.
Functions $A_{0},B_{0},B_{1}$ contain both infinite and finite parts while 
$C_{0}, C_{11}, C_{12}$ - functions are purely finite.  
Our calculation has been done in unitary and (for a cross check) in Feynman gauge.
The result for - $i \,\delta Y$ -  is both renormalization scale invariant and finite.

%%%%%%%%%%% BIBLIOGRAPHY %%%%%%%%%%%%%%%%%
\bibliography{DarkMatter-Biblio}{}
\bibliographystyle{utcaps}
%\bibliographystyle{plain}
%%%%%%%%%%%%%%%%%%%%%%%%%%%%%%%%

\end{document}